**Master's Thesis**

# Quantized Polar Code Decoders: Analysis and Design

Joachim Neu

Munich, September 2018

Supervisor: Dr. Gianluigi Liva

Master's Thesis

Institute for Communications Engineering
Technical University of Munich (TUM)

Title: Quantized Polar Code Decoders: Analysis and Design
Author: Joachim Neu

Supervisor: Dr. Gianluigi Liva

Munich, September 2018

# Acknowledgements

I thank my supervisor Gianluigi Liva for his excellent academic guidance and cheerful temper which rendered working on this thesis both productive and fun. I thank Mustafa Coşkun for being my peer in this endeavor and for his patience in bouncing countless ideas back and forth. I thank Rüdiger Urbanke who hosted me for part of this work at École Polytechnique Fédérale de Lausanne (EPFL) under the Summer@EPFL program. Finally, I am deeply indebted to Gerhard Kramer who guided and supported me throughout my time as a master's student at Technische Universität München (TUM).

# Abstract

Applications of massive machine-type communications, such as sensor networks, smart metering, 'internet-of-things', or process and factory automation, are forecast to have great economic impact in the next five to ten years. Low-complexity energy- and cost-efficient communication schemes are essential to enable large-scale deployments. To target these requirements, we study decoding of polar codes with coarse quantization in the short block length regime. In particular, we devise schemes to mitigate the impact of coarse quantization, which has not been adequately explored in prior works.

We introduce the 3-level quantized successive cancellation (SC) and SC list (SCL) decoder. Coarse quantization of log-likelihood ratios (LLRs) leads to quantization of path metrics (PMs). Quantized PMs severely impact the list management of SCL decoders, and hence cause performance degradation. Two mitigation strategies are presented: 1.) Selecting the winning codeword from the decoder's list based on maximum-likelihood (ML) rather than PM. 2.) Utilizing statistical knowledge about the reliability of bit estimates in each decoding step to improve list management.

We demonstrate the effectiveness of our techniques in simulations. In particular, our enhancements prove useful in the low code rate regime, where theory available in the literature predicts pronounced losses caused by quantization. Furthermore, we put our work into perspective by comparing it to finer quantization and partially unquantized schemes. This yields insights and recommendations as to which quantization schemes offer the best cost-benefit ratio for practical implementation.

# Contents

# 1. Introduction

In this chapter, (a) we sketch the big picture of current trends and developments in wireless communications which necessitate the analysis of polar code decoding using coarsely quantized signals (Section 1.1), (b) we recapitulate prior works and results in this context (Section 1.2), (c) we state our main research objectives and results (Section 1.3), and, (d) we outline the remainder of this thesis (Section 1.4).

## 1.1. Why Study Quantized Polar Decoding?

Past decades have witnessed an exponential increase in data rate transferred in the wireless medium, largely due to a massive increase in bandwidth per terminal. While bandwidth per terminal saturates, future expansion is expected to take place in the form of a drastic increase in number of terminals. Applications of *massive machine-type communications* (mMTC) such as sensor networks, smart metering, 'internet-of-things', or process and factory automation drive this demand [Boc+14; Boc+16; Dut+17]. Some analysts forecast these applications to have an impact of up to 10 trillion USD annually on the world economy by the year 2025 [PST18; Man+15]. While such numbers should be taken with a grain of salt, they highlight the economic relevance of the technology. These applications have requirements different from those of mobile broadband: (1) *Low data rates* and packet sizes due to the type of data (*e.g.*, sensor readings, telecommands). (2) *Low latency* for real-time applications. (3) *High reliability*, as latency might not permit retransmissions. (4) *Low cost* to make large-scale deployments affordable. (5) *Low energy consumption* to enable battery-powered operation and/or energy harvesting. In this thesis, we analyze and engineer communication (sub-)systems that target these requirements in the following way:

- *Short block lengths* of data frames in the range of a few tens to hundreds of bits are considered, to account for (1) low data rates and (2) low latency. No special attention is given to (3) high reliability, *i.e.*, we do not consider the *ultra-reliable machine-type communications* (uMTC)/*ultra-reliable low latency communications* (URLLC) regime [Boc+16; Shi+18]. Instead, we target frame error rates of $10^{-3}$, as typical for mobile broadband scenarios.

- *Polar codes* are used to protect transmitted data frames against corruption caused by channel noise. Polar codes are competitive channel codes for short block lengths [Liv+16; Shi+18] and recently receive increasing interest from practitioners after having been standardized for *fifth generation* (5G) mobile networks [BCL18]. Theorists appreciate polar codes for their conceptual simplicity and elegance, and good amenability to rigorous mathematical analysis. In June 2018, it was announced that Erdal Arıkan, arguably the pioneer of polar coding, would receive the information theory community's highest distinction, the Shannon Award 2019.

- *Quantization* is inevitable in digital communication systems, yet much theory is developed under the assumption of full-precision rather than quantized signals. To achieve (4) low cost and (5) low energy consumption, the complexity of the communications equipment needs to be kept low. *Coarse quantization* with accordingly low resolution of signal levels reduces the amount of components (and hence the complexity, cost and energy consumption) required for the decoder logic.

We highlighted how the study of coarsely quantized polar decoding fits well into the big picture of economic and application developments and requirements, is a prerequisite in demand for adequate technical solutions, and attractive from a theoretical perspective.

## 1.2. Related Work

Polar coding techniques were pioneered by N. Stolte [Sto02] and E. Arıkan [Arı09]. The seminal result that polar codes achieve capacity under *successive cancellation* (SC) decoding for the large and practically relevant class of binary-input memoryless symmetric channels is due to E. Arıkan [Arı09], who also coined the term *polar codes.*

The fundamental problem of polar code design is to estimate or approximate certain probability distributions and to calculate (or estimate) functionals thereof, which give an indication of reliable input bits to the polar transform that can be trusted with payload data. This task is complicated because the involved distributions are typically continuous (at least in the case of channels with continuous output alphabet). Initially, Monte-Carlo simulation was proposed [Arı09] to estimate the relevant functionals. Later, a density evolution [RU01; CRU01; Chu+01] based technique was introduced [MT09; Mor10]. However, for implementation it usually requires a discretization of the involved distributions. As pointed out in [TV13], this leads to the following predicament: An exact implementation of density evolution even for channels with finite output alphabet has exponential complexity in the codeword length and is hence intractable. Approximating the involved operations (*e.g.*, by rounding) could reduce the complexity, but it is at first unclear how it distorts the result and to which degree the latter can still be trusted. Both [TV13; Ped+11] devise techniques that enable low-complexity code design while controlling the error introduced by approximations. Note that some authors speak of 'quantization' in the context of discretizing the distributions, *i.e.*, for an 'implementation detail' of the density evolution method, which has no correspondence in the communication system under investigation. We avoid this use, as we utilize 'quantization' in a different context, to mean an intentional signal processing step taking place in the analyzed/engineered coding system. These two uses should not be confused.

State-of-the-art polar coding systems use *successive cancellation list* (SCL) decoding [TV15], commonly in *log-likelihood-ratio* (LLR) domain [BPB15]. To boost the performance of polar codes in combination with SCL decoding, concatenation schemes have been proposed, such as with *cyclic redundancy checks* (CRC) [TV15], and with *parity checks* [WQJ16]. Furthermore, polar codes with dynamically frozen bits ('polar subcodes') [TM16] show promising performance under SCL decoding. Alternative decoding algorithms, such as *belief-propagation* (BP) decoding [Elk+18; Doa+18] are being investigated, as well as variants of decoding algorithms that are better suited for implementation, *e.g.*, due to parallelization [Sar+16; Has+17].

The effect of quantization (of decoder input and/or stored values and computations within the decoder) has been studied before [Gal63; RU01; LPK12; Mei+15] in the context of *low-density parity-check* (LDPC) codes [Gal63]. If sufficient information about the channel output is retained, coarse quantization of the LLR messages passing within a BP decoder for LDPC codes does little harm. The reason is that the (unquantized or finely quantized) channel output informs the decoder in each round of the iterative decoding process, exerting a correcting force on LLR values that deviate from the ideal value, be it because of wrong decoder decisions or quantization error. Decoding polar codes on the other hand is equivalent to message passing over a tree structure, where only the lowest layer is informed by the channel outputs, and upper layers are not subject to its correcting influence. Hence, the results and intuitions from LDPC codes cannot be directly transferred to polar codes.

In the context of polar coding, several authors comment on quantization in passing [Gia+16; BPB15; Ler+13] as they proceed to discuss hardware implementation details of polar decoders. They agree that uniform quantization with in the range of 4 to 6 bit suffices to achieve close to full-precision error correcting capability. This observation is typically made from experimentation as the authors' objective is to match full-precision decoding rather than to give an account of how (deliberate) coarse quantization affects performance. Other works [SCN14; SN14] devise a density evolution based technique to study the effects of quantization on polar code decoders (and to optimize the involved quantizer), however, also with the intent to match full-precision decoding rather than to examine (and potentially mitigate) the influence of coarse quantization.

To the best of our knowledge, the only work so far that studies the effect of coarse quantization on polar decoding is by S. H. Hassani and R. Urbanke [HU12a; HU12b; Has13]. They study an SC decoder with alphabet of cardinality three. This 'extreme' form of quantization is nonetheless appealing for theorists, as it represents 'the most extreme' quantization (hard thresholding, *i.e.*, using two quantization levels, seems to not exhibit channel polarization) and the resulting decoder resembles that for the *binary erasure channel* (BEC). The authors seem to suggest as conclusion of their analysis that the effect of coarse quantization is not dramatic:

"We show that even very coarsely quantized decoding algorithms lead to excellent performance. More concretely, we show that under successive decoding with an alphabet of cardinality only three, the decoder still has a threshold and this threshold is a sizable fraction of capacity." [HU12a, Abstract]

As evidence they present Figure 1.1, where there is a gap between capacity of unquantized decoding (red) and 3-level quantized SC decoding (black), but it seems small and perhaps tolerable in light of the huge complexity reduction of quantized decoding.

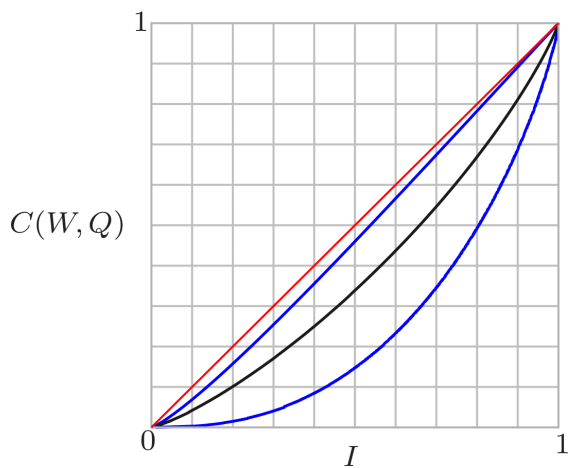

Figure 1.1.: Figure taken from [HU12a, Fig. 1]: "The maximum achievable rate, call it $C(W, Q)$, of a simple three message decoder, called the decoder with erasures, as a function of the capacity of the channel for different channel families. [...] [T]he [red] curve corresponds to the family of binary erasure channels (BEC) where the decoder with erasures is equivalent to the original SC decoder and, hence, the maximum achievable rate is the capacity itself. [...] The [black] curve corresponds to the family of binary additive white Gaussian channels (BAWGN). [...]"

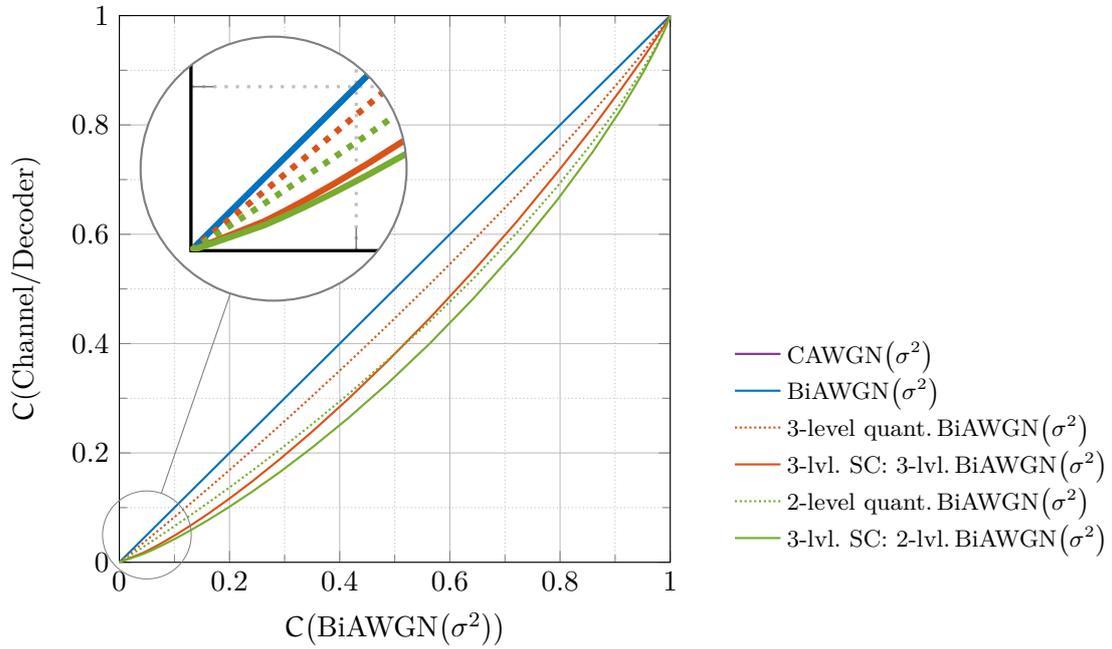

(a) Capacities as a function of $C\big(\mathrm{BiAWGN}(\sigma^2)\big)$ of equivalent $\mathrm{BiAWGN}(\sigma^2)$

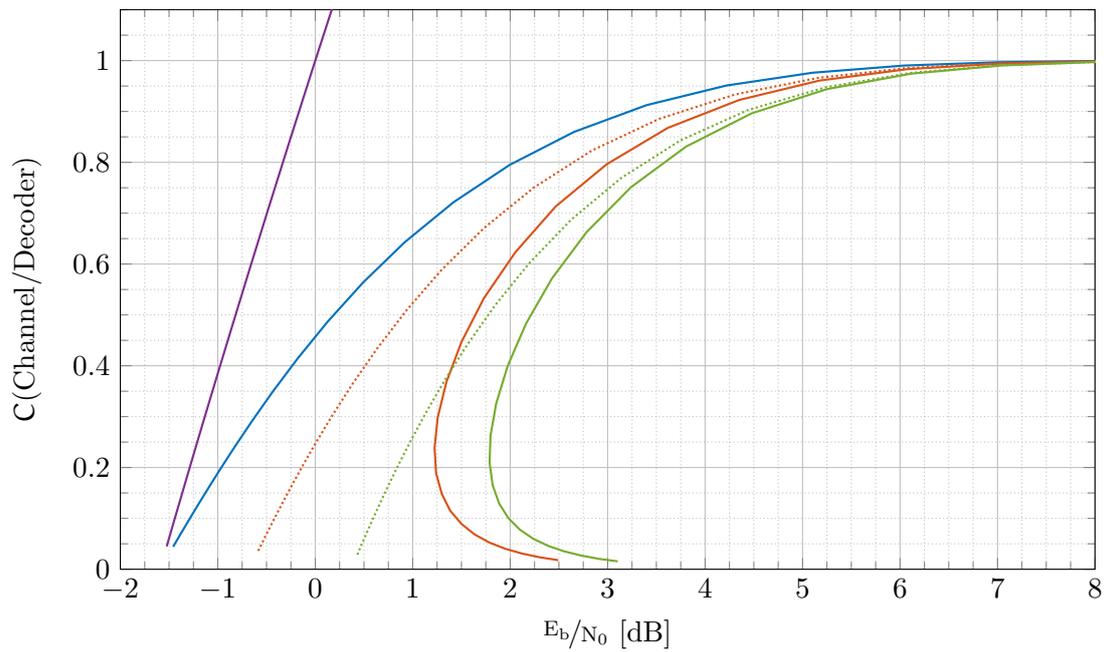

(b) Capacities as a function of $E_b/N_0$

Figure 1.2.: Reproduction of Figure 1.1, (a), and reparametrization in $E_b/N_0$, (b)

We reproduce their plot in Figure 1.2(a). Here, —— corresponds to the black curve of Figure 1.1, —— corresponds to the red curve, the remaining curves have no correspondence. Note that in [HU12a], 3-level quantization is used for LLR values within the SC decoder, but channel output LLRs are effectively 2-level quantized as one of the three quantization levels remains unused. As a result, —— should be compared to the capacity of the hard threshold BiAWGN, ⋯⋯. Adjusting the channel output LLR quantization in such a way that the third level is utilized, improves not only the capacity of the resulting quantized channel (cf. ⋯⋯ vs. ⋯⋯) but also of the 3-level quantized SC decoder over that respective channel (cf. —— vs. ——).

To evaluate the performance of channel coding schemes in terms of energy efficiency, it is common to plot rates and capacities parametrized in $E_b/N_0$, which measures energy per transmitted information bit in relation to the energy contained in the channel noise. Figure 1.2(a) is reparametrized in $E_b/N_0$ in Figure 1.2(b). Several observations are due here: At rate $R = 1/2$, 3-level quantization of the BiAWGN output causes a loss of 0.7 dB (cf. —— vs. ⋯⋯). Furthermore, 2-level quantization causes an additional loss of 0.8 dB (cf. ⋯⋯ vs. ⋯⋯), resulting in a total loss of more than 1.5 dB with respect to unquantized BiAWGN (cf. —— vs. ⋯⋯). SC decoding the 2-level and 3-level quantized channel output with 3-level quantized LLR messages leads to a loss of 2.0 dB and 1.4 dB, respectively, with respect to the unquantized BiAWGN capacity, *i.e.*, even for asymptotically large block lengths (cf. —— vs. ——, —— vs. ——).

At low rates, the losses increase slightly due to channel output quantization (cf. —— vs. ⋯⋯ and ⋯⋯), but drastically due to the increasing suboptimality of coarsely quantized SC decoding (cf. 'bending' of —— and —— 'to the right' in Figure 1.2(b)). At $R = 0.1$, *e.g.*, the 3-level quantized SC decoder performs more than 2.7 dB away from unquantized BiAWGN (cf. —— vs. ——) and still 1.9 dB away from the 3-level quantized BiAWGN (cf. ⋯⋯ vs. ——). In Figure 1.2(a), this effect is only indicated by the differences in slope of the curves as they approach the point $(0, 0)$.

We conclude that the losses in $E_b/N_0$ due to the use of coarse quantization in the SC decoder are sizable both at moderate code rates ($R \approx 1/2$), but in particular so at low code rates ($R \approx 0.1$). These impacts and low-complexity mitigation schemes have not yet been exhaustively studied in the available literature.

## 1.3. Research Objectives and Results

In Section 1.1, we motivate the need for low-complexity polar decoding based on coarsely quantized signals. We survey work and results related to this topic in Section 1.2, and conclude that the negative performance impact of coarse quantization and low-complexity mitigation schemes have not been investigated to the desirable extent and depth in present literature.

Hence, we set out to analyze and propose polar coding techniques for short block lengths that employ coarse quantization. The primary focus is not in the design of coding schemes for 'real' applications, but in exploring the limits imposed by quantization on 'traditional' decoders, and in devising low-complexity extensions or modifications that allow to go past them. We usually consider the polar code as given (as is the case, *e.g.*, after standardization) and modify solely the decoding procedure. For most of our work we adopt 3-level quantization as in [HU12a; HU12b; Has13]. This is suitable for our analysis as it represents the 'most extreme' form of quantization (2-level quantization is not feasible as channel polarization does not take place then) and hence promises the most unobstructed insight into the effects of quantization. Results and intuitions can be transferred to finer quantization.

We devise a number of quantized decoding techniques and analyze and empirically evaluate their performance in the short block length regime, usually in comparison to their unquantized counterparts[1] (cf. Chapter 3):

- *Three-level quantized SC decoding:* The losses are considerable and well in line with the predictions derived from the asymptotic results (cf. Figure 1.2).

- *Three-level quantized SCL decoding:* Gains considerable with respect to unquantized SC decoding, but as list decoding benefits unquantized decoding to a similar degree, the gap between (un)quantized decoding remains unaltered.

- *Three-level quantized SCL decoding with ML-among-list step:* Quantization of LLRs and path metrics in SCL decoding effect that the quantized SCL decoder does not declare the winning codeword from its list based on likelihood. Theory and experimentation suggest that path metrics in quantized SCL decoders cannot reliably guide the selection of the winning codeword. To overcome this, the winning codeword is declared in a subsequent maximum-likelihood-among-list (ML-among-list) step. Restoring the ML-among-list decision rule this way leads to considerable performance gains.

At this point it becomes clear that three factors influence the performance of quantized polar decoding: (1) Properties of the code itself, *e.g.*, its minimum distance, (2) quantization renders it difficult to select the most likely codeword from the decoder's list based on path metric at the end of the decoding process, (3) quantization causes that the correct codeword is more often inadvertently removed from the list during the decoding process. As (1) involves code design, it is out of the scope of this thesis. Point (2) is addressed by the aforementioned additional ML-among-list step. Hence, we target (3) in that we devise techniques to enhance the list management in SCL (cf. Chapter 4):

- *Expected path metric updates:* Due to LLR quantization, instantaneous reliability information of bit estimates in the decoder is lost. However, statistical knowledge, *e.g.*, mean reliability, is obtained through analysis and leveraged to inform decoding decisions, which leads to sizable performance improvements.

- *Expected path metric updates with contradiction counting:* Some quantities can be used as indicators of the instantaneous reliability of bit estimates (and hence be used to refine the statistical knowledge about the reliability), *e.g.*, the number of contradictions encountered in a decoding step. These behave profoundly different from LLRs in that computing and storing them has strictly lower complexity. We use contradiction counts to refine expected path metric updates. The gains encourage the examination of similar quantities in future works.

---

[1] Note that for the sake of brevity we cannot anticipate and state the exact circumstances leading to and the precise quantities of performance gains/losses here. Instead, we utter only qualitative statements of whether a technique was found to have a rather small/large positive/negative impact. You find the precise analyses in the respective succeeding sections of this thesis.

Subsequently, we evaluate the robustness of our findings and techniques by benchmarking them with respect to alternative methods or in new scenarios (cf. Chapter 5):

- *Low code rates:* In devising and analyzing quantized decoding techniques, we worked with codes of rate $R = 1/2$. However, Figure 1.2 suggests that low code rates are a more challenging regime where solutions are more sought after. We test our methods for $R = 37/256 \approx 0.145$ and confirm considerable gains.

- *Finer quantization:* Three-level quantization might seem 'extreme' from a practitioner's point of view. Hence, we compare 3- to 7-level quantization. Our simulation results suggest that few quantization bits suffice to match the performance of full-precision decoding and confirm according earlier findings in the literature.

- *The effect of a single quantized decoding step:* To gauge the impact of a single step of quantized decoding operations, we compare full 3-level quantized decoding with a scheme where the first decoding step uses full-precision and subsequent steps are 3-level quantized. This is equivalent to 'virtual' multi-stage decoding. Our findings suggest that finer quantization benefits early decoding stages more, while coarse quantization 'suffices' for later decoding stages.

In the course of our work, we touch upon a few loosely-related side topics:

- *Confidence intervals and termination criteria* for code simulations are presented.

- *A genetic algorithm for polar code design for SCL decoding* is sketched, which recovers Reed-Muller codes when they exist for given code parameters. Thus, it independently confirms current design approaches and suggests a natural extension for code parameters where no Reed-Muller codes exist.

- *Quantization threshold selection* strategies are in particular required for schemes based on 7-level quantization and virtual multi-stage decoding. We demonstrate that when the suboptimal behavior of quantized decoders is taken into account, the systems outperform those with quantization thresholds determined solely based on channel capacity considerations.

- The 3-level quantized decoder is capacity-achieving for BECs. Hence, insights from 3-level quantized SCL decoding of BECs might provide hints for the analysis of SCL decoding or the design of polar codes for list decoding. We provide empirical evidence suggesting a connection between the amount of mutual information 'lost' in information bits and the number of paths in the SCL decoder's list, and suggest starting points for further investigation.

To facilitate our inquiry, we created artifacts (using the Julia programming language, [Bez+17]) that we intend to make publicly available as open source software soon:

- A software library for performant manipulation of discretized probability distributions, as occur, *e.g.*, in density evolution.

- A generic (*e.g.*, data type agnostic) implementation of density evolution and related methods (*e.g.*, polar code design routines) that can be used in the design of quantized and unquantized polar coding systems.

- A versatile (*e.g.*, data type agnostic) and performant implementation of SCL decoding which is easily reused for various instances of quantized polar decoding.

## 1.4. Organization of this Thesis

The 'core' of this thesis consists of Chapters 3 through 5, which introduce most of the novel techniques and findings. This thesis is structured in analogy to the objectives put forward in Section 1.3:

- Chapter 1 motivates and positions our research in the context of current trends and developments in wireless communications.

- Chapter 2 settles our notation and revisits preliminaries of polar codes and their decoding. Furthermore, a genetic algorithm for polar code design for SCL decoding is briefly presented, and confidence intervals for code simulations are discussed.

- Chapter 3 introduces and evaluates 3-level quantized SC, SCL and SCL-ML decoding, and backs up the need for list enhancement techniques.

- Chapter 4 devises expected path metric updates, where statistical knowledge about the reliability of bit estimates is leveraged to enhance quantized decoders.

- Chapter 5 validates our findings in the low code rate regime, and compares them to competing approaches such as finer quantization.

- Chapter 6 establishes that and how the suboptimality of quantized decoders needs to be taken into account in quantization threshold selection.

- Chapter 7 concludes with recommendations for practitioners drawn from our findings, and potential directions for future research.

- Appendix A reports observations about 3-level quantized SCL decoding of BECs.

# 2. Preliminaries and Notation

We expect the reader to be well familiar with basic concepts of communications (*e.g.*, error probabilities, common channel models), coding theory (*e.g.*, linear block codes, maximum-likelihood decoding) and information theory (*e.g.*, entropy, mutual information, capacity, achievable rates), as well as with polar codes and their decoding. Consequently, the objective of this chapter is to establish a common terminology and notation rather than providing a comprehensive introduction.

Two sections stand out in level of detail as they have a more educational flavor: Section 2.5 introduces density evolution with a strict distinction between (a) the transformation of random variables and (b) the 'implementation' of this transformation on a chosen parametric representation of the involved distributions. Section 2.7 discusses confidence intervals for error rate estimation and termination criteria for code simulations, a methodological detail that we find worth pointing out.

## 2.1. Notational Conventions

For variables and constants we use the following notation based on 'type' and dimensionality of the involved object: Scalar constants, values and random variable realizations are denoted as lowercase regular Roman or Greek letters (*e.g.*, $u$, $c$, $\lambda$, $\sigma$). The corresponding random variables use uppercase letters (*e.g.*, $U$, $C$, $\Lambda$, $\Sigma$). Vector constants, values and random variable realizations are denoted as lowercase bold Roman or Greek letters (*e.g.*, $\boldsymbol{u}$, $\boldsymbol{c}$, $\boldsymbol{\lambda}$). The corresponding random variables use uppercase letters (*e.g.*, $\boldsymbol{U}$, $\boldsymbol{C}$, $\boldsymbol{\Lambda}$). Matrix constants and values are denoted as uppercase bold Roman or Greek letters (*e.g.*, $\boldsymbol{B}$, $\boldsymbol{F}$, $\boldsymbol{G}$, $\boldsymbol{H}$, $\boldsymbol{\Sigma}$), but appear seldom and almost exclusively in Section 2.2 and Appendix A. A confusion between matrix values and vector random variables is avoided by context. Random matrices are not used. Sets are conventionally denoted as calligraphic uppercase letters, *e.g.*, the domain of random variables $X$ and $\Lambda$ or variables $x$ and $\lambda$ is denoted as $\mathcal{X}$ and $\mathcal{L}$, respectively.

By $[i{:}j] \triangleq \{i, i+1, \ldots, j-1, j\}$ we denote the set of integers from $i$ to $j$ inclusively. Given a set of indices $\mathcal{I}$ (with implicit order of elements) we denote by $\boldsymbol{x}_{\mathcal{I}} \triangleq (x_i | i \in \mathcal{I})$ the vector of coefficients of column vector $\boldsymbol{x} \triangleq (x_0, \ldots, x_{n-1})^{\mathsf{T}}$ indexed by $\mathcal{I}$, and by $\boldsymbol{M_I} \triangleq [\boldsymbol{m_i} | i \in \mathcal{I}]$ the matrix of columns of matrix $\boldsymbol{M} \triangleq [\boldsymbol{m_0}, \dots, \boldsymbol{m_{n-1}}]$ indexed by $\mathcal{I}$. Furthermore, $\boldsymbol{x_i^j} \triangleq \boldsymbol{x_{[i:j-1]}}$, $\boldsymbol{x^j} \triangleq \boldsymbol{x_{[0:j-1]}}$, and $\boldsymbol{x_{\sim i}} \triangleq \boldsymbol{x_{[0:n-1] \setminus \{i\}}}$. These definitions carry over analogously to random variables and to columns of matrices.

Functions are usually denoted by $f$ with a suitably chosen subscript. Continuous probability densities are denoted as $p_{XY|Z}$ and $q_{XY|Z}$ or $p(x, y|z)$ and $q(x, y|z)$, discrete densities use $P_{XY|Z}$ and $Q_{XY|Z}$ or $P(x, y|z)$ and $Q(x, y|z)$. The aforementioned describe the conditional joint distribution of random variables $X$ and $Y$ given $Z$. We use $P_{XY|Z}$ and $P(x, y|z)$, respectively, depending on whether (a) the distribution as an object is itself subject to a mathematical operation (*e.g.*, in density evolution) or (b) the distribution is used as a function to evaluate probabilities, in the specific context. By $(P_X P_Y)$ we denote the distribution that is the product of the marginal distributions $P_X$ and $P_Y$, *i.e.*, $(P_X P_Y)(x, y) \triangleq P_X(x) P_Y(y)$. Expected value and variance of a random variable $X$ are customarily indicated as $\mathsf{E}[X]$ and $\mathsf{Var}[X]$, the probability of an event $E$ as $\mathsf{Pr}[E]$.

We use shorthands for common phrases, *w.l.o.g.* (without loss of generality), *RHS* (right-hand side of an (in)equality or equation), *w.r.t.* (with respect to), *w.p.* (with probability), *iff* (if and only if).

For information measures we use the conventional symbols and definitions [Gal68; CT06], *i.e.*, $\mathsf{H}(X|Z)$ for *discrete entropy* of discrete random variables $X$ given $Z$, $\mathsf{h}(X|Z)$ for *continuous entropy* of continuous random variables $X$ given $Z$, $\mathsf{I}(X; Y|Z)$ for *mutual information* of random variables $X$ and $Y$ given $Z$. All logarithms $\log(.)$ are base-2 unless explicitly stated otherwise. By $\mathsf{C}\left(P_{Y|X}\right)$ we denote the *capacity* of a channel with *channel law* $P_{Y|X}$ describing the distribution of the *channel output* $y \in \mathcal{Y}$ given the *channel input* $x \in \mathcal{X}$. A *rate* or *achievable rate* is denoted by $\mathsf{R}\left(P_{Y|X}\right)$. In addition to the channel law, other arguments can characterize the precise circumstances for which the capacity or achievable rate is specified (*e.g.*, block length or block error probability). We use abbreviations for the following decoders:

- *ML decoder* denotes the *maximum-likelihood decoder* [RU08].

- *SC decoder* refers to the *successive cancellation decoder* for polar codes [Sto02; Arı09], introduced in Section 2.3 and refined in Section 3.1.

- *SCL decoder* refers to the *successive cancellation list decoder* for polar codes [TV15], introduced in Section 2.4 and refined in Section 3.2.

- *SCL-ML decoder* refers to the *successive cancellation list decoder with ML-among-list selection* for polar codes, introduced in Section 3.3.

Different notions of *error probability/error rate* are used throughout this document, for which we use variations of the symbol $\mathsf{P}_e$. Commonly, we consider *block error probabilities/frame error rates* (FER), denoted by $\mathsf{P}_{e,B}$. A prefix can specify the precise error event for which the rate/probability is given:

- *PM-FER* ('Path-Metric-FER') refers to the FER of an SCL decoder which declares the winning codeword based on its path metric (precise definition follows in Section 2.4).

- *List-FER* is used for list decoders such as the SCL decoder, where it denotes the probability that the correct codeword is not included in the resulting list.

- *LML-FER* ('List-Maximum-Likelihood-FER') refers to the FER of a list decoder, such as the SCL decoder, which declares the winning codeword based on an ML decision among the resulting list.

The FER of interest is implicit from the context where no prefix is provided. For ML and SC decoders 'the' FER is unequivocal. PM-FER and list-FER are commonly of interest for SCL decoders, LML-FER is of interest for SCL-ML decoders.

Estimated quantities are typically decorated with a hat, *e.g.*, $\hat{\mathsf{P}}_e$ is an estimated error probability. Approximated objects such as functions are usually decorated with a tilde, *e.g.*, $\tilde{f}$ is an approximation of the function $f$. The Kronecker product is denoted by $\otimes$, the modulo-2 sum by $\oplus$. The indicator function $\mathbb{1}_{(.)}$ evaluates to 1 if the expression $(.)$ is satisfied and 0 otherwise; $\mathbf{0}$ denotes the all-zero vector. Optimizers, *i.e.*, solutions to optimization problems, are denoted by superscript asterisk, *e.g.*, $\delta^*$ is the optimal $\delta$.

We write $\boldsymbol{X} \sim \mathcal{N}_{\mathbb{R}}(\boldsymbol{\mu}, \boldsymbol{\Sigma})$ or $\boldsymbol{X} \sim \mathcal{N}_{\mathbb{C}}(\boldsymbol{\mu}, \boldsymbol{\Sigma})$ to indicate that $\boldsymbol{X}$ is a multivariate real Gaussian or multivariate circularly-symmetric complex Gaussian random variable with mean $\boldsymbol{\mu}$ and covariance matrix $\boldsymbol{\Sigma}$, respectively. The complementary Gaussian cumulative distribution function is given as

$$Q(x) \triangleq \int_x^\infty \frac{1}{\sqrt{2\pi}} \exp\left(-\frac{1}{2}t^2\right) \mathrm{d}t, \tag{2.1}$$

and its inverse as $Q^{-1}(x)$. We write $X \sim \mathrm{Bern}(\rho)$ to indicate that $X$ is a Bernoulli random variable with $P_X(1) = \rho$, $P_X(0) = 1 - \rho \triangleq \bar{\rho}$. The support $\mathrm{supp}(X)$ of a random variable $X$ is the subset of its alphabet $\mathcal{X}$ where $P_X(x)$ is strictly positive, *i.e.*, $\mathrm{supp}(X) \triangleq \{x \in \mathcal{X} \,|\, P_X(x) > 0\}$.

Finally, we fix the terminology regarding fundamental channel models. Further models are introduced as necessary throughout the text. We consider the class of *binary-input memoryless symmetric* (BMS) channels. This means, *memoryless:* for a sequence of $n$ channel uses, the distribution of the output $Y_i$ in time instance $i$ depends solely on the input $X_i$, and the channel law does not vary with $i$, *i.e.*,

$$P_{\boldsymbol{Y}|\boldsymbol{X}}(\boldsymbol{y}|\boldsymbol{x}) = \prod_{i=0}^{n-1} P_{Y|X}(y_i|x_i), \tag{2.2}$$

*binary-input:* codeword bits $C$ from $\mathcal{C} = \{0,1\}$ are modulated to input symbols $X$ from $\mathcal{X} = \{-1, +1\}$ using $X = (-1)^C$, and *symmetric:* the distributions $P_{Y|X}(y|-1)$ and $P_{Y|X}(y|+1)$ (and hence also $P_{Y|C}(y|0)$ and $P_{Y|C}(y|1)$) are identical up to a permutation on the output alphabet $\mathcal{Y}$. Where the channel under consideration is clear from the context, we use the conventional abbreviation $W$ rather than the full channel law $P_{Y|C}$. Symmetric channels achieve their capacity under uniform input distribution $C \sim \text{Bern}(1/2)$ (cf. [Gal68, p. 94, Theorem 4.5.2]), hence their capacity can be computed as $\mathsf{C}(W) = \mathrm{I}(Y; C)$ using the definition of mutual information.

Figure 2.1 illustrates the BMS *binary-input additive white Gaussian noise* (BiAWGN) channel. In each channel use, a bit $C$ is modulated to $X$ and an independent Gaussian noise sample $N$ is added, where $N \sim \mathcal{N}_{\mathbb{R}}(0, \sigma^2)$, to form the output $Y \triangleq X + N$. We denote the channel with its parameter $\sigma^2$ as BiAWGN($\sigma^2$). The channel law $p_{Y|X}$ is given through the Gaussian probability density function of the noise as

$$p_{Y|X}(y|x) = \frac{1}{\sqrt{2\pi\sigma^2}} \exp\left(-\frac{(y-x)^2}{2\sigma^2}\right). \tag{2.3}$$

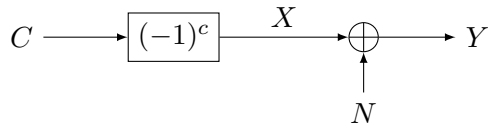

Figure 2.1.: Signal flow diagram of the BiAWGN($\sigma^2$) channel

The BiAWGN models *binary phase-shift keying* (BPSK) over a *continuous-input complex additive white Gaussian noise* (CAWGN) channel with noise power $\sigma^2$ per real-valued signal dimension, *i.e.*, total noise power $2\sigma^2$ for real and imaginary dimension. Denote this CAWGN as CAWGN($\sigma^2$), for which we have the famous result [Sha48],

$$\mathsf{C}\Big(\text{CAWGN}\big(\sigma^2\big)\Big) = \log(1 + \mathsf{SNR}), \tag{2.4}$$

where $\mathsf{SNR}$ denotes the *signal-to-noise ratio*, *i.e.*, the ratio of average signal energy per symbol $\mathrm{E_s} \triangleq \mathsf{E}\Big[|X|^2\Big]$ to single-sided noise power spectral density $\mathrm{N_0} \triangleq 2\sigma^2$. When the BiAWGN($\sigma^2$) is viewed as a CAWGN($\sigma^2$) with BPSK input constraint, the SNR is

naturally given as

$$\mathrm{E_s/N_0} \triangleq \mathsf{SNR} = \frac{1}{2\sigma^2}. \tag{2.5}$$

If a coding scheme of information bit rate $R$ is used over the channel, we can use the reparametrization $\mathrm{E_b} = \frac{\mathrm{E_s}}{R}$ to obtain

$$\mathrm{E_b/N_0} \triangleq \frac{1}{R}\mathrm{E_s/N_0}. \tag{2.6}$$

Note that $\mathrm{E_s/N_0}$ and $\mathrm{E_b/N_0}$ are usually not given as fractions but in logarithmic scale (decibels, dB) using the conversion

$$\mathrm{E_b/N_0}|_{\mathrm{dB}} \triangleq 10\log_{10}(\mathrm{E_b/N_0}). \tag{2.7}$$

Lastly, there is the BMS *binary(-input) error and erasure channel* (BEEC) with both discrete input $x \in \mathcal{X} = \{\mathtt{0},\mathtt{1}\}$ and output $y \in \mathcal{Y} = \{\mathtt{0},\mathtt{E},\mathtt{1}\}$. The channel law of the BEEC$(p,e)$ reads

$$P_{Y|X}(y|x) = \begin{cases} x & \text{w.p. } 1-p-e, \\ \mathtt{E} & \text{w.p. } e, \\ 1-x & \text{w.p. } p, \end{cases} \tag{2.8}$$

and is illustrated in Figure 2.2.

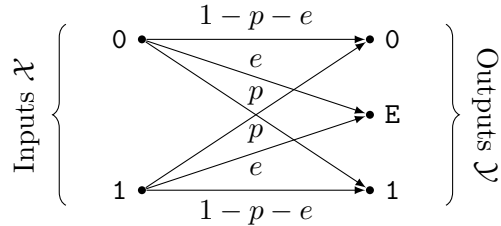

Figure 2.2.: Channel law of the BEEC$(p,e)$ channel

For a binary-input channel with $p_{Y|X}$ we can compute the *log-likelihood ratio* (LLR)

$$f_{\mathrm{LLR}}(y) \triangleq \log\left(\frac{p_{Y|X}(y|0)}{p_{Y|X}(y|1)}\right) \tag{2.9}$$

for a channel output $y \in \mathcal{Y}$, which we usually denote by $\lambda \triangleq f_{\mathrm{LLR}}(y)$, where $\lambda \in \mathcal{L}$. The definition is analogous for a channel with channel law $P_{Y|X}$. The LLR is a sufficient statistic for estimating whether 0 or 1 was transmitted (cf. Neyman-Pearson lemma

[Ash07]). LLRs for $\text{BiAWGN}(\sigma^2)$ and $\text{BEEC}(p, e)$ are

$$f_{\text{LLR,BiAWGN}(\sigma^2)}(y) \triangleq \frac{2}{\sigma^2}y, \quad f_{\text{LLR,BEEC}(p,e)}(y) \triangleq \begin{cases} \log\left(\frac{1-p-e}{p}\right) \triangleq +\Delta & \text{if } y = \mathsf{0}, \\ 0 & \text{if } y = \mathsf{E}, \\ \log\left(\frac{p}{1-p-e}\right) = -\Delta & \text{if } y = \mathsf{1}. \end{cases} \quad (2.10)$$

LLRs might be (un)quantized, which we denote by $\lambda^{(\text{q})}$ and $\lambda^{(\text{unq})}$, respectively.

## 2.2. Polar Codes

Polar codes were pioneered in [Sto02; Arı09]. We closely follow the notation of [BPB15; MT09; HU12a; Arı09]. Note that in this subsection modulation is considered part of the channel, *i.e.*, the channel law describes the relation between codeword bits $\boldsymbol{c}$ and channel output symbols $\boldsymbol{y}$.

The polar code construction is summarized as follows: Given are $n$ uses of a BMS $p_{Y|C}$. Using a special linear transform in the $n$-dimensional vector space over the binary field, the *polar transform* $\boldsymbol{G}_m$, a new channel $p_{\boldsymbol{Y}|\boldsymbol{U}}(\boldsymbol{y}|\boldsymbol{u}) \triangleq p_{\boldsymbol{Y}|\boldsymbol{C}}(\boldsymbol{y}|\boldsymbol{G}_m\boldsymbol{u})$ emerges. So called *synthetic channels* $p_{\boldsymbol{Y}\boldsymbol{U}^i|U_i}(\boldsymbol{y}, \boldsymbol{u}^i|u_i)$ are suitably defined, such that (a) they *polarize*, *i.e.*, as $n \to \infty$, a fraction $\mathsf{C}\!\left(p_{Y|C}\right)$ of the channels has mutual information close to 1 (essentially noise-free, used for information) and a fraction $1 - \mathsf{C}\!\left(p_{Y|C}\right)$ of the channels has mutual information close to 0 (essentially useless, *frozen* to predefined values) [Arı09, Theorem 1], and (b) there exists an *efficient* decoder, the *successive cancellation decoder*, cf. Section 2.3, which requires a number of operations proportional to $n \log n$.

We proceed with the details concerning the aforementioned steps: Fix $m$ for the desired block length $n = 2^m$. The polar transform $\boldsymbol{G}_m$ is defined as

$$\boldsymbol{c} = \boldsymbol{G}_m\boldsymbol{u}, \qquad \boldsymbol{G}_m \triangleq \boldsymbol{F}^{\otimes m}\boldsymbol{P}_m^{(\text{bitrev})}, \qquad \boldsymbol{F} \triangleq \begin{bmatrix} 1 & 1 \\ 0 & 1 \end{bmatrix}, \qquad (2.11)$$

where $\boldsymbol{F}$ is called the *polarization kernel*, $\boldsymbol{F}^{\otimes m} \triangleq \boldsymbol{F} \otimes \boldsymbol{F}^{\otimes(m-1)}$, $\boldsymbol{F}^{\otimes 0} \triangleq \begin{bmatrix} 1 \end{bmatrix}$ denotes the *Kronecker power*, and $\boldsymbol{P}_m^{(\text{bitrev})}$ is known as the *bit-reversal permutation* defined as follows: Let $\boldsymbol{v}$ and $\boldsymbol{w}$ be binary vectors of length $n$. Index their coefficients $v_i$ and $w_i$ as $v_{(b_{m-1}...b_0)}$ and $w_{(b_{m-1}...b_0)}$ using the binary sequence $(b_{m-1} \ldots b_0)$ that corresponds to the binary representation $\text{bin}_m(i)$ of $i$ of length $m$, *e.g.*, $m = 3$, $i = 3$, $(b_2 b_1 b_0) = (011)$. Then, $\boldsymbol{w} = \boldsymbol{P}_m^{(\text{bitrev})}\boldsymbol{v}$ iff $\forall (b_{m-1} \ldots b_0) \in \{0,1\}^m \colon w_{(b_0...b_{m-1})} = v_{(b_{m-1}...b_0)}$.

Using the polar transformation $\boldsymbol{c} = \boldsymbol{G}_m \boldsymbol{u}$, $n$ uses of BMS $p_{Y|C}$, *i.e.*, of the channel

$$p_{\boldsymbol{Y}|\boldsymbol{C}}(\boldsymbol{y}|\boldsymbol{c}) = \prod_{i=0}^{n-1} p_{Y|C}(y_i|c_i), \qquad (2.12)$$

are transformed into a new channel

$$p_{\boldsymbol{Y}|\boldsymbol{U}}(\boldsymbol{y}|\boldsymbol{u}) \triangleq p_{\boldsymbol{Y}|\boldsymbol{C}}(\boldsymbol{y}|\boldsymbol{G}_m \boldsymbol{u}). \qquad (2.13)$$

Using $p_{\boldsymbol{Y}|\boldsymbol{U}}(\boldsymbol{y}|\boldsymbol{u})$, $n$ synthetic channels,

$$p_{\boldsymbol{Y}\boldsymbol{U}^i|U_i}\big(\boldsymbol{y}, \boldsymbol{u}^i\big|u_i\big) \triangleq \sum_{\boldsymbol{u}_{i+1}^n \in \{0,1\}^{n-i-1}} \frac{1}{2^{n-1}} p_{\boldsymbol{Y}|\boldsymbol{U}}(\boldsymbol{y}|\boldsymbol{u}), \qquad (2.14)$$

are defined, which lend themselves to successive decoding, as decoding $u_0$ requires only $\boldsymbol{y}$, decoding $u_1$ requires $\boldsymbol{y}$ and the previously decoded $u_0$, decoding $u_2$ requires $\boldsymbol{y}$ and the previously decoded $\boldsymbol{u}^2$, and so forth. Decoding $u_i$ requires $\boldsymbol{y}$ and the previously decoded $\boldsymbol{u}^i$, while future $\boldsymbol{u}_{i+1}^n$ are unknown and hence treated as $\text{Bern}(1/2)$.

The synthetic channels polarize, *i.e.*, as $n \to \infty$, a fraction $\mathsf{C}\big(p_{Y|C}\big)$ of them has mutual information close to 1 (essentially noise-free, call them *good channels*) and a fraction $1 - \mathsf{C}\big(p_{Y|C}\big)$ has mutual information close to 0 (essentially useless, call them *bad channels*) [Arı09, Theorem 1]. The decoding of good channels will usually succeed, *i.e.*, $\hat{u}_i = u_i$ with high probability, under the assumption that the previous $\hat{\boldsymbol{u}}^i$, which are used in successively decoding $u_i$, were decoded correctly. We use good channels to convey information, call them *information bits*, and denote the set of indices of good channels $\mathcal{I}$. The decoding of bad channels is hopeless, as $\hat{u}_i$ takes a value (almost) independent of $u_i$. To ensure that these channels do not jeopardize the successive decoding process, we *freeze* them to a predefined value, usually 0, and always 'decode' $\hat{u}_i = 0$, rather than attempting to convey information through them. Call them *frozen bits*, and denote the set of indices of bad channels $\mathcal{F}$. Note that

$$\mathcal{F} = \mathcal{I}^{\mathsf{c}} \triangleq [0{:}n-1] \setminus \mathcal{I}, \qquad |\mathcal{I}| \triangleq k = nR, \qquad (2.15)$$

with *code rate* $R \triangleq \frac{k}{n}$. As $n \to \infty$, $\frac{|\mathcal{I}|}{n} \to \mathsf{C}\big(p_{Y|C}\big)$ (cf. polarization) and thus $R \to \mathsf{C}\big(p_{Y|C}\big)$, hence polar codes are *capacity achieving*. The overall process of a polar coded communication system is illustrated in Figure 2.3. The successive cancellation decoder is presented in the following Section 2.3. How to choose $\mathcal{I}$ is subject to polar code design techniques presented in Section 2.6.

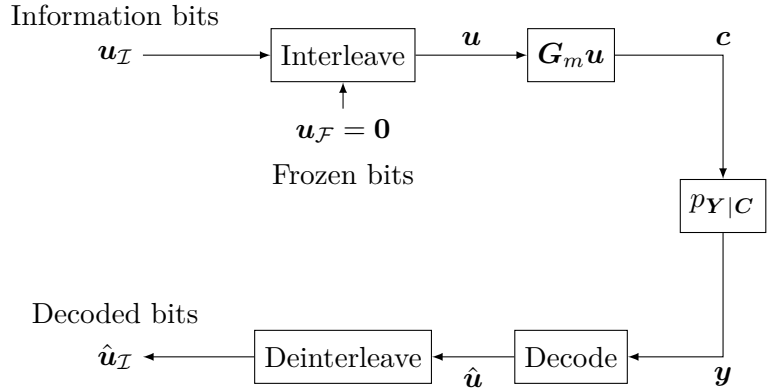

Figure 2.3.: Block diagram of a polar coded communication system

## 2.3. Successive Cancellation Decoding

We present the *successive cancellation* (SC) decoder for polar codes in LLR domain in this section, following the notation of [BPB15; MT09; HU12a].

As suggested in the previous section, the SC decoder estimates the inputs $u_i$ to the synthetic channels $p_{\boldsymbol{Y}\boldsymbol{U}^i|U_i}(\boldsymbol{y}, \boldsymbol{u}^i|u_i)$ successively, *i.e.*, $u_0$ is decoded from $\boldsymbol{y}$, $u_1$ is decoded from $\boldsymbol{y}$ and the previously decoded $u_0$, $u_2$ is decoded from $\boldsymbol{y}$ and the previously decoded $\boldsymbol{u}^2$, and so forth, $u_i$ is decoded from $\boldsymbol{y}$ and the previously decoded $\boldsymbol{u}^i$. Only information bits ($\mathcal{I}$) are actually decoded, frozen bits ($\mathcal{F}$) have a predefined value (usually 0) which the decoder returns as 'decoding result'. To estimate $\hat{u}_i$, the LLR $\lambda_i$ of the $i$-th bit,

$$\lambda_i \triangleq \log\left(\frac{p_{\boldsymbol{Y}\boldsymbol{U}^i|U_i}(\boldsymbol{y}, \boldsymbol{u}^i|0)}{p_{\boldsymbol{Y}\boldsymbol{U}^i|U_i}(\boldsymbol{y}, \boldsymbol{u}^i|1)}\right), \tag{2.16}$$

needs to be obtained. Since $\hat{\boldsymbol{u}}^i = \boldsymbol{u}^i$ with high probability (the values were either decoded from very reliable channels or predefined for frozen channels),

$$\lambda_i = \log\left(\frac{p_{\boldsymbol{Y}\boldsymbol{U}^i|U_i}(\boldsymbol{y}, \hat{\boldsymbol{u}}^i|0)}{p_{\boldsymbol{Y}\boldsymbol{U}^i|U_i}(\boldsymbol{y}, \hat{\boldsymbol{u}}^i|1)}\right), \qquad \text{with high probability.} \tag{2.17}$$

Due to the recursive structure of the polar transform, (a) the computation of $\lambda_i$ can be defined recursively [BPB15; MT09] [Arı09, Sec. VIII], giving rise to an equivalent message passing procedure over a suitably defined tree, and (b) intermediary results of the recursion can be reused in the computation of multiple LLRs, leading to low complexity of the decoder. The recursion reads [BPB15]

$$\lambda_i \triangleq \lambda_m^i, \tag{2.18}$$

$$\lambda_s^{2i} \triangleq f_{\boxplus}\left(\lambda_{s-1}^{2i-(i \bmod 2^{s-1})}, \lambda_{s-1}^{2^s+2i-(i \bmod 2^{s-1})}\right), \tag{2.19}$$

$$\lambda_s^{2i+1} \triangleq f_{\bullet}\left(\lambda_{s-1}^{2i-(i \bmod 2^{s-1})}, \lambda_{s-1}^{2^s+2i-(i \bmod 2^{s-1})}, u_s^{2i}\right), \tag{2.20}$$

$$\lambda_0^i \triangleq \log\left(\frac{p_{Y|C}(y_i|0)}{p_{Y|C}(y_i|1)}\right), \tag{2.21}$$

$$u_m^i \triangleq \hat{u}_i, \tag{2.22}$$

$$u_s^{2i-(i \bmod 2^s)} \triangleq u_{s+1}^{2i} \oplus u_{s+1}^{2i+1}, \tag{2.23}$$

$$u_s^{2^s+2i-(i \bmod 2^s)} \triangleq u_{s+1}^{2i+1}, \tag{2.24}$$

$$u_0^i \triangleq \hat{c}_i, \tag{2.25}$$

where

$$f_{\boxplus}(\lambda_1, \lambda_2) \triangleq 2\tanh^{-1}\left(\tanh\left(\frac{\lambda_1}{2}\right)\tanh\left(\frac{\lambda_2}{2}\right)\right) \tag{2.26}$$

$$f_{\bullet}(\lambda_1, \lambda_2, u) \triangleq (-1)^u \lambda_1 + \lambda_2. \tag{2.27}$$

The recursive nature of the computations of $\lambda_s^i$ and $u_s^i$ within an SC decoder becomes apparent from the example illustrated in Figures 2.4 and 2.5 for $m = 3$.

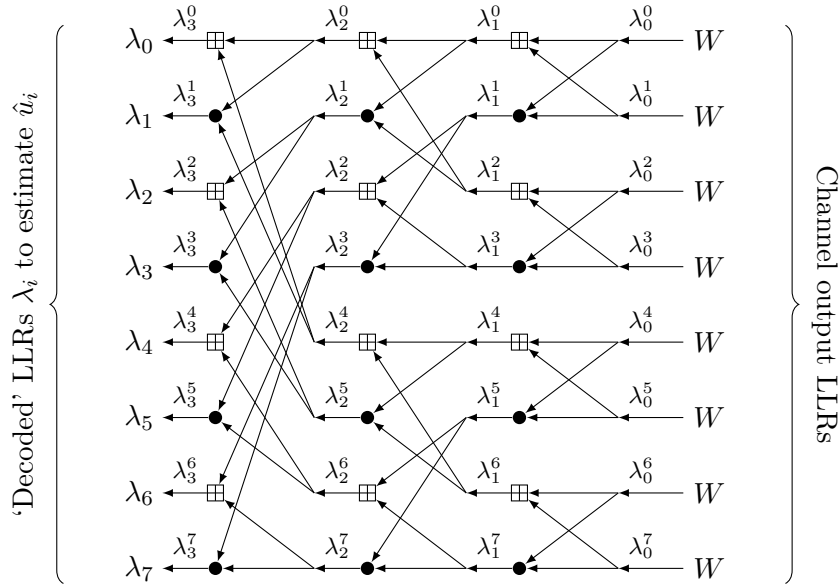

Figure 2.4.: Recursive structure of LLR computation in an SC decoder for $m = 3$, implements (2.18), (2.19), (2.20), and (2.21)

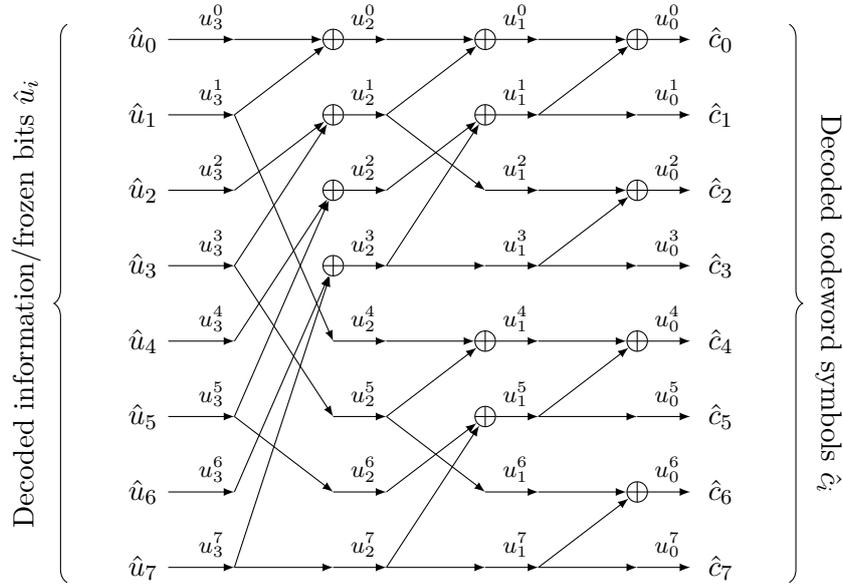

Figure 2.5.: Recursive structure of codeword reconstruction (required for LLR recursion) in SC decoding for $m = 3$, implements (2.22), (2.23), (2.24), and (2.25)

The functions $f_{\boxplus}$ and $f_{\bullet}$ are variants of what is called *check node operation* and *variable node operation* in belief-propagation decoding of low-density parity-check codes [RU08]. We define binary operations,

$$\lambda_1 \boxplus \lambda_2 \triangleq 2 \tanh^{-1}\left( \tanh\left( \frac{\lambda_1}{2} \right) \tanh\left( \frac{\lambda_2}{2} \right) \right), \qquad \text{'check node operation'} \qquad (2.28)$$

$$\lambda_1 \bullet \lambda_2 \triangleq \lambda_1 + \lambda_2, \qquad \text{'variable node operation'} \qquad (2.29)$$

such that $f_{\boxplus}$ and $f_{\bullet}$ can be reduced to these two basic operations,

$$f_{\boxplus}(\lambda_1, \lambda_2) \triangleq \lambda_1 \boxplus \lambda_2 \qquad (2.30)$$

$$f_{\bullet}(\lambda_1, \lambda_2, u) \triangleq ((-1)^u \lambda_1) \bullet \lambda_2. \qquad (2.31)$$

The *min-approximation* is commonly used for $\boxplus$,

$$\lambda_1 \boxplus \lambda_2 \approx \text{sign}(\lambda_1) \, \text{sign}(\lambda_2) \min\{|\lambda_1|, |\lambda_2|\}. \qquad (2.32)$$

While the difference in computational complexity between the 'perfect' and min-approximate $\boxplus$ is tangible, the impact on FER is minor (cf. Figure 2.6). Hence, we employ the min-approximate $\boxplus$ throughout this thesis.

Figure 2.4 exemplifies that the recursive LLR computation determined by (2.18), (2.19),

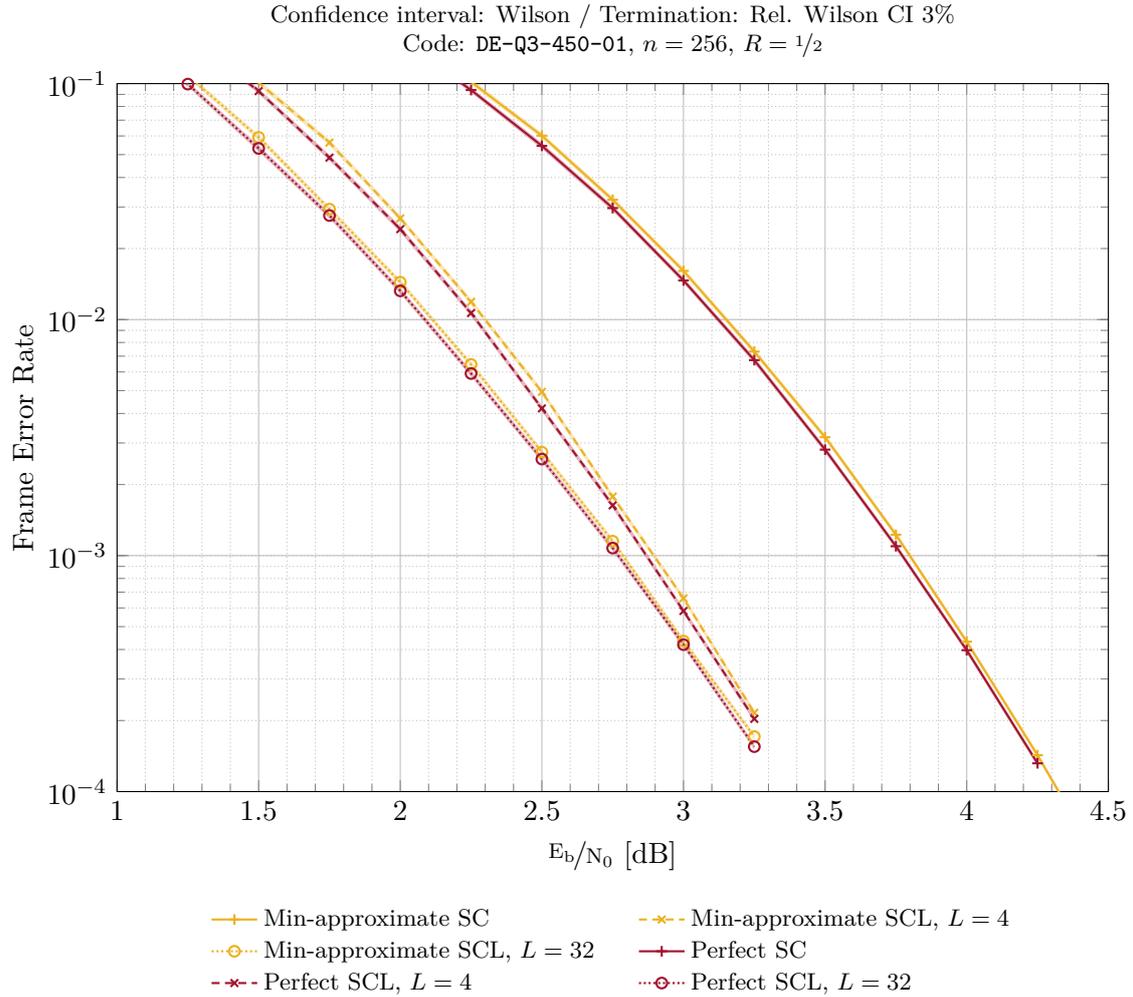

Confidence interval: Wilson / Termination: Rel. Wilson CI 3%
Code: `DE-Q3-450-01`, $n = 256$, $R = \frac{1}{2}$

— Min-approximate SC
— Min-approximate SCL, $L = 4$
— Min-approximate SCL, $L = 32$
— Perfect SC
— Perfect SCL, $L = 4$
— Perfect SCL, $L = 32$

Figure 2.6.: FER difference between perfect SC/SCL (using (2.28) and (2.29))) and min-approximate SC/SCL (using (2.32) and (2.29)) is negligible

(2.20) and (2.21) results in a tree structure [HU12a] such as illustrated for $\lambda_3$, $m = 3$, in Figure 2.7. The *decoding tree* for the $i$-th bit with LLR $\lambda_i$ is constructed as follows: The $n = 2^m$ channel outputs are the leaf nodes of a perfect binary tree of height $m$. The interior nodes are annotated with either $\boxplus$ or $\bullet$ depending on the binary expansion $\text{bin}_m(i) \triangleq (b_{m-1} \dots b_0) \in \{0,1\}^m$ of $i$ of length $m$. An interior node of depth $d$ is annotated with $\boxplus$ if $b_d = 0$ and annotated with $\bullet$ if $b_d = 1$. Computing $\lambda_i$ as part of SC decoding according to the recursion specified by (2.18), (2.19), (2.20) and (2.21) can be implemented as message-passing procedure over the respective decoding tree: Channel output LLRs are fed into the leafs of the tree. Each interior node applies the operation it is annotated with on the two incoming LLR messages and passes its result up the

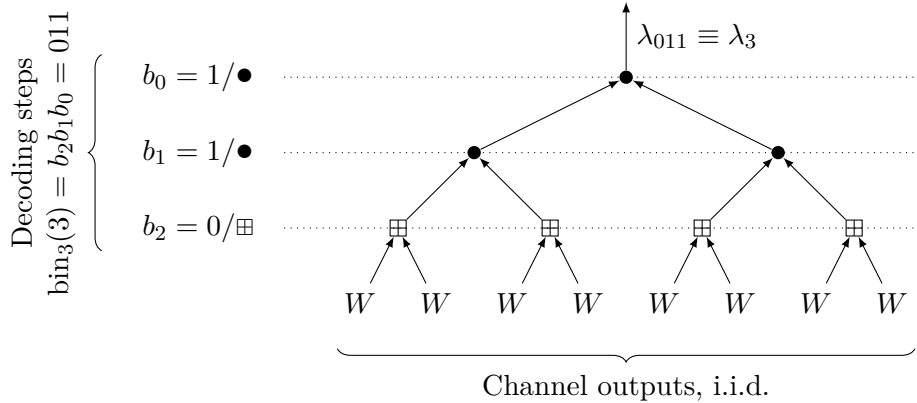

Figure 2.7.: Example decoding tree for recursive computation of $\lambda_3$ for $m = 3$ according to (2.18), (2.19), (2.20), and (2.21), cf. Figure 2.4

tree. The message output by the root node is $\lambda_i$.

## 2.4. Successive Cancellation List Decoding

The *successive cancellation list* (SCL) decoder for polar codes was first introduced in [TV15] and restated in LLR domain in [BPB15]. Giving a complete formal description here would be very lengthy and redundant given the aforementioned literature and the fact that the inner workings of SCL beyond the basics of SC decoding are not central to our work. Hence, we provide only a short overview of the idea behind SCL decoding. The SC decoder proceeds as follows: Compute $\lambda_0$ (cf. Figure 2.4), estimate $\hat{u}_0$ (either from $\lambda_0$ if $0 \in \mathcal{I}$, or using predefined value if $0 \in \mathcal{F}$), propagate $\hat{u}_0$ for future LLR computations (cf. Figure 2.5); compute $\lambda_1$, estimate $\hat{u}_1$, propagate $\hat{u}_1$; and so forth, until $\hat{\boldsymbol{u}}$ has been obtained.

The central idea of SCL decoding is the following: Rather than estimating $\hat{u}_i$ once and for all, whenever it comes to estimating a bit, 'copy' the SC decoder including all its internal state and input, and have one copy work with $\hat{u}_i = 0$ and one with $\hat{u}_i = 1$. Every such SC decoder instance corresponds to a *path* through its and its ancestors' past decisions $\hat{\boldsymbol{u}}^i$. So far, this is of little use, as the complexity of this brute force search grows exponentially. As mitigation, we constrain ourselves to never use more than the *list size* $L$ number of SC instances at the same time. Whenever we temporarily created more than the allowed $L$ paths, we remove paths until we are left with $L$ paths.

Which paths to keep and which to remove? To decide this, the *path metric* (PM) is introduced. The PM $\mathrm{P_{M_\ell}}$ of path $\ell$ is like a 'karma score': Whenever the path or its ancestors take a decision $\hat{u}_i$ opposed to what is suggested by the respective $\lambda_i$, some penalty points are added to the PM. The greater the discrepancy between $\hat{u}_i$ and $\lambda_i$, *i.e.*, the greater the magnitude of $\lambda_i$, the more penalty points are awarded. The PM is inherited by the copies of a path. Whenever the list needs truncation, the paths with worst PM, *i.e.*, highest number of penalty points, are removed. This process is illustrated in Figure 2.8 by means of a toy example with $n = 3$ and $L = 2$.

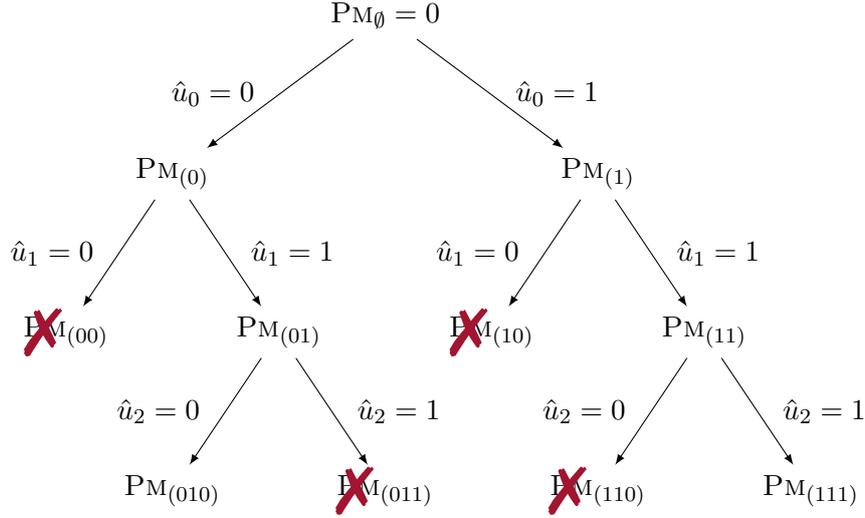

Figure 2.8.: Toy example of SCL decoding process with $n = 3$ and $L = 2$: Starting with the empty path ($\mathrm{PM}_\emptyset$), repeatedly, all possible continuations of surviving paths from the previous layer are compared using their updated PMs. The list of paths is truncated based on PM as necessary to keep within list size $L = 2$. For this example it is assumed that $\mathrm{PM}_{(00)}, \mathrm{PM}_{(10)} > \max\{\mathrm{PM}_{(01)}, \mathrm{PM}_{(11)}\}$ and $\mathrm{PM}_{(011)}, \mathrm{PM}_{(110)} > \max\{\mathrm{PM}_{(010)}, \mathrm{PM}_{(111)}\}$. The list of paths output by the decoder is $\{(010), (111)\}$, among which the path with smallest PM is declared the decoded codeword.

In the following we give basic technical details about the PM: The SCL decoder starts out with one SC instance which corresponds to the 'empty path' before commencing decoding. It is initialized with $\mathrm{PM}_\emptyset \triangleq 0$. In each $i$-th iteration of the decoder and for each path $\ell$, it computes $\lambda_{\ell,i}$ and produces the path's two possible continuations $\ell_0$ and $\ell_1$ with $\hat{u}_i = 0$ and $\hat{u}_i = 1$, respectively. Their updated PMs are

$$\mathrm{PM}_{\ell_0} \triangleq f_{\mathrm{PMU}}(\mathrm{PM}_\ell, \lambda_{\ell,i}, 0), \tag{2.33}$$

$$\mathrm{PM}_{\ell_1} \triangleq f_{\mathrm{PMU}}(\mathrm{PM}_\ell, \lambda_{\ell,i}, 1), \tag{2.34}$$

using the *path metric update function*

$$f_{\mathrm{PMU}}(\mathrm{PM}, \lambda, u) \triangleq \mathrm{PM} + \ln(1 + \exp(-(-1)^u \lambda)). \tag{2.35}$$

The path metric update function $f_{\mathrm{PMU}}$ is chosen (cf. [BPB15, Theorem 1, Lemma 1]) such that once decoding is complete, for each path $\ell$ its $\mathrm{PM}_\ell$ is related to the plausibility of the decoded sequence $\hat{\boldsymbol{u}}_{\ell,0}^n$, *i.e.*,

$$\mathrm{PM}_\ell = -\ln\Big(\mathrm{Pr}\Big[\boldsymbol{U} = \hat{\boldsymbol{u}}_{\ell,0}^n \,\Big|\, \boldsymbol{Y} = \boldsymbol{y}\Big]\Big). \tag{2.36}$$

Hence, selecting the codeword with best PM from the list output by SCL and declaring it as decoded codeword is equivalent to performing ML among the list.

Based on this property, a lower bound on the FER of ML decoding (referred to as 'ML-LB') of the respective polar code is computed from the list output by SCL: If the correct codeword has a lower likelihood than any of the (wrong) codewords in the list, at least then even the ML decoder errs. The unquantized SCL decoder's PM-FER performance usually saturates to this ML-LB for list sizes around $L = 32$, which provides a characterization of the ML decoder's performance and shows that SCL matches it.

## 2.5. Density Evolution

'Density evolution' is a formal tool which is used extensively throughout 'modern' coding theory literature to analyze the performance of iterative message passing (decoding) algorithms (cf. [RU08, §3.9, §4.5] and references therein). It allows to track how probability distributions (or *densities*) within the decoder *evolve* over time as the algorithm proceeds (hence the name *density evolution*). As will become clear in the following, SC and SCL decoding of polar codes can be viewed as such a message passing procedure. Consequently, variants of density evolution are employed in this thesis. It is therefore worth introducing density evolution in a solid and sufficiently general form, such that we can refer to this foundation where appropriate in the remainder of this thesis and spare repetitions.

We proceed in three steps: *First,* we recapitulate transformations of random variables, by means of a simple toy example. This is the basic operation at the core of density evolution. We draw a clear distinction between (a) the 'abstract' rules and mechanics inherent in density evolution independent of the representation chosen for the involved probability distributions, and (b) how the 'concrete' density evolution procedure follows from a particular choice of representation for the involved probability distributions. Making this clear distinction eliminates the need to repeat (a) whenever we use a new

'flavor' of density evolution; instead, it suffices to clarify the employed representation, and the reader can immediately proceed to (b). *Secondly*, we show how SC decoding can be seen as a message passing algorithm. *Finally*, we show how density evolution is used to analyze the SC decoder.

Note that SC decoding is an integral part of SCL decoding, as illustrated in the previous section. Hence, density evolution is also useful for analyzing (certain aspects of) the SCL decoder. However, in the following we illustrate the utility of density evolution analysis by taking the example of SC decoding.

**Transformations of Random Variables: The Core of Density Evolution**

At the very core of density evolution is the following task: Given is a function $f$ which is fed some data $V$ to produce a result $W \triangleq f(V)$. We do not know the precise value of $V$ (else this was merely a computational task), but we know $P_V$. Then, what is $P_W$? To have a more precise toy example, turn to the function

$$f(x, y) = (-1)^x + y, \tag{2.37}$$

which we present with data $(x, y)$ distributed as

$$(x, y) = \begin{cases} (1, 0) & \text{w.p. } 1/10, \\ (2, 0) & \text{w.p. } 2/10, \\ (3, 0) & \text{w.p. } 3/10, \\ (3, 1) & \text{w.p. } 4/10. \end{cases} \tag{2.38}$$

By evaluating $z \triangleq f(x, y)$ for all possible tuples $(x, y)$ and summing up the probabilities that lead to the same output $z$, we obtain the distribution

$$z = \begin{cases} 0 & \text{w.p. } 2/10 + 4/10 = 6/10, \\ -1 & \text{w.p. } 1/10 + 3/10 = 4/10. \end{cases} \tag{2.39}$$

The procedure for obtaining $P_Z$ from $P_{XY}$ is mathematically formulated as

$$P_Z(z) = \sum_{\{(x,y) \mid f(x,y)=z\}} P_{XY}(x, y), \qquad \text{or} \qquad P_Z(z) = \Pr\left[f^{-1}[z]\right], \tag{2.40}$$

where $f^{-1}[z] = \{(x, y) \in \text{supp}(P_{XY}) \mid f(x, y) = z\}$ denotes the preimage of $z$ under $f$. The second formulation carries over to the general case of deriving $P_W$ from $P_V$ for $W \triangleq f(V)$ and both discrete or continuous $V$ and $W$. Recall that if $f$ is a *measurable* *function*, then the probability of an event 'about $W$' can be reduced to the probability of the preimage of that event under $f$. This is necessary for $W$ to be a random variable. Note that here we are usually concerned with discrete random variables where the technical subtleties are less involved, and we can simply use

$$P_W(w) = \Pr\Big[f^{-1}[w]\Big], \qquad \text{with} \qquad f^{-1}[w] = \{v \in \text{supp}(P_V) \,|\, f(v) = w\}. \qquad (2.41)$$

To highlight the fact that $f$ transforms the distribution $P_V$ into the distribution $P_W$ in the way described by (2.41), we write equivalently with slight abuse of notation

$$P_W = f(P_V). \qquad (2.42)$$

Equation (2.41) gives meaning to (2.42) in the general case of arbitrary discrete distributions $P_V$. It describes how the density $P_V$ evolves into $P_W$ by means of the processing step $f$, and thus captures this essential part of density evolution. To apply (2.41) or (2.42) and perform density evolution on any discrete distributions, it suffices to define the operation $f \colon \text{supp}(V) \to \text{supp}(W)$.

However, sometimes it is not necessary to assume arbitrary discrete distributions $P_V$ and $P_W$ and to track their support set and the probabilities of all elements therein. To see this, assume a new example: Let $X \sim \text{Bern}(\rho_X)$, $Y \sim \text{Bern}(\rho_Y)$, X independent of Y, and $Z \triangleq f(X,Y) \triangleq X \oplus Y$. Then,

$$z = \begin{cases} 0 & \text{w.p. } \rho_X \rho_Y + \overline{\rho}_X \overline{\rho}_Y, \\ 1 & \text{w.p. } \overline{\rho}_X \rho_Y + \rho_X \overline{\rho}_Y. \end{cases} \qquad (2.43)$$

Equivalently, $Z \sim \text{Bern}(\rho_Z)$ with $\rho_Z = \overline{\rho}_X \rho_Y + \rho_X \overline{\rho}_Y$. Again, we write

$$P_Z = f(P_{XY}) = f(P_X P_Y), \qquad (2.44)$$

but in this case there is a simpler way than (2.41) to express how $P_Z$ relates to $P_X$ and $P_Y$ (and thus give meaning to (2.44)), namely through the relation between $\rho_X$, $\rho_Y$ and $\rho_Z$. The parameters $\rho_X$, $\rho_Y$ and $\rho_Z$ of the Bernoulli distributions represent these distributions in the sense that they uniquely specify them. Thus, after fixing this representation, we can unambiguously describe how the density $P_Z$ evolves from $P_X$ and $P_Y$ by means of the processing step $f$ as

$$\rho_Z = \tilde{f}(\rho_X, \rho_Y) \triangleq \overline{\rho}_X \rho_Y + \rho_X \overline{\rho}_Y. \qquad (2.45)$$

It is important to notice the difference between $f \colon \text{supp}(X) \times \text{supp}(Y) \to \text{supp}(Z)$ and $\tilde{f} \colon [0,1] \times [0,1] \to [0,1]$. The first is agnostic to the representation chosen for $P_X$, $P_Y$ and $P_Z$; it merely describes the actual operation of $f$ on pairs of binary numbers. It can be used in conjunction with (2.41) to perform one step of density evolution. The second uses the representation of a Bernoulli random variable via its parameter $\rho$ to simplify and speed up the density evolution computations for (2.44). But it is tailored to this particular choice of $X$, $Y$ and $f$.

The 'raw' transformation of $X$ and $Y$ via $f$ into $Z$ is illustrated as a tree in Figure 2.9(a). The 'equivalent' transformation after choosing a parametrization of the probability distributions is illustrated in Figure 2.9(b). Conveniently, since $f$ and $\tilde{f}$ are symmetric, *i.e.*, $f(x, y) = f(y, x)$, $\tilde{f}(\rho_X, \rho_Y) = \tilde{f}(\rho_Y, \rho_X)$, the order of edges into $f$ and $\tilde{f}$ does not matter. This is commonly the case where density evolution is applied.

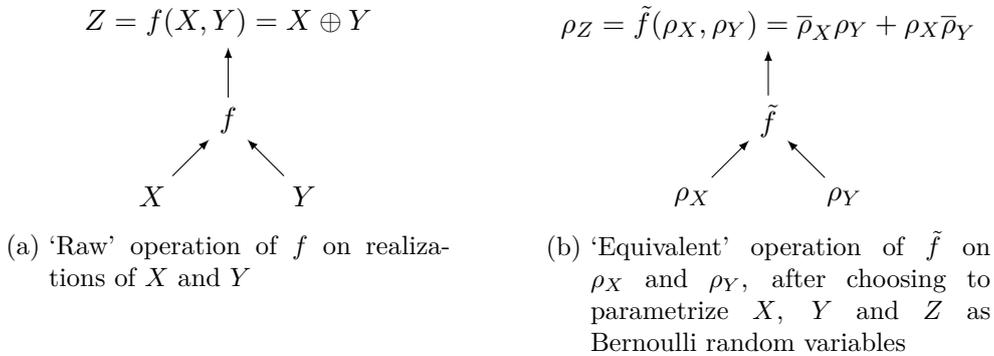

(a) 'Raw' operation of $f$ on realizations of $X$ and $Y$

(b) 'Equivalent' operation of $\tilde{f}$ on $\rho_X$ and $\rho_Y$, after choosing to parametrize $X$, $Y$ and $Z$ as Bernoulli random variables

Figure 2.9.: Transformation of random variables $X$ and $Y$ via $f$ into $Z$

In this section, we recapitulated transformations of random variables, which is about obtaining $P_W$ from $P_V$ for $W \triangleq f(V)$ (cf. (2.42)). This is the fundamental building block of density evolution analysis, which consists of repeated application of such transformations. We presented the general formulation for arbitrary discrete distributions (cf. (2.41)). Throughout this thesis, we mostly work with discretized distributions and hence use this formulation. We also showed that for some $V$ and $f$ the parameters of $P_W$ can be computed directly from the parameters of $P_V$ through some $\tilde{f}$. This usually provides a much more efficient implementation of density evolution, if possible for $V$ and $f$ at hand. Next, we show how SC decoding can be seen as a message passing algorithm and how density evolution is used to analyze it.

**Successive Cancellation Decoding as Message Passing Algorithm**

Recall from Section 2.3, that the SC decoder proceeds as follows: Successively for each $i$-th bit, it estimates its decision LLR $\lambda_i$ based on previous observations $\hat{\boldsymbol{u}}^i$ using (2.17). It then determines $\hat{u}_i$ according to the sign of $\lambda_i$, and proceeds to the next $i + 1$. The computation of $\lambda_i$ is recursive, using (2.18), (2.19), (2.20) and (2.21), based on two fundamental operations $\boxplus$ and $\bullet$, cf. (2.28) and (2.29). The recursion can be illustrated as a tree, cf. Figure 2.7. The tree of a further reduced toy example is given in Figure 2.10. If channel LLRs originate from leaf nodes, and each inner node applies its operation to the two values it receives and passes the result up to its parent node, then this message passing procedure computes $\lambda_i$ as output of the root node of the tree.

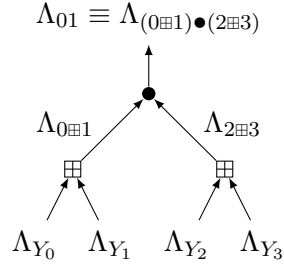

Figure 2.10.: Example decoding tree for recursive computation of $\lambda_{01}$ for $m = 2$ according to (2.18), (2.19), (2.20), and (2.21), cf. Figures 2.7 and 2.4

**Analyzing Successive Cancellation Decoding using Density Evolution**

The reconstruction of the recursive computation of $\lambda_i$ in an SC decoder as message passing procedure over a suitably chosen tree (cf. Figure 2.10) neglects the role of previous decisions $\hat{\boldsymbol{u}}^i$. To be able to analyze this message passing procedure using density evolution, it is necessary to remove the dependency on $\hat{\boldsymbol{u}}^i$. This is achieved using two assumptions common in density evolution applications:

1. It is assumed that the all-zero codeword was transmitted. Hence, the channel outputs $Y_i$ follow the distribution $p_{Y|C}(y|0)$, and the desired correct decoder output would be $\hat{\boldsymbol{u}} = \boldsymbol{0}$. As long as the decoder has not erred yet, $\hat{\boldsymbol{u}}^i = \boldsymbol{0}$.

2. Individual computations and decisions for each $i$-th bit are decoupled: A 'stronger' genie-aided SC decoder is assumed (we show in Section 2.6.1 that in fact this genie-aided SC decoder has the same FER as the 'standard' SC decoder), where after computing $\lambda_{i-1}$ and estimating and fixing $\hat{u}_{i-1}$, the true value of $u_{i-1}$ is revealed to the decoder by a genie and used in computing subsequent $\lambda_i$. In this genie-aided SC decoder, $\hat{\boldsymbol{u}}^i = \boldsymbol{u}^i$, hence the computation of $\lambda_i$ is not disturbed by errors propagating from prior wrong decisions.

On the basis of these two assumptions, $\hat{\boldsymbol{u}}^i = \boldsymbol{u}^i = \boldsymbol{0}$ for all $i$. Then, the recursion on $\hat{\boldsymbol{u}}$ expressed by (2.22), (2.23), (2.24), and (2.25) (cf. Figure 2.5) can be neglected, and the tree structure as in Figure 2.10 is in fact appropriate.

By analogy of Figures 2.9 and 2.10, it becomes clear that the distributions of the random variables $\Lambda$ representing the LLR messages $\lambda$ being passed on in the decoding tree, can be expressed using (2.41) and (2.42),

$$p_{\Lambda_{0\boxplus1}} = p_{\Lambda_{Y_0}} \boxplus p_{\Lambda_{Y_1}}, \tag{2.46}$$

$$p_{\Lambda_{2\boxplus3}} = p_{\Lambda_{Y_2}} \boxplus p_{\Lambda_{Y_3}}, \tag{2.47}$$

$$p_{\Lambda_{(0\boxplus1)\bullet(2\boxplus3)}} = p_{\Lambda_{0\boxplus1}} \bullet p_{\Lambda_{2\boxplus3}}, \tag{2.48}$$

where it is assumed that (under the all-zero codeword assumption) all $Y_i$ and thus all $\Lambda_{Y_i}$ are independent. Here, we make use of the notation introduced with (2.42) and (2.44), namely, $p_Z = p_X \boxplus p_Y$ means that $(X, Y) \sim p_X p_Y$ is transformed using $Z = X \boxplus Y$ into $Z \sim p_Z$. One could already perform these transformations and hence obtain the involved distributions. But a key ingredient simplifies this process tremendously: Not only is $Y_i$ independent of $Y_j$ for $i \neq j$, and hence $\Lambda_{Y_i}$ independent of $\Lambda_{Y_j}$, but all $Y_i$ (and hence all $\Lambda_{Y_j}$) are identically distributed, for all $i$, $p_{\Lambda_{Y_i}} = p_{\Lambda_Y}$. Hence,

$$p_{\Lambda_\boxplus} \triangleq p_{\Lambda_{0\boxplus1}} = p_{\Lambda_{2\boxplus3}} = p_{\Lambda_Y} \boxplus p_{\Lambda_Y}, \tag{2.49}$$

and thus,

$$p_{\Lambda_{\boxplus\bullet}} \triangleq p_{\Lambda_{(0\boxplus1)\bullet(2\boxplus3)}} = p_{\Lambda_{0\boxplus1}} \bullet p_{\Lambda_{2\boxplus3}} = p_{\Lambda_\boxplus} \bullet p_{\Lambda_\boxplus}. \tag{2.50}$$

The following recursion holds,

$$p_{\Lambda_\emptyset} \triangleq p_{\Lambda_Y}, \tag{2.51}$$

$$p_{\Lambda_{\boldsymbol{S}\boxplus}} = p_{\Lambda_{\boldsymbol{S}}} \boxplus p_{\Lambda_{\boldsymbol{S}}}, \tag{2.52}$$

$$p_{\Lambda_{\boldsymbol{S}\bullet}} = p_{\Lambda_{\boldsymbol{S}}} \bullet p_{\Lambda_{\boldsymbol{S}}}, \tag{2.53}$$

where $\boldsymbol{S} \in \{\boxplus, \bullet\}^l$ is a sequence of $\boxplus$ or $\bullet$ symbols of length $l \geq 0$. Note the direct correspondence between $p_{\Lambda_{\boxplus\bullet}}$ and $p_{\Lambda_{01}}$: For each $i$-th bit with binary representation $\text{bin}_m(i)$ of $i$ of length $m$, the distribution $p_{\Lambda_{\text{bin}_m(i)}}$ of its LLR $\lambda_{\text{bin}_m(i)}$ can be obtained using (2.51), (2.52) and (2.53), as the distribution $p_{\Lambda_{\boldsymbol{S}}}$, where $\boldsymbol{S} \in \{\boxplus, \bullet\}^m$ is the sequence of $\boxplus$ or $\bullet$ symbols that corresponds to $\text{bin}_m(i)$ by replacing a 0 with $\boxplus$ and a 1 with $\bullet$. Using this recursive density evolution procedure, all distributions $p_{\Lambda_i} \equiv p_{\Lambda_{\text{bin}_m(i)}}$ of bit estimate LLR values $\lambda_i$ can be obtained in an efficient way. The $p_{\Lambda_i}$ characterize the behavior of the SC decoder, *e.g.*, they allow to obtain the probability that the SC decoder takes a wrong bit decision in step $i$, under the assumption that previous bit decisions were correct. More details on the error probabilities are given in Section 2.6.1.

Three remarks are due here:

- In few special cases we parametrize the involved distributions as Gaussian distributions and then express the evolution of the distributions using the evolution of their respective parameters, mean and variance (cf. Figure 2.9). In iterative decoding literature, this is referred to as *Gaussian approximation*, cf. [RU08].

- Usually however, we represent the probability distributions as finely discretized (but ultimately discrete) distributions. Then, we use (2.41) and (2.42) to compute the transformations (cf. Figure 2.9(a)).

- Note that in this case, density evolution is agnostic as to 'what' the messages are that are being passed around, as long as ⊞ and ● are defined for them and the process follows the tree structure of SC decoding, cf. Figure 2.10. This allows a decoupling of the 'abstract' density evolution algorithm (to obtain distributions of messages being passed) from the 'concrete' 'interpretation' or 'implementation' of the messages and the operations on them. Both in analysis and implementation, we make heavy use of this abstraction.

## 2.6. Polar Code Design

In Sections 2.2 and 2.3 we hint to the question of polar code design. The polar code design problem is the following: Given $n = 2^m$ and $k$ and for a specific decoder (*e.g.*, SC or SCL decoder with some list size $L$) and channel $p_{Y|X}$, pick the information bit locations $\mathcal{I}$ such that the resulting FER is as low as possible. Note that we do not optimize over the values of the frozen bits (*i.e.*, $\boldsymbol{u}_\mathcal{F} = \boldsymbol{0}$) or the polarization kernel (*i.e.*, $\boldsymbol{F} = \begin{bmatrix} 1 & 1 \\ 0 & 1 \end{bmatrix}$), or design for time-varying channels or channels with memory.

For SC decoding, a method based on density evolution gives good results. It is due to [MT09] and presented in Section 2.6.1. As density evolution cannot capture all aspects of SCL decoding, several other heuristics/approaches exist to design for SCL:

1. For BiAWGN and variants thereof (*e.g.*, quantized BiAWGN, introduced in Section 2.8): Use the same method as for SC decoding, but at large SNR. The reasoning behind this heuristic: At large SNR, the FER under SC decoding is dominated by the code's minimum distance, which governs the FER under ML decoding and, since SCL usually approaches ML performance at reasonably low $L$, also under SCL decoding.

2. Use Reed-Muller codes. [MHU14]

3. Treat the SCL decoder as a black box and use a metaheuristic to find an $\mathcal{I}$ that minimizes its FER. Such an approach is presented in Section 2.6.2. We observe that it returns a Reed-Muller code for $(n, k)$ such that a Reed-Muller code exists, providing empirical support for the aforementioned design rule. Due to the consistency with the aforementioned design rule, we believe that it extends the design rule's 'spirit' to $(n, k)$ where no Reed-Muller code exists.

### 2.6.1. Density Evolution

Note that all channels and decoders considered here are symmetric such that it suffices to study the performance for the all-zero codeword. Furthermore, the FER of (a) the SC decoder (cf. Section 2.3) and (b) the genie-aided SC decoder, where a genie provides the true value of $\boldsymbol{u}^i$ for the computation of $\lambda_i$, are the same, for the following reason: It is clear that whenever (b) fails to decode, (a) also fails to decode. We prove the opposite direction. Assume a frame error has occurred for decoder (a). Let $j$ be the position of the first incorrect bit. Then $\hat{\boldsymbol{u}}^j = \boldsymbol{u}^j$. Hence, decoders (a) and (b) used the same input to compute $\lambda_j$. Consequently, decoder (b) must have made the same bit error at position $j$ and thus also produced a frame error event. Hence, we can use density evolution to analyze the FER of the genie-aided SC decoder for the all-zero codeword, which is equivalent to the FER of the SC decoder for any codeword.

Given the channel $p_{Y|X}$, density evolution (cf. Section 2.5) yields the distributions $p(\lambda_i)$, $i \in [0{:}n-1]$, of LLRs computed by the genie-aided SC decoder for decoding the $i$-th bit $u_i$. The 'probability of erroneously decoding the $i$-th bit', $\mathsf{P}_\mathrm{e}(p_{\Lambda_i})$, is given as

$$\mathsf{P}_\mathrm{e}(p_{\Lambda_i}) \triangleq \Pr[\{\Lambda_i < 0\}] + \frac{1}{2}\Pr[\{\Lambda_i = 0\}], \tag{2.54}$$

since under all-zero codeword assumption the decoder errs if either the LLR suggests $\hat{u}_i = 1$ (negative LLR) or the LLR is undecided (LLR zero) and thus the decoder flips a coin. The following union upper bound on the FER $\mathsf{P}_\mathrm{e,B}$ holds [MT09, eq. (3)]:

$$\mathsf{P}_\mathrm{e,B}^{(\mathrm{ub})} \triangleq \sum_{i \in \mathcal{I}} \mathsf{P}_\mathrm{e}(P_{\Lambda_i}) \geq \mathsf{P}_\mathrm{e,B} \tag{2.55}$$

To minimize $\mathsf{P}_\mathrm{e,B}^{(\mathrm{ub})}$ and thus minimize $\mathsf{P}_\mathrm{e,B}$, the information bits $\mathcal{I}$ with $|\mathcal{I}| = k$ are chosen as the indices of the $k$ smallest $\mathsf{P}_\mathrm{e}(P_{\Lambda_i})$.

This method usually gives good results and can be generalized to decoders with LLR alphabets other than the reals, *e.g.*, quantized decoders, cf. Section 3. We used it to design several codes used throughout this thesis, which we list for reference in Section 2.6.3.

### 2.6.2. Genetic Algorithm

Metaheuristics are optimization methods that are heuristic (*i.e.*, rely on intuition rather than theoretical performance guarantees) and generic (*i.e.*, work well for a large class of optimization problems without much fine-tuning). Genetic algorithms are a class of evolutionary-biology-inspired metaheuristics. They resemble the evolution of a *population* of *individuals* (candidate solutions) over *generations*, where the evolutionary *fitness* of an individual is determined by the cost function. From one generation to the next, individuals recombine their *genes* (assignment of values to the parameters of the cost function) or die out, based on their fitness. After gradual improvement of the cost function value from generation to generation, the individuals concentrate around the optimum. While the success of this method is in general not guaranteed, in practice it works well for many optimization problems if the number of generations and individuals and the recombination mechanism based on fitness are chosen suitably.

We used a genetic algorithm to design polar codes with short block lengths, $m \in \{6, 7, 8\}$, for the SCL decoder with usually $L = 32$. For the sake of brevity, we only give a sketch of our approach. The populations typically consisted of 160 to 240 individuals. The information bits $\mathcal{I}$ of an individual were represented as a binary vector of length $n$ with exactly $k$ ones. The population was initialized with random individuals. The PM-FER of SCL decoding at a chosen design $E_b/N_0$ served as cost function and fitness.

The best 20% survived from one generation to the next, the remaining 80% were discarded and new individuals were generated based on the individuals of the old generation in the following way: Alternatingly, a new individual was proposed either (a) as mutation of a single individual of the previous generation, or (b) as recombination of two individuals of the previous generation. A proposed individual was added to the new generation if it had not been seen before. Individuals of the old generation with better fitness were more likely to be chosen as starting point for mutations or recombinations, the selection probability of the $j$-th individual with FER $P_{e,B}(j)$ was proportional to $P_{e,B}(j)^{-3}$. Mutation of a single individual meant swapping a random number (typically between 1 and 10) of its information and frozen bits. Recombination of two individuals with $\mathcal{I}_1$ and $\mathcal{I}_2$, respectively, meant keeping the $\mathcal{I}_1 \cap \mathcal{I}_2$ that were identical in both parent individuals and randomly selecting some of the $\mathcal{I}_1 \cup \mathcal{I}_2 \setminus (\mathcal{I}_1 \cap \mathcal{I}_2)$ that only one of the parents had, and repeating this random selection until the resulting $|\mathcal{I}| = k$.

Codes with performance competitive to that of codes designed with density evolution emerged as early as the tenth generation. Little to no improvement was observed after the hundredth generation. Different cost functions can be used, *e.g.*, the list-FER or ML-LB in the case of SCL decoding. Often times, the genetic algorithm converged to a Reed-Muller code, if such existed for the parameters $(n, k)$, or codes that agreed with Reed-Muller codes in all but one or two information bit positions and with numerically indistinguishable FER at the design $E_b/N_0$. This provides independent evidence that there is a good fit between Reed-Muller codes and SCL decoding. Note that our method is only applicable to short block lengths, as the computational cost of the decoder and the number of generations required for the algorithm to converge grow with $n$.

### 2.6.3. Reference Codes

Throughout this thesis, we usually consider the polar code as given, *i.e.*, we improve various aspects of the decoder but leave the polar code (in particular its information bits $\mathcal{I}$) untouched. In this section, we provide a list of polar codes which we use and refer to in empirical performance evaluations.

`DE-Q3-450-01`

• Parameters: $n = 256$, $k = 128$, $R = {}^1\!/{}_2$ • Design method: density evolution for 3-level quantized channel and decoder with capacity-maximizing quantization threshold, at $E_b/N_0 = 4.5\,\mathrm{dB}$ • Information bits $\mathcal{I}$ depicted in Figure 2.11, frozen bits all set to 0

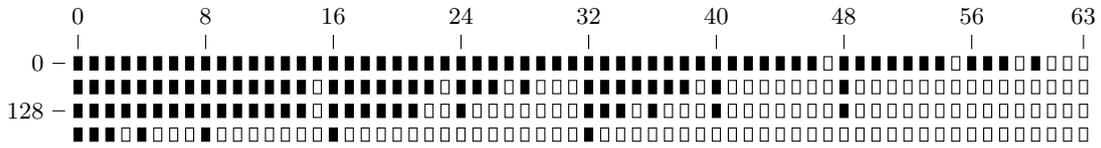

Figure 2.11.: Information bits (□) and frozen bits (■) of `DE-Q3-450-01`

`DE-Q3-450-02`

• Parameters: $n = 128$, $k = 64$, $R = {}^1\!/{}_2$ • Design method: density evolution for 3-level quantized channel and decoder with capacity-maximizing quantization threshold, at $E_b/N_0 = 4.5\,\mathrm{dB}$ • Information bits $\mathcal{I}$ depicted in Figure 2.12, frozen bits all set to 0

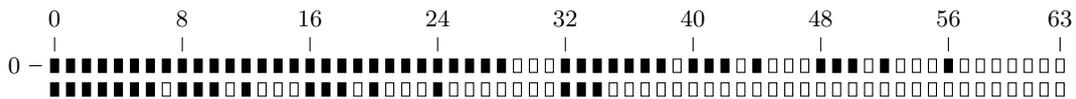

Figure 2.12.: Information bits (□) and frozen bits (■) of `DE-Q3-450-02`

`DE-MSD-450-01`

• Parameters: $n = 256$, $k = 128$, $R = 1/2$ • Design method: density evolution for the virtual multi-stage decoding scheme described in Section 5.3, using the technique described in Section 6.3 to obtain the quantization thresholds $\delta_{cn}^* = 2.0$ and $\delta_{vn}^* = 2.8$, joint optimization of quantization thresholds and information bits, at $E_b/N_0 = 4.5\,\mathrm{dB}$ • Information bits $\mathcal{I}$ depicted in Figure 2.13, frozen bits all set to 0

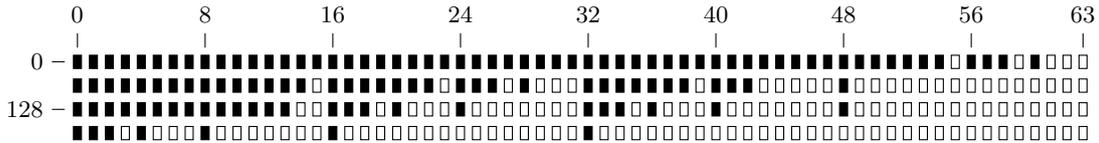

Figure 2.13.: Information bits (□) and frozen bits (■) of `DE-MSD-450-01`

`RM-37-256`

• Parameters: $n = 256$, $k = 37$, $R = 37/256 \approx 0.145$ • Design method: Reed-Muller code • Information bits $\mathcal{I}$ depicted in Figure 2.14, frozen bits all set to 0

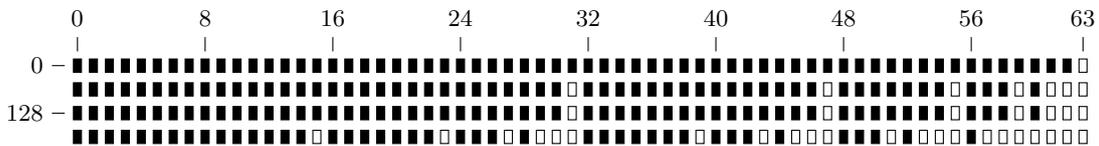

Figure 2.14.: Information bits (□) and frozen bits (■) of `RM-37-256`

## 2.7. Code Simulation Basics

The performance of codes and decoders proposed in this thesis is evaluated numerically using Monte Carlo method based simulations. To this end, a message to be communicated is drawn uniformly at random, encoded using the proposed code, transmitted over the channel using the probabilistic channel law, and decoded using the proposed decoder. Whenever the decoded message is equivalent to the transmitted message, the scheme was successful; otherwise an error event has occurred. This procedure is repeated $N$ times while the number of errors $N_{\mathrm{error}}$ is counted.

Subsequently, the error rate $\hat{P}_e \triangleq \frac{N_{\mathrm{error}}}{N}$ is estimated. Natural questions arising in this context concern the reliability of the error rate estimate and the number of repetitions required to reach a predefined level of reliability. *Confidence intervals* (CI) can be used to gauge the reliability of an error rate estimate. Based upon this, criteria can be defined which terminate repetition once a required level of reliability is reached.

### 2.7.1. Confidence Intervals

Common confidence intervals used in estimation of the parameter (called 'success probability') of a Bernoulli random variable are the *Wald CI* and the *Wilson CI* [Ash07; JBS02; Ham14]. Both require to fix a *confidence level* $\alpha$ first. In this thesis, $\alpha = 0.95 = {}^{19}/_{20}$, *i.e.*, on average in 19 out of 20 times a confidence interval is specified, the true parameter value lies within the confidence interval. In other words, the chances that the estimated parameter value deviates from the true parameter value by more than the confidence interval is less than 5%. Based on $\alpha$, an auxiliary constant $z \triangleq Q^{-1}\left(\frac{1-\alpha}{2}\right)$ is calculated, cf. (2.1). For $\alpha = 0.95$ we obtain $z = 1.96$.

#### Wald Confidence Interval

The Wald CI estimates the error rate as [Ash07]

$$\hat{P}_e \pm z \sqrt{\frac{\hat{P}_e\left(1 - \hat{P}_e\right)}{N}} \quad \equiv \quad \frac{N_{error}}{N} \pm \frac{z}{N} \sqrt{\frac{N_{error}(N - N_{error})}{N}}. \tag{2.56}$$

In particular for $P_e$ close to 0 or 1 and small numbers of errors $N_{error}$ and experiments $N$, the Wald CI is inaccurate and a refined CI, the Wilson CI, should be used.

#### Wilson Confidence Interval

The Wilson CI estimates the error rate as [JBS02]

$$\frac{\hat{P}_e + \frac{z^2}{2N}}{1 + \frac{z^2}{N}} \pm \frac{z}{1 + \frac{z^2}{N}} \sqrt{\frac{\hat{P}_e\left(1 - \hat{P}_e\right)}{N} + \frac{z^2}{4N^2}}$$

$$\equiv \quad \frac{N_{error} + \frac{z^2}{2}}{N + z^2} \pm \frac{z}{N + z^2} \sqrt{\frac{N_{error}(N - N_{error})}{N} + \frac{z^2}{4}}. \tag{2.57}$$

We use a slight simplification in that we continue to use the naive estimator $\hat{P}_e = \frac{N_{error}}{N}$,

$$\hat{P}_e \pm \frac{z}{1 + \frac{z^2}{N}} \sqrt{\frac{\hat{P}_e\left(1 - \hat{P}_e\right)}{N} + \frac{z^2}{4N^2}}$$

$$\equiv \quad \frac{N_{\text{error}}}{N} \pm \underbrace{\frac{z}{N + z^2} \sqrt{\frac{N_{\text{error}}(N - N_{\text{error}})}{N} + \frac{z^2}{4}}}_{\triangleq f_{\text{CI}(z)}(N_{\text{error}}, N)} . \tag{2.58}$$

### 2.7.2. Termination Criteria

CIs can be used to gauge the reliability of the error rate estimate based on the number of experiments $N$ and the number of errors $N_{\text{error}}$. In particular, during the simulation we check whether a desired level of reliability has been achieved and terminate as soon as this is the case. Of special interest is the *relative CI*,

$$\frac{f_{\text{CI}(z)}(N_{\text{error}}, N)}{\hat{\mathsf{P}}_{\text{e}}}, \tag{2.59}$$

which gauges the relative error of the estimate $\hat{\mathsf{P}}_{\text{e}}$. Conveniently, imposing a maximum permissible relative CI is equivalent to imposing a minimum number of observed error events $N_{\text{error,min}}$, where $N_{\text{error,min}}$ is almost independent of the error rate but depends practically only on the desired relative CI (cf. Table 2.1). This justifies the widespread practice of simulating up to a predefined number of error events.

|           | 20% | 10% | 5%   | 3%   | 1%    |
|-----------|-----|-----|------|------|-------|
| $10^{-3}$ | 97  | 385 | 1537 | 4266 | 38379 |
| $10^{-5}$ | 97  | 386 | 1538 | 4270 | 38417 |
| $10^{-7}$ | 97  | 386 | 1538 | 4270 | 38417 |
| $10^{-9}$ | 97  | 386 | 1538 | 4270 | 38417 |

Table 2.1.: Minimum number of observed error events $N_{\text{error,min}}$ to ensure a certain relative CI (horizontal) at a certain error rate (vertical)

Throughout this thesis we typically plot error rates (lines of different colors, dash patterns and markers) together with their Wilson CI (semi-transparent band). A common simulation termination criterion is relative Wilson CI of 3% or 5%.

## 2.8. Quantization

In this thesis we restrict ourselves to *mid-tread uniform quantization*, *i.e.*, schemes with an odd number of levels that are spaced uniformly and symmetrically around and including 0. An *N-level quantization scheme* is associated with a set of *labels* $\mathcal{Q}_N \triangleq \{0, \pm 1, \ldots, \pm \frac{N-1}{2}\} \subseteq \mathbb{Z}$ (denoted in monospaced font), $|\mathcal{Q}_N| = N$, each of which has a *reconstruction value* $x_q \triangleq 2\delta q$ according to the chosen *quantization threshold* $\delta$.

Any $x \in \mathbb{R}$ is quantized to the nearest reconstruction value using the *quantizer*

$$f_{Q(N,\delta)} \colon \mathbb{R} \to \mathcal{Q}_N, x \mapsto \underset{q \in \mathcal{Q}_N}{\arg\min} |x - x_q|, \qquad (2.60)$$

with ties broken towards 0. This partitions $\mathbb{R}$ into *decision regions*

$$\mathcal{R}_q = \left\{ x \in \mathbb{R} \,\middle|\, f_{Q(N,\delta)}(x) = q \right\} = \begin{cases} (-\infty, -(N-2)\delta) & \text{if } q = -\frac{N-1}{2}, \\ \left[ 2\delta(q - \frac{1}{2}), 2\delta(q + \frac{1}{2}) \right) & \text{if } q \in \{-1, \ldots, -\frac{N-3}{2}\}, \\ [-\delta, +\delta] & \text{if } q = 0, \\ \left( 2\delta(q - \frac{1}{2}), 2\delta(q + \frac{1}{2}) \right] & \text{if } q \in \{+1, \ldots, +\frac{N-3}{2}\}, \\ (+(N-2)\delta, +\infty) & \text{if } q = +\frac{N-1}{2}, \end{cases} \qquad (2.61)$$

with *decision boundaries* at $\pm(2i - 1)\delta$ for $i \in \left\{ 1, \ldots, \frac{N-1}{2} \right\}$. For examples of 3-level and 7-level quantization, cf. Figure 2.15.

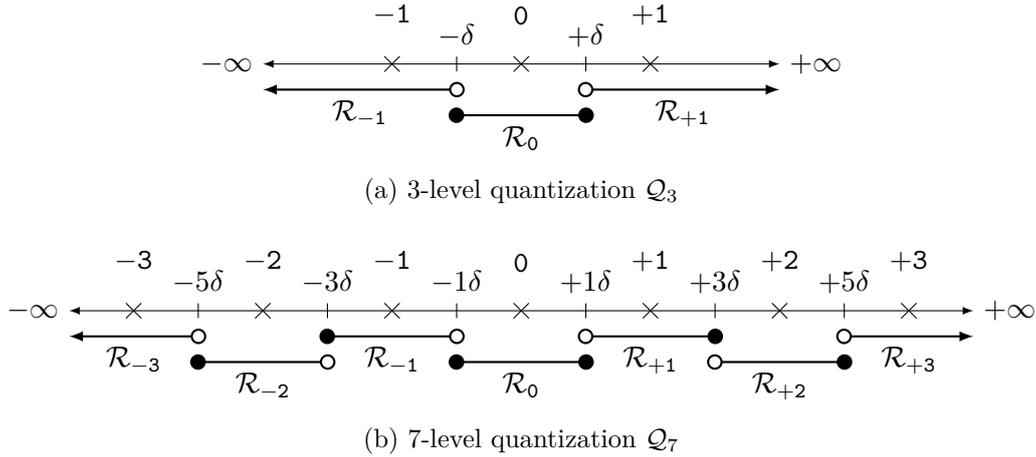

(a) 3-level quantization $\mathcal{Q}_3$

(b) 7-level quantization $\mathcal{Q}_7$

Figure 2.15.: Labels (monospaced font), reconstruction values (crosses), decision regions $\mathcal{R}_q$, and decision boundaries $\pm k\delta$, $k$ odd, for 3- and 7-level quantization with quantization threshold $\delta$

In this thesis, we assume that uniform quantization is applied to obtain quantized LLRs $\tilde{\boldsymbol{\lambda}}$ from the unquantized LLRs $\boldsymbol{\lambda}$ computed from the channel output $\boldsymbol{y}$ using the BiAWGN's LLR function. We refer to this scheme as *quantized BiAWGN channel*, $Q(N, \delta)$-BiAWGN($\sigma^2$), illustrated in Figure 2.16. The advantage is that $\tilde{\boldsymbol{\lambda}}$ are uniformly spaced so that there is a natural match between (uniformly) quantized channel output LLRs and quantized messages processed within a quantized decoder. The study of non-uniform quantization schemes is beyond the scope of this thesis.

In the case of 3-level quantization, we pick $\delta$ as to maximize the capacity of the result-ing $Q(3, \delta)$-BiAWGN$(\sigma^2)$. In Chapter 6 we discuss quantization threshold selection in detail, and argue why capacity maximization is suitable for 3-level quantization. Furthermore, we adopt a modification: We view the $Q(3, \delta)$-BiAWGN$(\sigma^2)$ as an equivalent BEEC$(p, e)$ with suitable $p$ and $e$ chosen such that the channel laws are identical. Instead of the reconstruction values $(x_{-1}, x_0, x_{+1}) = (-2\delta, 0, +2\delta)$ as described above for uniform quantization with threshold $\delta$, we use $(x_{-1}, x_0, x_{+1}) = (-\Delta, 0, +\Delta)$:

- Where we feed the quantized channel output into an unquantized decoder, we use $\Delta = \log\left(\frac{1-p-e}{p}\right)$. Hence, we use the exact 'ideal' LLRs for the BEEC$(p, e)$, and thus all losses seen in comparison to unquantized channel/decoder can be attributed to the quantized channel output rather than to suboptimal decoding.

- Where we feed the quantized channel output into a 3-level quantized decoder, we use $\Delta = 1$. This simplifies implementation, while simulations suggest that the performance is robust to such a choice, as long as the path metric updates (cf. (2.35) and Figure 3.2) for all messages $-1$, $0$, and $+1$ are well distinct.

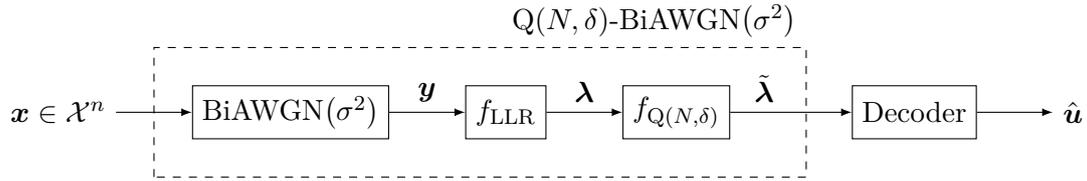

Figure 2.16.: The concatenation of BiAWGN channel, LLR computation, and quantizer gives rise to an equivalent $Q(N, \delta)$-BiAWGN$(\sigma^2)$ channel

# 3. Enhanced Quantized Polar Decoders

The objective of this chapter is to study the effect of 3-level quantization on the decoding of polar codes, and to devise and motivate strategies to mitigate the impact of coarse quantization. In Section 2.3 we revisit the unquantized SC decoder. In Section 3.1 we generalize it to the $\mathcal{L}$-SC decoder which operates using an arbitrary message alphabet $\mathcal{L}$, $e.g.$, 3-level quantized LLRs. We show that the empirical performance penalty from 3-level quantization is in line with the penalty predicted from theory. Like we do in Section 2.4 for the unquantized decoder, we subsequently introduce the $\mathcal{L}$-SCL decoder in Section 3.2 as list decoder extension of the $\mathcal{L}$-SC decoder. Similar to the unquantized case, $\mathcal{L}$-SCL improves considerably upon plain $\mathcal{L}$-SC performance at the cost of linearly (in the list size) increasing complexity. Due to LLR and subsequently also PM quantization, the order of paths in quantized SCL's list according to PM does not reflect the likelihood order. Hence in Section 3.3, we modify $\mathcal{L}$-SCL into $\mathcal{L}$-SCL-ML where the winning codeword is not chosen based on PM but through a successive ML-among-list step. In some situations, the $\mathcal{L}$-SCL-ML already approaches the performance of the unquantized decoder. In other situations, cf. Section 3.4, the poor list-FER of $\mathcal{L}$-SCL restricts the gains from $\mathcal{L}$-SCL-ML and necessitates further list enhancement techniques, which are developed in Chapter 4.

## 3.1. Quantized SC Decoder

In Chapter 2 (cf. Sections 2.3 and 2.5) we revisit how the recursive structure of (unquantized) SC decoding can be used to interpret SC decoding as message-passing procedure over suitably chosen decoding trees (cf. Figure 2.7). In a decoder, it is implemented as a sequence of check node operations ($\boxplus$) and variable node operations ($\bullet$) applied to LLR values in order to successively obtain estimates for the information bits from channel output LLRs. This is a natural point to 'hook into' the SC decoding procedure to devise quantized variants of it: One isolates the SC algorithm ($i.e.$, the sequence of operations) from the data type of the LLR values and the check node and variable node operations defined for it ($i.e.$, the 'meaning' or implementation of these operations).

An $\mathcal{L}$-SC decoder follows the sequence of operations of SC decoding described in Sections 2.3 and 2.5, but uses LLR values from a set $\mathcal{L}$ for which two basic operations,

$$\boxplus \colon \mathcal{L} \times \mathcal{L} \to \mathcal{L}, \qquad \text{'check node operation'} \tag{3.1}$$

$$\bullet \colon \mathcal{L} \times \mathcal{L} \to \mathcal{L}, \qquad \text{'variable node operation'} \tag{3.2}$$

are defined. All other operations occurring in SC decoding, like variable node operations depending on whether the neighboring check node operation in the factor graph 'forwards' a 0 or 1, are reduced to the two basic operations (3.1) and (3.2).

The breakdown into SC algorithm and LLR data type is not only of theoretical value, where it makes clear what belongs to the SC algorithm (*i.e.*, cannot be changed without modifying the 'nature' of SC) and what belongs to the representation and implementation of LLR values (*i.e.*, depends, *e.g.*, on whether LLRs are quantized or not). It is also of immense practical value because a software implementation of an $\mathcal{L}$-SC decoder is agnostic w.r.t. $\mathcal{L}$, $\boxplus$ and $\bullet$, *i.e.*, it can be re-used as unquantized or quantized decoder upon 'plugging in' an appropriate definition for $\mathcal{L}$, $\boxplus$ and $\bullet$. Implemented in a programming language providing a sufficiently developed concept of data types (*e.g.*, Julia [Bez+17]), the difference between an unquantized SC decoder, an approximate unquantized SC decoder (*e.g.*, using the min-approximation for $\boxplus$), and a quantized SC decoder consists of only very few lines of code.

We introduce some $\mathcal{L}$-SC decoders used throughout our work in the following. The unquantized SC decoder is referred to as $\mathcal{L}_\infty$-SC decoder with $\mathcal{L}_\infty \triangleq \mathbb{R}$, and min-sum approximated check node and variable node operations defined as

$$\lambda_1 \boxplus \lambda_2 \triangleq \operatorname{sign}(\lambda_1) \operatorname{sign}(\lambda_2) \min\{|\lambda_1|, |\lambda_2|\}, \tag{3.3}$$

$$\lambda_1 \bullet \lambda_2 \triangleq \lambda_1 + \lambda_2, \tag{3.4}$$

for $\lambda_1, \lambda_2 \in \mathcal{L}_\infty$. A derivative of the $\mathcal{L}_\infty$-SC decoder is the $\mathcal{L}_{\widetilde{\infty}}$-SC decoder where LLRs are very finely quantized using a function $f_{Q,\widetilde{\infty}}$ that discards all but three bits of the mantissa of the LLR value in IEEE 754 floating-point *binary64*/'*double*' format [IEEE 754-2008]. The performance of decoders using $\mathcal{L}_{\widetilde{\infty}}$ is indistinguishable from those using $\mathcal{L}_\infty$, but $\mathcal{L}_{\widetilde{\infty}} \triangleq \left\{ \widetilde{x} \,\middle|\, \exists x \in \mathbb{R} : \widetilde{x} = f_{Q,\widetilde{\infty}}(x) \right\}$ is discrete and thus the decoder is straight forward to analyze using quantized density evolution (cf. Section 2.5, (2.41)). The check node and variable node operations for $\lambda_1, \lambda_2 \in \mathcal{L}_{\widetilde{\infty}}$ are defined as

$$\lambda_1 \boxplus \lambda_2 \triangleq f_{Q,\widetilde{\infty}}(\operatorname{sign}(\lambda_1) \operatorname{sign}(\lambda_2) \min\{|\lambda_1|, |\lambda_2|\}), \tag{3.5}$$

$$\lambda_1 \bullet \lambda_2 \triangleq f_{Q,\widetilde{\infty}}(\lambda_1 + \lambda_2). \tag{3.6}$$

The 3-level quantized $\mathcal{L}_3$-SC decoder uses one of three LLR values from the alphabet $\mathcal{L}_3 \triangleq \{0, \pm 1\}$, with check node and variable node operations defined analogous to the min-sum approximation but clipped to $\mathcal{L}_3$. Lookup tables for $\boxplus$ and $\bullet$ over $\mathcal{L}_3$ are depicted in Table 3.1.

| $\boxplus$ | $-1$ | $0$ | $+1$ |
|---|---|---|---|
| $-1$ | $+1$ | $0$ | $-1$ |
| $0$ | $0$ | $0$ | $0$ |
| $+1$ | $-1$ | $0$ | $+1$ |

(a) Check node operation in $\mathcal{L}_3$

| $\bullet$ | $-1$ | $0$ | $+1$ |
|---|---|---|---|
| $-1$ | $-1$ | $-1$ | $0$ |
| $0$ | $-1$ | $0$ | $+1$ |
| $+1$ | $0$ | $+1$ | $+1$ |

(b) Variable node operation in $\mathcal{L}_3$

Table 3.1.: Check node operation (a) and variable node operation (b) in $\mathcal{L}_3 = \{0, \pm 1\}$

Figure 3.1 shows the performance of unquantized $\mathcal{L}_\infty$-SC decoding over unquantized BiAWGN (cf. —, -+-) and 3-level quantized BiAWGN (cf. —, -+-) as well as the performance of quantized $\mathcal{L}_3$-SC decoding over 3-level quantized BiAWGN (cf. —, -+-). Two block lengths are considered, $n = 256$ (cf. —, —, —) and $n = 128$ (cf. -+-, -+-, -+-), for which codes `DE-Q3-450-01` and `DE-Q3-450-02`, both rate $R = 1/2$, have been designed using density evolution (cf. Sections 2.6.1 and 2.6.3). For both $n \in \{128, 256\}$ and FER $10^{-3}$, a loss of $0.8\,\mathrm{dB}$ in $\mathrm{E_b/N_0}$ is caused by channel output quantization (cf. — vs. —, -+- vs. -+-) and a further loss of $1.2\,\mathrm{dB}$ in $\mathrm{E_b/N_0}$ is caused by quantized decoding (cf. — vs. —, -+- vs. -+-). These losses are in the order of what was to be expected based on the asymptotic analysis of 3-level quantization in [HU12a; HU12b; Has13] (cf. Figure 1.2).

Due to the reduced cost of coarsely quantized operations vs. finely quantized operations in terms of integrated circuit complexity, surface area, and energy consumption, a coarsely quantized SCL decoder could come at comparable cost of a finely quantized SC decoder. To make this comparison, we introduce and analyze quantized SCL decoding in the following section.

## 3.2. Quantized SCL Decoder

Analogous to the extension of unquantized SC decoding into unquantized SCL decoding in Section 2.4, the $\mathcal{L}$-SC decoder is extended into the $\mathcal{L}$-SCL decoder. The only adjustment concerns the interface between LLRs and PMs, where the PM update function needs to be compatible with the LLR type. For clarity, let $\mathcal{P}$ denote the set of possible PM values, *i.e.*, the data type of the PM employed in the $\mathcal{L}$-SCL decoder. Note that throughout this work, $\mathcal{P} \triangleq \mathbb{R}$, but we use $\mathcal{P}$ to make clear when functions take PMs as argument. The path metric update function $f_{\mathrm{PMU}} : \mathcal{P} \times \mathcal{L} \times \{0, 1\} \to \mathcal{P}$ computes the new PM $\text{PM}_{\ell'} \triangleq f_{\text{PMU}}(\text{PM}_\ell, \lambda, u)$ of a path $\ell'$ which results from a path $\ell$ with PM $\text{PM}_\ell$ by appending the symbol $u$ for the current $i$-th bit whose LLR was computed to $\lambda$.

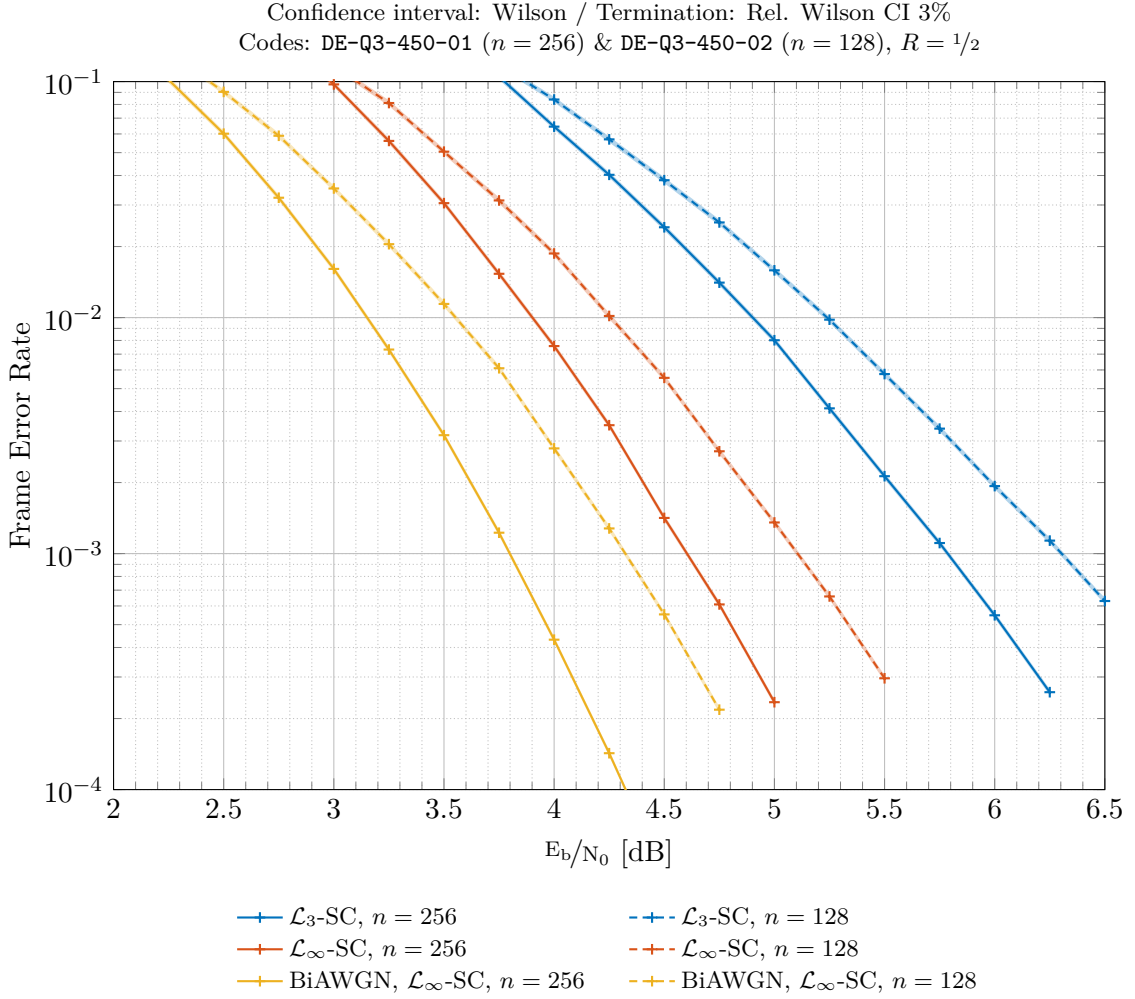

Figure 3.1.: FER of $\mathcal{L}_\infty/\mathcal{L}_3$-SC (varying block length $n \in \{128, 256\}$ and code), applied to $Q(3, \delta^*_{\text{cap}})$-BiAWGN($\sigma^2$) (default) or BiAWGN($\sigma^2$) (where stated)

For the unquantized $\mathcal{L}_\infty$-SCL decoder and the quasi-unquantized $\mathcal{L}_{\widetilde{\infty}}$-SCL decoder, the PM update function is given in the usual form,

$$f_{\text{PMU}}(\text{PM}, \lambda, u) \triangleq \text{PM} + \ln(1 + \exp(-(-1)^u \lambda)), \tag{3.7}$$

which we approximate as piecewise linear function with three pieces (cf. Figure 3.2),

$$
f_{\mathrm{PMU}}(\mathrm{PM}, \lambda, u) \approx \tilde{f}_{\mathrm{PMU}}(\mathrm{PM}, \lambda, u) \triangleq \mathrm{PM} +
\begin{cases}
(-1)^{1-u}\lambda & \text{if } (-1)^u\lambda < -2\ln(2), \\
\frac{1}{2}(-1)^{1-u}\lambda + \ln(2) & \text{if } |\lambda| \leq 2\ln(2), \\
0 & \text{if } (-1)^u\lambda > +2\ln(2).
\end{cases}
\tag{3.8}
$$

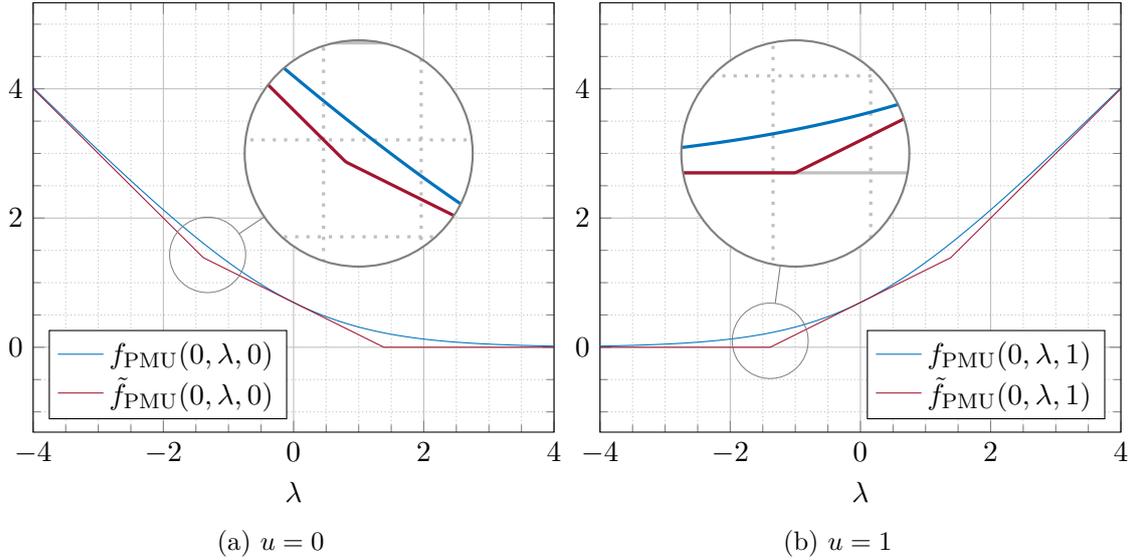

(a) $u = 0$          (b) $u = 1$

Figure 3.2.: PM update function $f_{\mathrm{PMU}}(\mathrm{PM}, \lambda, u)$ and its approximation $\tilde{f}_{\mathrm{PMU}}(\mathrm{PM}, \lambda, u)$ for $u = 0$ (cf. (a)) and $u = 1$ (cf. (b)); PM adds an offset (here: $\mathrm{PM} = 0$)

For quantized LLRs, *e.g.*, $\lambda \in \mathcal{L}_3$, their reconstruction value $x_\lambda$ is used for PM update,

$$
f_{\mathrm{PMU}} \colon \mathcal{P} \times \mathcal{L}_3 \times \{0, 1\} \to \mathcal{P}, (\mathrm{PM}, \lambda, u) \mapsto \tilde{f}_{\mathrm{PMU}}(\mathrm{PM}, x_\lambda, u).
\tag{3.9}
$$

Note that quantized LLRs undergo severe distortion due to rounding and clipping. Hence they provide only rough proxies for the true log-likelihood ratios. This carries over to PMs: While in the unquantized decoder the order among PMs preserves the order in likelihood, this is not the case if PM computation is based on imprecise LLRs. Furthermore, as a result of coarse LLR quantization PMs become de-facto quantized. Both effects render PMs little useful for discriminating paths, as we see in the following. Figures 3.3 and 3.4 show simulation results for unquantized $\mathcal{L}_\infty$-SCL decoding over unquantized BiAWGN (yellow) and 3-level quantized BiAWGN (orange), as well as the performance of quantized $\mathcal{L}_3$ decoding over 3-level quantized BiAWGN (blue), for $n = 256$ and $n = 128$, respectively, and different list sizes $L \in \{1, 32, 128\}$. All scenarios

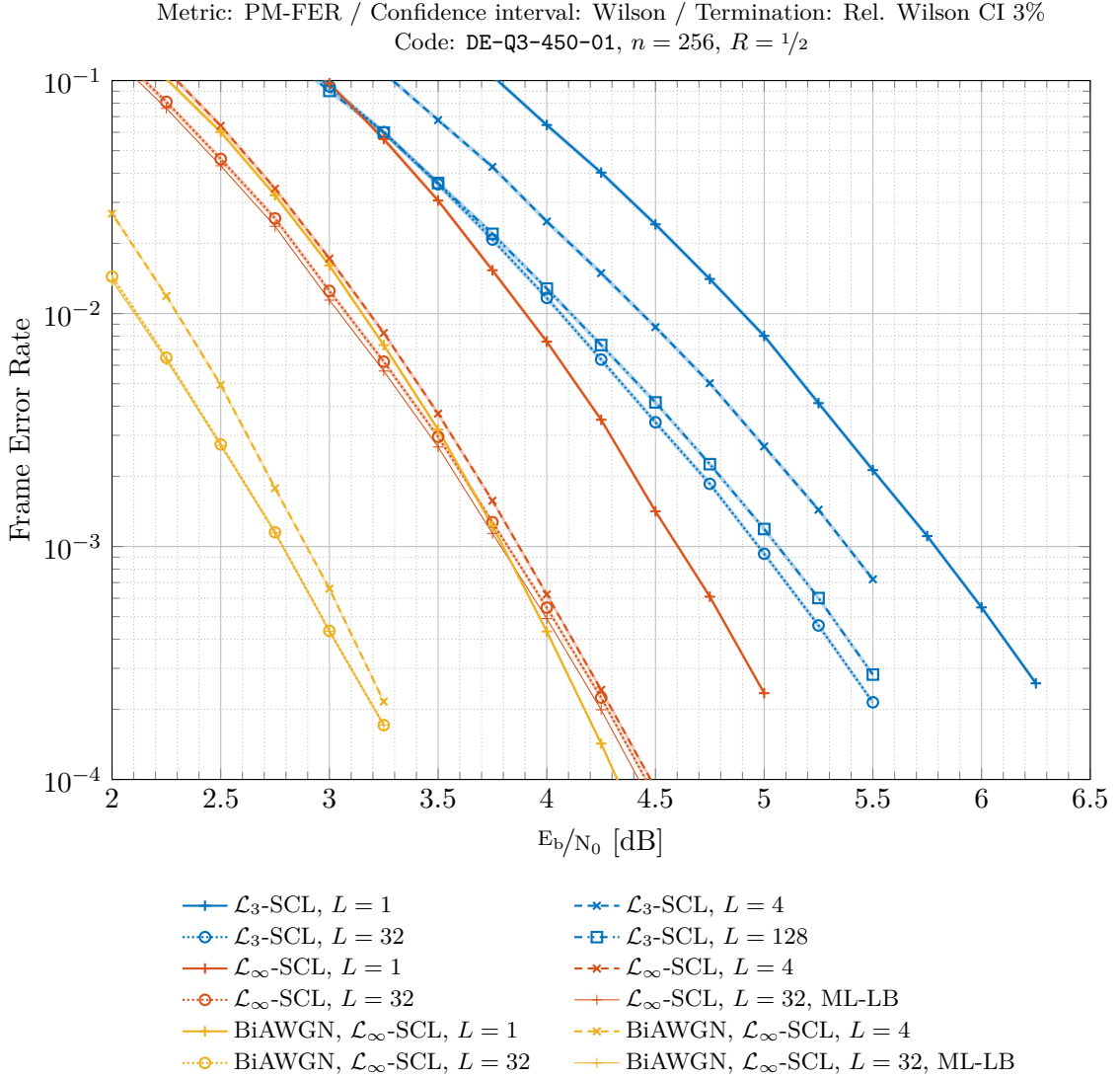

Figure 3.3.: PM-FER of $\mathcal{L}_\infty/\mathcal{L}_3$-SCL (varying list size $L$), applied to $Q(3, \delta^*_{\mathrm{cap}})$-BiAWGN$(\sigma^2)$ (default) or BiAWGN$(\sigma^2)$ (where stated)

benefit from SCL decoding vs. SC decoding (cf. ━╋━ vs. ⋯◦⋯, ━╋━ vs. ⋯◦⋯, ━╋━ vs. ⋯◦⋯), the gains are generally more pronounced for $n = 256$ than for $n = 128$.

The gains from list decoding saturate for $L = 32$, larger list sizes do not lead to further improvements: $\mathcal{L}_\infty$-SCL saturates to its ML-LB for unquantized and 3-level quantized BiAWGN (cf. ⋯◦⋯ vs. ─┼─, ⋯◦⋯ vs. ─┼─). (For $n = 128$ and $\mathcal{L}_\infty$-SCL over 3-level quantized BiAWGN, there remains a gap between PM-FER and ML-LB for $L = 32$, but the SCL performance does not improve for larger $L$, not shown in the plots.) Similarly for $\mathcal{L}_3$-SCL over 3-level quantized BiAWGN. Here, larger lists can even lead to performance

Metric: PM-FER / Confidence interval: Wilson / Termination: Rel. Wilson CI 3%
Code: `DE-Q3-450-02`, $n = 128$, $R = \frac{1}{2}$

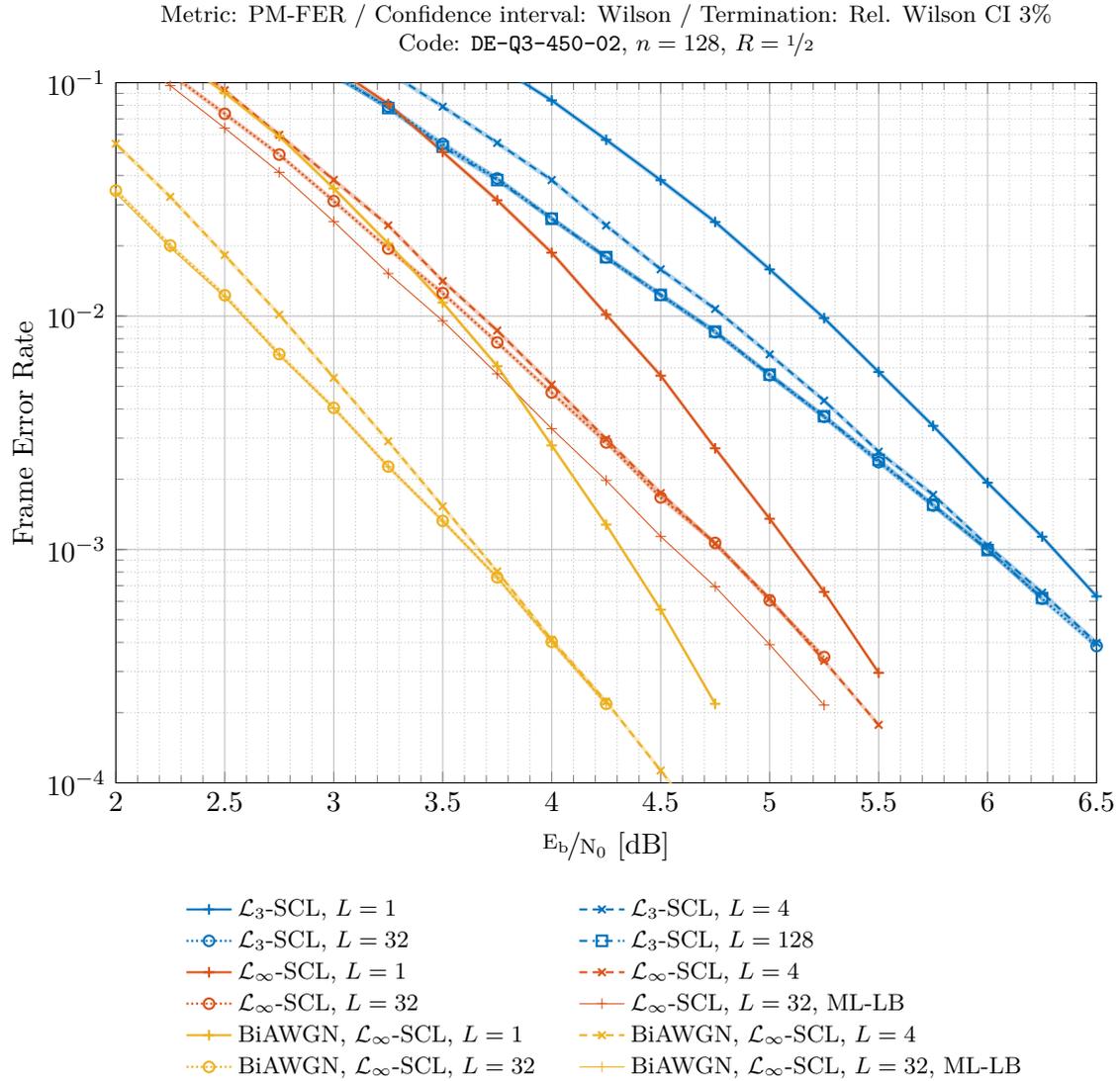

Figure 3.4.: PM-FER of $\mathcal{L}_\infty/\mathcal{L}_3$-SCL (varying list size $L$), applied to $Q(3, \delta^*_{\text{cap}})$-BiAWGN$(\sigma^2)$ (default) or BiAWGN$(\sigma^2)$ (where stated)

degradation because of PM quantization (cf. ⊶⊸⊶ vs. ⊢⊟⊣): The larger the list, the more likely will it contain two or more paths with identical best PM. The decoder flips a coin to decide the winning path, which, more often than not with increasing list size, will be wrong. Observe that a fairly small list size $L = 4$ suffices to achieve close to best possible performance (cf. ⊶✕⊶ vs. ⊶⊸⊶, ⊢✕⊣ vs. ⊶⊸⊶, ⊢✕⊣ vs. ⊶⊸⊶). The only exception is $\mathcal{L}_3$-SCL over 3-level quantized BiAWGN for $n = 256$, where $L = 4$ achieves only half of the gain of $L = 32$ (cf. ⊢✕⊣ vs. ⊶⊸⊶).

As expected, $\mathcal{L}_3$-SCL decoding improves over $\mathcal{L}_3$-SC decoding, 0.8 dB in $E_b/N_0$ for $n =$

256 and $0.3\,\mathrm{dB}$ in $\mathrm{E_b/N_0}$ for $n = 128$, both at FER $10^{-3}$ (cf. ⌀⋯ vs. ⊹⊹). However, the same gains are obtained for $\mathcal{L}_\infty$-SCL vs. $\mathcal{L}_\infty$-SC over 3-level quantized BiAWGN (cf. ⌀⋯ vs. ⊹⊹), so that the performance gap between $\mathcal{L}_3$- and $\mathcal{L}_\infty$-decoding remains unaltered $1.2\,\mathrm{dB}$ (cf. ⌀⋯ vs. ⌀⋯, ⊹⊹ vs. ⊹⊹). Note that for unquantized BiAWGN, $\mathcal{L}_\infty$-SCL leads to even higher gains over $\mathcal{L}_\infty$-SC (cf. ⌀⋯ vs. ⊹⊹).

Confidence interval: Wilson / Termination: Rel. Wilson CI 3% PM-FER
Codes: `DE-Q3-450-01` ($n = 256$) & `DE-Q3-450-02` ($n = 128$), $R = \frac{1}{2}$

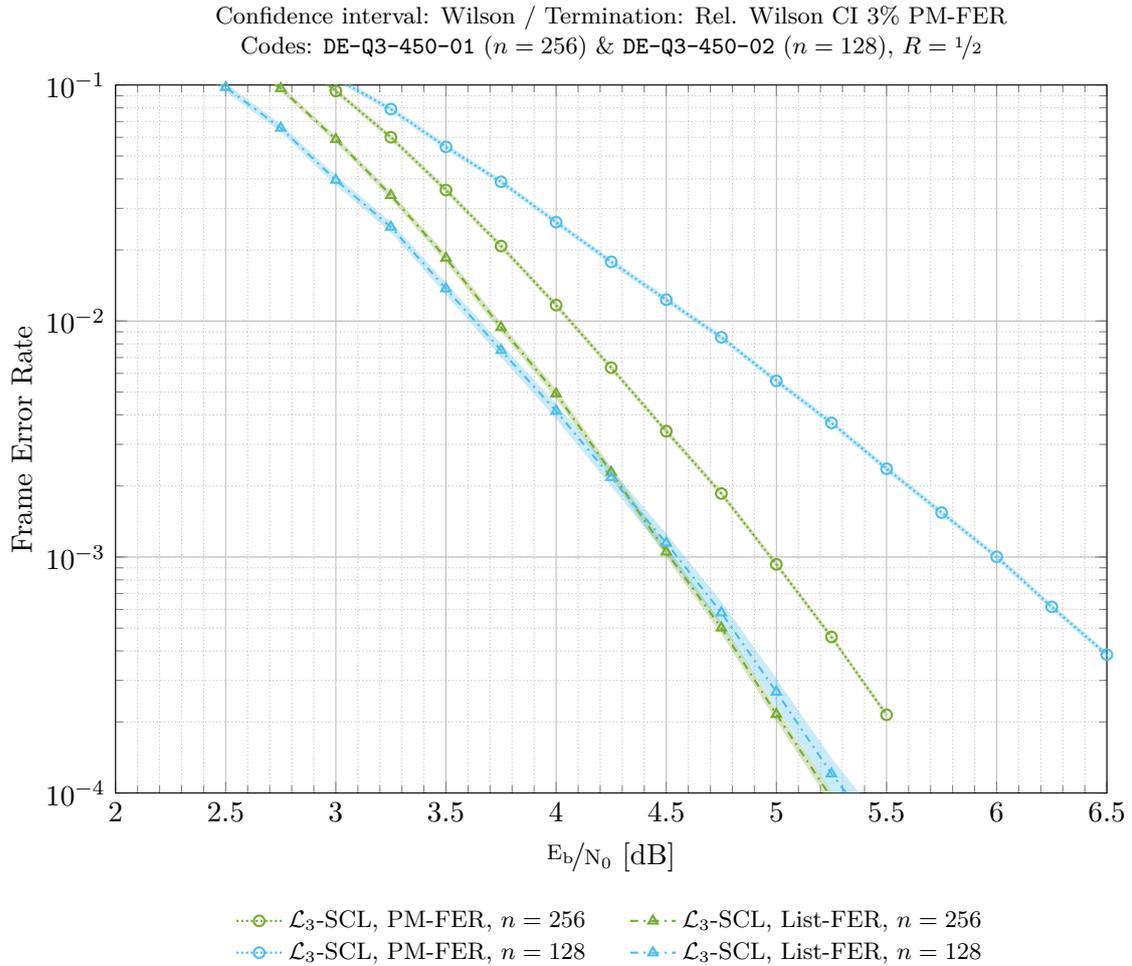

Figure 3.5.: PM-/List-FER of $\mathcal{L}_3$-SCL ($L = 32$, varying code and block length $n$), applied to Q(3, $\delta^*_{\mathrm{cap}}$)-BiAWGN($\sigma^2$)

Figure 3.5 shows the PM-FER and list-FER of $\mathcal{L}_3$-SCL decoding for $L = 32$ and varying codes, *i.e.*, the probability of successful decoding when picking the winning codeword from the list based on PM and the probability that the correct codeword is contained in the list, respectively. Observe the gap at FER $10^{-3}$ of $0.5\,\mathrm{dB}$ in $\mathrm{E_b/N_0}$ for $n = 256$ (cf. ⌀⋯ vs. ⊹⋯) and of $1.5\,\mathrm{dB}$ in $\mathrm{E_b/N_0}$ for $n = 128$ (cf. ⌀⋯ vs. ⊹⋯). This suggests that often the true codeword is contained in the list resulting from SCL, but it is not declared winning according to PM. For unquantized SCL (*i.e.*, as in [BPB15; TV15]) selection based on PM is equivalent to selection based on likelihood. As reasoned before in this section, this is not the case for quantized SCL. Therefore we restore likelihood-based selection from the SCL decoder's list in the following section, which reclaims some of the gap between PM-FER and list-FER.

## 3.3. Quantized SCL Decoder with Likelihood-Based Selection Among List

Recall the *maximum likelihood* (ML) decision rule, which chooses the codeword $\hat{\boldsymbol{c}}$ from a codebook $\mathcal{C}_{\mathrm{code}}$ maximizing the probability of the observed channel output sequence $\boldsymbol{y}$ according to the channel law $P_{\boldsymbol{Y}|\boldsymbol{C}}$,

$$\hat{\boldsymbol{c}} = \arg\max_{\boldsymbol{c} \in \mathcal{C}_{\mathrm{code}}} P(\boldsymbol{y}|\boldsymbol{c}). \tag{3.10}$$

In general, ML decoding is not feasible due to exponential complexity. Yet, given a small list $\mathcal{C}_{\mathrm{list}} \subseteq \mathcal{C}_{\mathrm{code}}$ of candidate codewords, ML can be used to pick the most probable among them, in complexity linear in $|\mathcal{C}_{\mathrm{list}}|$. This is used to enhance the $\mathcal{L}$-SCL decoder into the $\mathcal{L}$-SCL-ML decoder, which uses an additional ML-among-list step rather than PM to select the winning codeword based on the list output by SCL.

Figure 3.6 shows the performance of $\mathcal{L}_3$-SCL and $\mathcal{L}_3$-SCL-ML compared to $\mathcal{L}_\infty$-SCL over the 3-level quantized BiAWGN for $n = 128$. While $\mathcal{L}_3$-SCL PM-FER performance saturates at $L = 32$ (cf.  vs. ), the additional ML-among-list step of $\mathcal{L}_3$-SCL-ML leads to a gain of 1.1 dB and 1.2 dB in $\mathrm{E_b/N_0}$ for $L = 32$ and $L = 128$, respectively, at FER $10^{-3}$ (cf.  vs. ,  vs. ). For $L = 32$ and FER $10^{-3}$, the performance of $\mathcal{L}_3$-SCL-ML is already close (less than 0.1 dB in $\mathrm{E_b/N_0}$) to that of $\mathcal{L}_\infty$-SCL (cf.  vs. ), for $L = 128$ it is practically indistinguishable (cf.  vs. ).

The situation for $n = 256$ shown in Figure 3.7 is similar but different: PM-FER performance of $\mathcal{L}_3$-SCL still saturates at $L = 32$ (cf.  vs. ). The additional ML-among-list step of $\mathcal{L}_3$-SCL-ML leads to a gain of 0.4 dB and 0.8 dB in $\mathrm{E_b/N_0}$ for $L = 32$ and $L = 128$, respectively, at FER $10^{-3}$ (cf.  vs. ,  vs. ). However, $\mathcal{L}_3$-SCL-ML does not saturate to the performance of $\mathcal{L}_\infty$-SCL easily (cf.  vs. ), but is limited by the list-FER of $\mathcal{L}_3$-SCL (cf.  vs. ). Note that as $L$ increases, the list-FER of $\mathcal{L}_3$-SCL likely improves further and so would the performance of $\mathcal{L}_3$-SCL-ML, but the comparison between list-FER at $L = 32$ and $L = 128$ suggests that this improvement is slow in $L$ and hence comes at great (perhaps unbearable) cost due to large list sizes (cf.  vs. ,  vs. ).

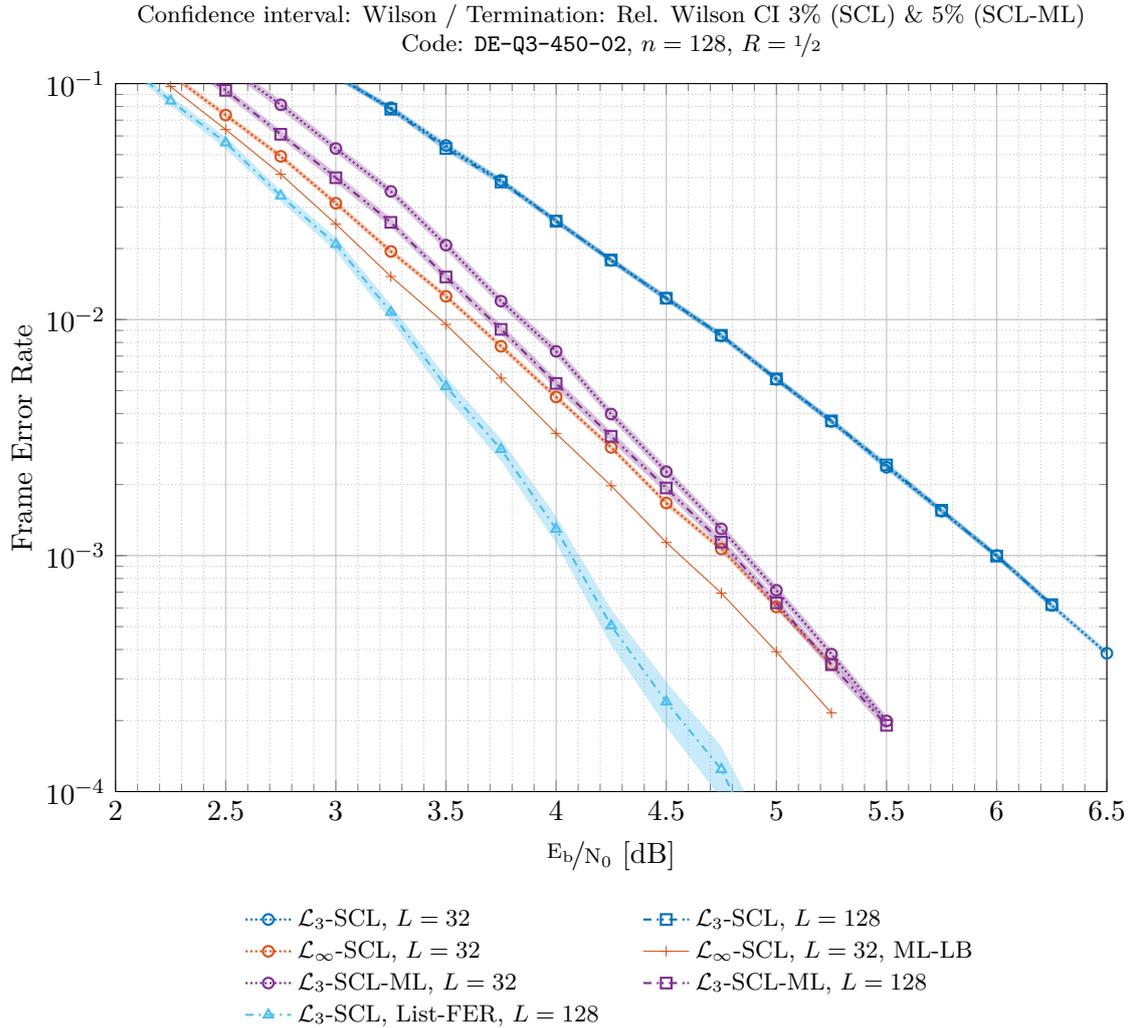

Confidence interval: Wilson / Termination: Rel. Wilson CI 3% (SCL) & 5% (SCL-ML)
Code: `DE-Q3-450-02`, $n = 128$, $R = \frac{1}{2}$

- ···◯··· $\mathcal{L}_3$-SCL, $L = 32$
- ··◻·· $\mathcal{L}_3$-SCL, $L = 128$
- ···◯··· $\mathcal{L}_\infty$-SCL, $L = 32$
- —+— $\mathcal{L}_\infty$-SCL, $L = 32$, ML-LB
- ···◯··· $\mathcal{L}_3$-SCL-ML, $L = 32$
- ··◻·· $\mathcal{L}_3$-SCL-ML, $L = 128$
- – △ – $\mathcal{L}_3$-SCL, List-FER, $L = 128$

Figure 3.6.: PM-/LML-FER of $\mathcal{L}_\infty$/$\mathcal{L}_3$-SCL/-SCL-ML (varying list size $L$), applied to $Q(3, \delta^*_{\mathrm{cap}})$-BiAWGN($\sigma^2$)

In this section we showed that $\mathcal{L}_3$-SCL-ML (*i.e.*, $\mathcal{L}_3$-SCL with ML-based rather than PM-based selection of the winning codeword from the list) can lead to major performance improvements over $\mathcal{L}_3$-SCL. In some scenarios it matches the performance of the unquantized decoder (cf. Figure 3.6), in other scenarios its performance is limited by a poor list-FER of $\mathcal{L}_3$-SCL (cf. Figure 3.7). This leads to a need for list enhancement techniques which we discuss in detail in the following section. We reiterate the fact that an additional ML-among-list step would not benefit the unquantized SCL decoder, where selection based on PM guarantees ML-among-list [BPB15].

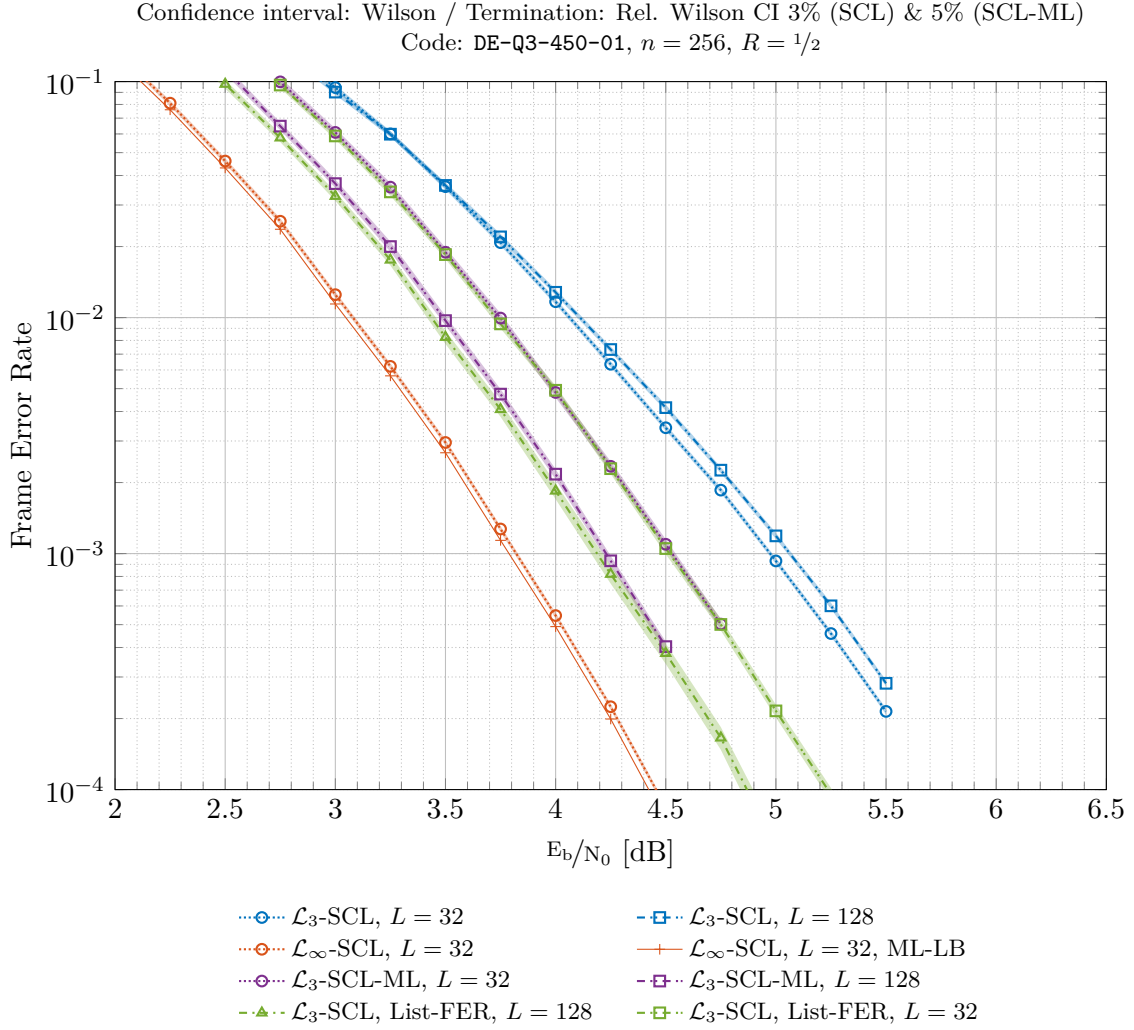

Confidence interval: Wilson / Termination: Rel. Wilson CI 3% (SCL) & 5% (SCL-ML)
Code: `DE-Q3-450-01`, $n = 256$, $R = \frac{1}{2}$

Legend:
- $\mathcal{L}_3$-SCL, $L = 32$
- $\mathcal{L}_\infty$-SCL, $L = 32$
- $\mathcal{L}_3$-SCL-ML, $L = 32$
- $\mathcal{L}_3$-SCL, List-FER, $L = 128$
- $\mathcal{L}_3$-SCL, $L = 128$
- $\mathcal{L}_\infty$-SCL, $L = 32$, ML-LB
- $\mathcal{L}_3$-SCL-ML, $L = 128$
- $\mathcal{L}_3$-SCL, List-FER, $L = 32$

Figure 3.7.: PM-/LML-FER of $\mathcal{L}_\infty/\mathcal{L}_3$-SCL/-SCL-ML (varying list size $L$), applied to $Q(3, \delta_{\mathrm{cap}}^*)$-BiAWGN($\sigma^2$)

## 3.4. Need for List Enhancement Techniques

In the previous section we show that $\mathcal{L}_3$-SCL-ML can improve upon the performance of $\mathcal{L}_3$-SCL up to either (a) the PM-FER of unquantized $\mathcal{L}_\infty$-SCL decoding, or (b) the list-FER of $\mathcal{L}_3$-SCL, whichever is higher. Point (a) is a property of the code, more precisely of its distance profile and in particular its minimum-distance, and hence by the scope of our work something we cannot improve (recall that we assume the code as a given). Point (b) however is something we can influence by modifications of the decoder, as can be seen from the fact that $\mathcal{L}_3$-SCL and $\mathcal{L}_\infty$-SCL show different list-FERs.

Based on our observations we identify three factors influencing the performance of $\mathcal{L}_3$-

SCL-ML: (1) Poor list-FER of $\mathcal{L}_3$-SCL suggests that during SCL decoding the correct codeword is all too easily inadvertently removed from the list, due to LLR/PM quantization and thus inaccurateness. (2) PM-based selection among list does not correspond to likelihood-based selection among list, due to LLR/PM quantization. (3) Likelihood is not indicative of the correct codeword, due to poor distance properties of the code [TV15; MHU14]. Changes to (3) are not within our scope, see (a) before. Factor (2) is mitigated through the additional ML-among-list step in $\mathcal{L}_3$-SCL-ML. Factor (1) is both within the scope of our work and not addressed by the enhancements presented in this section, see (b) before.

Therefore, we investigate 'list enhancement techniques' in the following Chapter 4, *i.e.*, techniques to improve the list-FER by enhancing the mechanisms based on which codewords are selected for the list output by SCL.

Note that the use of a concatenated CRC, besides being out of scope as it constitutes a modification of the code, would most probably not mitigate (1): From the fact that $\mathcal{L}_3$-SCL-ML closes up to the list-FER of $\mathcal{L}_3$-SCL using ML-among-list[1], we can tell that the problem at hand stems from poor list-FER (which CRCs cannot directly improve) rather than from path likelihood not being indicative of the correct codeword in the list. However, other mechanisms that guide the selection of paths to be included in the list during SCL decoding rather than in the end, as proposed, *e.g.*, in [TM16; WQJ16; Yua+18], could be beneficial but are not investigated in this thesis.

---

[1]In fact, $\mathcal{L}_3$-SCL-ML closes up to the list-FER of $\mathcal{L}_3$-SCL using ML-among-list unless it saturates to $\mathcal{L}_\infty$-SCL before that, but in this case the original objective of matching $\mathcal{L}_\infty$-SCL performance is already achieved.

# 4. List Enhancement Techniques

In Sections 3.2 and 3.3, we show how $\mathcal{L}_3$-SCL and $\mathcal{L}_3$-SCL-ML decoding can recover some of the losses incurred by $\mathcal{L}_3$-SC decoding due to 3-level quantization (cf. Section 3.1). However, as argued in Section 3.4, the usually fairly high list-FER of quantized decoders puts a limit on the performance gains of these approaches. Thus, in this chapter, we devise techniques with which the selection of codewords for the final list is improved and thus the list-FER of accordingly enhanced decoders is lowered. Finally, we show that $\mathcal{L}_3$-SCL-ML is able to capitalize on the improved list-FER.

The fundamental idea is the following: Conforming with (or violating) a reliable bit estimate should cause a higher benefit (or penalty) on a path's likelihood metric during decoding than for an unreliable bit estimate. A 3-level quantized decoder looses (almost[1]) all information about the reliability of its bit decisions because the $\mathcal{L}_3$ LLR values contain (almost) no magnitude information. However, this reliability information is not all instantaneous. In fact, some bits are statistically more reliable than others. In an unquantized decoder, LLR values, and thus PM updates, for such more reliable bits are 'on average' larger in magnitude than for less reliable bits, whereas the 3-level quantized decoder treats all bit decision as (almost) equally reliable when updating PMs. Knowledge about the statistical reliability of bits (obtained from analyzing an unquantized decoder) can be used to refine the PM updates in 3-level quantized decoders.

In Section 4.1 we show how to obtain and exploit statistical information about the reliability of the bit decisions. We extend this idea in Section 4.2 where contradiction counts are introduced as a low-complexity indicator of instantaneous reliability to refine the statistics-based reliability assessment. We briefly mention directions for further enhancements in line with this philosophy in Section 4.3.

## 4.1. Expected Path Metric Updates

The declared goal of this section is to use information about the statistical behavior of an unquantized decoder to modify the behavior of a quantized decoder in such a way that it mimics the (optimal) behavior of the unquantized decoder 'as closely as possible'. In particular, we observe in Section 3.2 that quantized LLRs lead to quantized PMs, rendering a reliable comparison of list entries w.r.t. likelihood more difficult. To combat this effect and thus improve the selection of codewords for the list, we set out to emulate the PM update step of an unquantized decoder 'as good as possible' using only the (fewer) LLR information available in a quantized decoder, cf. Figure 4.1. Incorporating reliability information beyond the magnitude of the LLRs leads to increased differentiation among paths w.r.t. likelihood.

---

[1] Strictly speaking there is *some* reliability information in the magnitude $|\pm 1|$ as compared to $|0|$; but very little compared to, *e.g.*, the elements of $\mathbb{R}$.

Recall from Section 2.4 how PM updates work in plain SCL decoding: Successively for each $i$-th bit $u_i$, the decoder computes for each path $\ell$ in the list the LLR $\lambda_{i,\ell}$ using the decoding trees introduced in Section 2.3. The path $\ell$ with its $\mathrm{PM}_\ell$ is extended into two new paths $\ell_0$ and $\ell_1$ by appending 0 and 1 to the original path and with $\mathrm{PM}_{\ell_j} = \mathrm{PM}_\ell + \mathrm{PM}'_{\ell_j}$ and PM increments $\mathrm{PM}'_{\ell_j} = f_{\mathrm{PM}}(\lambda_{i,\ell}, j)$, for $j \in \{0, 1\}$, respectively. The computation of LLR $\lambda_{i,\ell}$ and PM increments $\mathrm{PM}'_{\ell_0}$ and $\mathrm{PM}'_{\ell_1}$ is illustrated for an unquantized decoder (exemplarily for $i = 3$) in Figure 4.1(a).

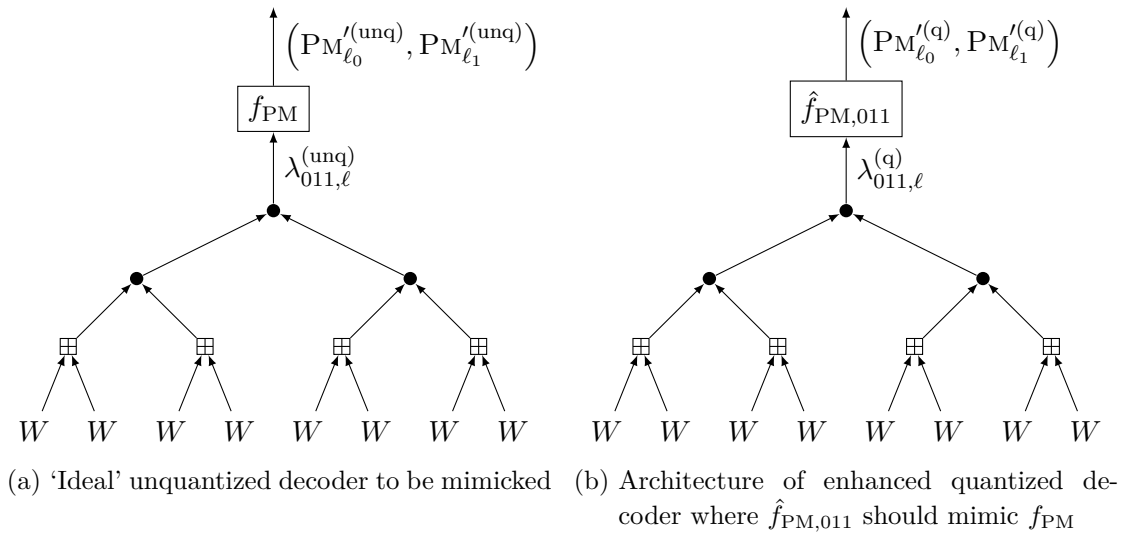

(a) 'Ideal' unquantized decoder to be mimicked

(b) Architecture of enhanced quantized decoder where $\hat{f}_{\mathrm{PM},011}$ should mimic $f_{\mathrm{PM}}$

Figure 4.1.: Decoding tree and PM increment computation of (a) unquantized decoder whose behavior is to be mimicked and (b) enhanced quantized decoder which should exhibit similar PM update behavior

The PM update of the unquantized SCL decoder is to be imitated by an enhanced quantized SCL decoder with the process depicted in Figure 4.1(b). The decoding tree is unchanged but uses quantized messages rather than unquantized messages to obtain the quantized LLR $\lambda^{(\mathrm{q})}_{011,\ell}$. Functions $\hat{f}_{\mathrm{PM},i}$ (which can be different for each $i$-th bit) are used in place of the original PM increment function $f_{\mathrm{PM}}$ to calculate the PM increments $PM'^{(q)}_{\ell_0}$ and $PM'^{(q)}_{\ell_1}$. The objective is to design all $\hat{f}_{PM,i}$ such that the PM increments $PM'^{(q)}_{\ell_j}$ of the enhanced quantized decoder are 'similar' to the PM increments $PM'^{(unq)}_{\ell_j}$ of the unquantized decoder.

Since the quantized decoder looses relevant information in the decoding trees due to coarse LLR quantization, it is generally not possible to imitate the unquantized decoder's PM update behavior exactly. In fact, in the 3-level quantized decoder each $\hat{f}_{PM,i}$ gets as input one out of the three LLR values in $\mathcal{L}_3$. Hence, while $f_{PM} \colon \mathbb{R} \to \mathbb{R} \times \mathbb{R}$, the $\hat{f}_{PM,i}$ are essentially lookup-tables with six entries.

To simplify notation, we define abbreviations for the first and second component of the PM increment functions marked by superscript (0) and (1), respectively,

$$f_{PM}(\lambda^{(unq)}_i) \triangleq \left( f^{(0)}_{PM}(\lambda^{(unq)}_i), f^{(1)}_{PM}(\lambda^{(unq)}_i) \right), \tag{4.1}$$

$$\hat{f}_{PM,i}(\lambda^{(q)}_i) \triangleq \left( \hat{f}^{(0)}_{PM,i}(\lambda^{(q)}_i), \hat{f}^{(1)}_{PM,i}(\lambda^{(q)}_i) \right), \tag{4.2}$$

which denote the PM increment functions for paths $\ell_0$ and $\ell_1$, respectively. Assume the distributions $P_{\Lambda^{(unq)}_i \Lambda^{(q)}_i}$ were known for all $i$ for a pair of an unquantized and a quantized decoder, both of which operate on the same channel output. Then, $P_{f^{(j)}_{PM}(\Lambda^{(unq)}_i) \Lambda^{(q)}_i}$ are obtained, and the mean squared error between $PM'^{(q)}_{\ell_j}$ and $PM'^{(unq)}_{\ell_j}$ is minimized by the conditional mean [Ash07, Section 18.2]

$$\hat{f}^{(j)}_{PM,i}(\lambda^{(q)}_i) \triangleq \mathsf{E}\left[ f^{(j)}_{PM}(\Lambda^{(unq)}_i) \,\middle|\, \Lambda^{(q)}_i = \lambda^{(q)}_i \right], \qquad \text{for } j \in \{0, 1\}. \tag{4.3}$$

For the sake of comprehensibility, the reader is invited to peek at Figure 4.4. It illustrates the conditional distributions of $\Lambda^{(unq)}_i$ and $f^{(j)}_{PM}(\Lambda^{(unq)}_i)$ given $\Lambda^{(q)}_i$ for a frozen and an information bit. In the sequel, we show how to obtain these distributions.

Envision a *super-decoder* which internally consists of two decoders, an unquantized and a quantized decoder, both operating on the same channel output using their respective LLRs. Every check node (variable node) operation performed by the super-decoder decomposes into a check node (variable node) operation each per constituent decoder. Yet, the messages they pass internally are correlated across decoders because of their shared input. The equivalent of the situation depicted in Figure 4.1 for two separate decoders is depicted in Figure 4.2 for the super-decoder (only decoding tree).

More formally, if the super-decoder is made up of an $\mathcal{L}_3$-SC and an $\mathcal{L}_\infty$-SC decoder, then the super-decoder is equivalent to an $\mathcal{L}_{(3,\infty)}$-SC decoder whose set of LLRs is the Cartesian product of $\mathcal{L}_3$ and $\mathcal{L}_\infty$, thus $\mathcal{L}_{(3,\infty)} \triangleq \mathcal{L}_3 \times \mathcal{L}_\infty$. Check node and variable node operations of the $\mathcal{L}_{(3,\infty)}$-SC decoder are defined component-wise through the check

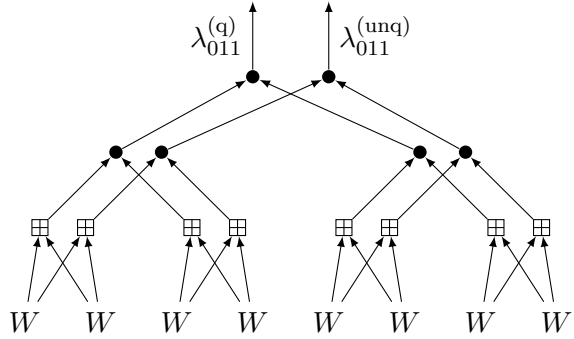

Figure 4.2.: Coupled decoding tree of super-decoder composed of a quantized and an unquantized decoder, computing $\lambda_{011}^{(\text{q})}$ and $\lambda_{011}^{(\text{unq})}$, respectively

node and variable node operations of the component decoders,

$$\lambda_1 \boxplus \lambda_2 = \left(\lambda_1^{(\text{q})}, \lambda_1^{(\text{unq})}\right) \boxplus \left(\lambda_2^{(\text{q})}, \lambda_2^{(\text{unq})}\right) \triangleq \left(\lambda_1^{(\text{q})} \boxplus \lambda_2^{(\text{q})}, \lambda_1^{(\text{unq})} \boxplus \lambda_2^{(\text{unq})}\right), \quad (4.4)$$

$$\lambda_1 \bullet \lambda_2 = \left(\lambda_1^{(\text{q})}, \lambda_1^{(\text{unq})}\right) \bullet \left(\lambda_2^{(\text{q})}, \lambda_2^{(\text{unq})}\right) \triangleq \left(\lambda_1^{(\text{q})} \bullet \lambda_2^{(\text{q})}, \lambda_1^{(\text{unq})} \bullet \lambda_2^{(\text{unq})}\right), \quad (4.5)$$

where $\lambda_1, \lambda_2 \in \mathcal{L}_{(3,\infty)}$. Such an $\mathcal{L}_{(3,\infty)}$-SC decoder performs decoding through message passing over *coupled decoding trees* such as illustrated in Figure 4.2.

For practical systems and numerical analysis it suffices to use a very finely quantized 'quasi-unquantized' $\mathcal{L}_{\widetilde{\infty}}$-SC decoder in place of the perfectly unquantized $\mathcal{L}_{\infty}$-SC decoder. The resulting super-decoder is an $\mathcal{L}_{(3,\widetilde{\infty})}$-SC decoder defined analogous to the $\mathcal{L}_{(3,\infty)}$-SC decoder. The $\mathcal{L}_{(3,\widetilde{\infty})}$-SC decoder has a discrete set of LLRs such that density evolution (cf. Section 2.5, in particular (2.41)) can be used to numerically obtain the distributions $P_{\Lambda_i}$ for all $i$. As $\Lambda_i \triangleq \left(\Lambda_i^{(\text{unq})}, \Lambda_i^{(\text{q})}\right)$, density evolution analysis of the $\mathcal{L}_{(3,\widetilde{\infty})}$-SC decoder recovers the desired distributions $P_{\Lambda_i^{(\text{unq})}\Lambda_i^{(\text{q})}}$.

Since both constituent decoders of the $\mathcal{L}_{(3,\widetilde{\infty})}$-SC decoder are supposed to operate on the same channel output, the distribution of the channel LLRs $\Lambda_{\text{ch}} = \left(\Lambda_{\text{ch}}^{(\text{q})}, \Lambda_{\text{ch}}^{(\text{unq})}\right)$ fed into the super-decoder has to reflect this coupling. Under the all-zero codeword assumption the output of a BiAWGN($\sigma^2$) is distributed as $Y \sim \mathcal{N}_{\mathbb{R}}(1, \sigma^2)$. Let $\tilde{f}_{\text{LLR},\mathcal{L}_3}$ and $\tilde{f}_{\text{LLR},\mathcal{L}_{\widetilde{\infty}}}$ be functions describing all processing steps from channel output to channel LLR for the $\mathcal{L}_3$-SC and $\mathcal{L}_{\widetilde{\infty}}$-SC decoder, respectively. Furthermore, let

$$\tilde{f}_{\text{LLR}}(y) \triangleq \left(\tilde{f}_{\text{LLR},\mathcal{L}_3}(y), \tilde{f}_{\text{LLR},\mathcal{L}_{\widetilde{\infty}}}(y)\right). \quad (4.6)$$

Then, $\Lambda_{\text{ch}} \triangleq \tilde{f}_{\text{LLR}}(Y)$ and $P_{\Lambda_{\text{ch}}} = \tilde{f}_{\text{LLR}}(P_Y)$, which can be evaluated using density evolution tools, cf. (2.41) and (2.42) in Section 2.5.

Note that all distributions $P_{\Lambda_i}$ obtained from density evolution are subject to the all-

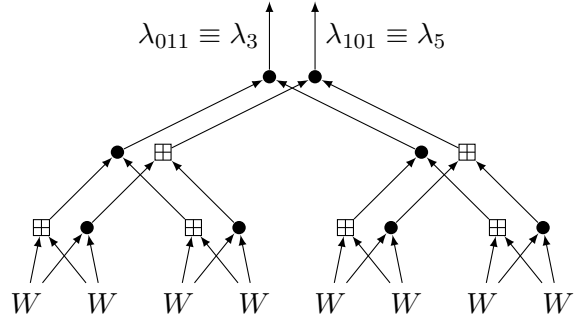

Figure 4.3.: Example of a coupled decoding tree which is equivalent to the approach
described in [MT09, Theorem 2] for obtaining $P_{\Lambda_3\Lambda_5}$ using density evolution

zero codeword assumption. To derive the distributions $Q_{\Lambda_i}$ under $U_i \sim \mathrm{Bern}(1/2)$ for
information bits $i \in \mathcal{I}$, and $U_i = 0$ for frozen bits $i \in \mathcal{F}$, the distributions $P_{\Lambda_i}$ need to
be *symmetrized* using

$$Q_{\Lambda_i}(\lambda_i) = \frac{1}{2}(P_{\Lambda_i}(\lambda_i) + P_{\Lambda_i}(-\lambda_i)), \qquad \text{for } i \in \mathcal{I}, \tag{4.7}$$

$$Q_{\Lambda_i}(\lambda_i) = P_{\Lambda_i}(\lambda_i), \qquad\qquad\qquad \text{for } i \in \mathcal{F}. \tag{4.8}$$

In the case of the $\mathcal{L}_{(3,\widetilde{\infty})}$-SC decoder this amounts to

$$Q_{\Lambda_i^{(\mathrm{unq})}\Lambda_i^{(\mathrm{q})}}\left(\lambda_i^{(\mathrm{unq})}, \lambda_i^{(\mathrm{q})}\right) = \frac{1}{2}\left(P_{\Lambda_i^{(\mathrm{unq})}\Lambda_i^{(\mathrm{q})}}\left(\lambda_i^{(\mathrm{unq})}, \lambda_i^{(\mathrm{q})}\right) + P_{\Lambda_i^{(\mathrm{unq})}\Lambda_i^{(\mathrm{q})}}\left(-\lambda_i^{(\mathrm{unq})}, -\lambda_i^{(\mathrm{q})}\right)\right)$$
$$\text{for } i \in \mathcal{I}, \tag{4.9}$$

$$Q_{\Lambda_i^{(\mathrm{unq})}\Lambda_i^{(\mathrm{q})}}\left(\lambda_i^{(\mathrm{unq})}, \lambda_i^{(\mathrm{q})}\right) = P_{\Lambda_i^{(\mathrm{unq})}\Lambda_i^{(\mathrm{q})}}\left(\lambda_i^{(\mathrm{unq})}, \lambda_i^{(\mathrm{q})}\right) \qquad \text{for } i \in \mathcal{F}. \tag{4.10}$$

We remark that we are not the first to generalize density evolution for joint distributions
over such coupled decoding trees. A similar procedure is devised in [MT09], however
over two different decoding trees belonging to the same decoder, in order to analyze the
joint distributions $P_{\Lambda_i \Lambda_j}$, $i \neq j$. Figure 4.3 provides an illustration of this method using
our symbolic formalism, which also provides a visualization for the ⊞⊞, ⊞●, ●⊞ and ●●
operations introduced by the authors.

Note furthermore that instead of density evolution we could have used the Monte Carlo
method to approximate the distributions $Q_{\Lambda_i^{(\mathrm{unq})}\Lambda_i^{(\mathrm{q})}}$. This is technically less involved
and works well as long as the cardinality of $\hat{\mathcal{L}}_{(3,\widetilde{\infty})}$ is reasonably small and one allows
for a sufficient number of trials. However, the density evolution method is more elegant
and insightful, and can be much more efficient if suitable representations are used for
the involved probability distributions.

For illustration, revisit Figure 4.4, which shows conditional distributions of $\Lambda_i^{(\mathrm{unq})}$ and $f_{\mathrm{PM}}^{(j)}(\Lambda_i^{(\mathrm{unq})})$ given $\Lambda_i^{(\mathrm{q})}$. From Figures 4.4(a) and 4.4(d) it is visible that the frozen bit $i = 128$ was not affected by symmetrization (cf. (4.10)) whereas the information bit $i = 127$ was (cf. (4.9)). The information bit is obviously very reliable (cf. Figures 4.4(e) and (f)): A bit estimate $\hat{u}_{127}$ in accordance with the LLR $\lambda_{127}^{(\mathrm{unq})}$ leads to zero PM penalty, a bit estimate contradicting the LLR leads to a sizable penalty. The frozen bit on the other hand is very unreliable: It is frozen to $u_{128} = 0$, hence the distribution of its LLR $\lambda_{128}^{(\mathrm{unq})}$ should favor large positive values. Instead, the distribution is close to symmetric, with only a subtle inclination towards positive LLRs. As a result, the PM increments only very slightly favor $\hat{u}_{128} = 0$ (cf. Figures 4.4(b) and (c)). (Note that for a conclusive judgement regarding the frozen bit, $Q_{\Lambda_{128}^{(\mathrm{q})}}$ needs to be taken into consideration.)

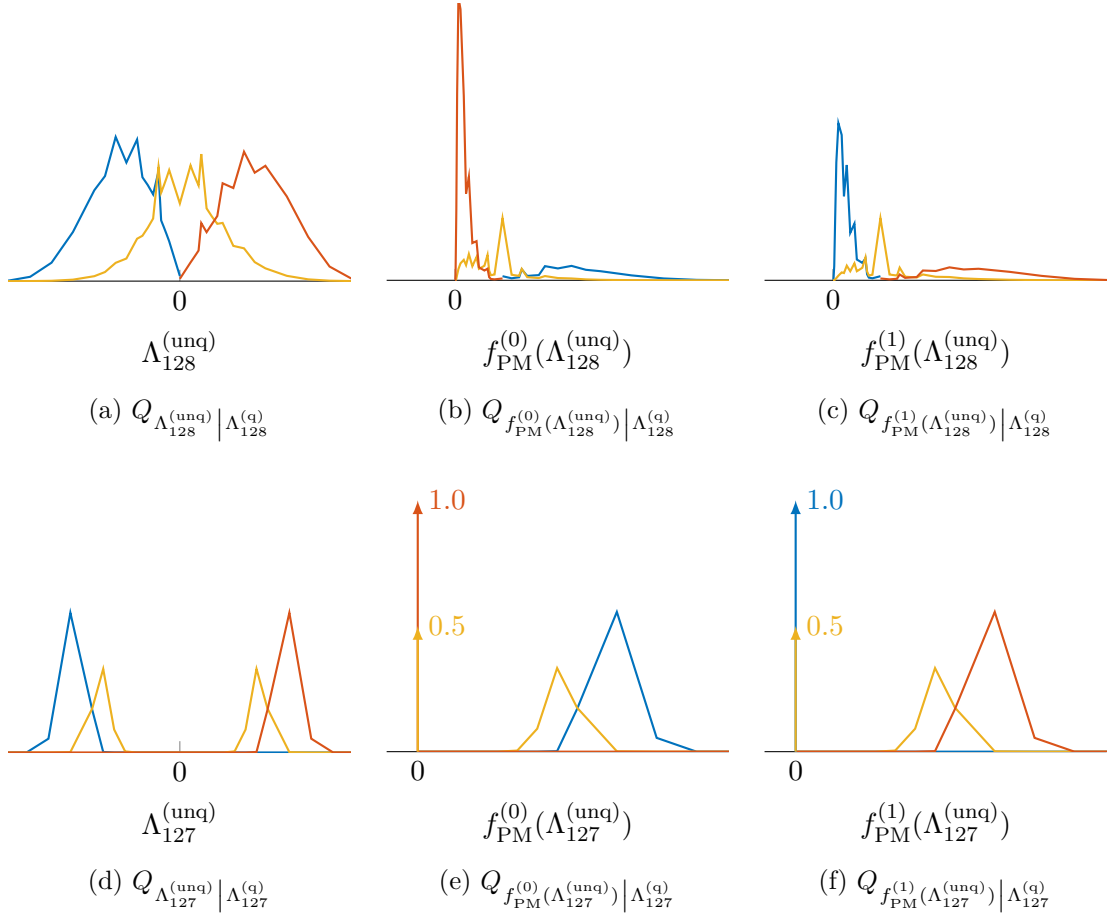

Figure 4.4.: Numerical example for illustration purposes with $m = 8$: Conditional distributions of $\Lambda_i^{(\mathrm{unq})}$ ((a), (d)), $f_{\mathrm{PM}}^{(0)}(\Lambda_i^{(\mathrm{unq})})$ ((b), (e)), and $f_{\mathrm{PM}}^{(1)}(\Lambda_i^{(\mathrm{unq})})$ ((c), (f)), given $\Lambda_i^{(\mathrm{q})}$ (with $\lambda_i^{(\mathrm{q})} = -1$: blue, $\lambda_i^{(\mathrm{q})} = 0$: yellow, $\lambda_i^{(\mathrm{q})} = +1$: orange), for a frozen bit $i = 128$ ((a), (b), (c)) and an information bit $i = 127$ ((d), (e), (f)), respectively

After the theoretical description of the *expected path metric updates* (EPMU) list enhancement technique in the first part of this section, we present our empirical results in the sequel. Using density evolution analysis of the $\mathcal{L}_{(3,\widetilde{\infty})}$-SC decoder, we obtained all $P_{\Lambda_i^{(\mathrm{unq})}\Lambda_i^{(\mathrm{q})}}$, symmetrized them according to (4.9) and (4.10) to derive all $Q_{\Lambda_i^{(\mathrm{unq})}\Lambda_i^{(\mathrm{q})}}$, and then designed the PM increment functions $\hat{f}_{\mathrm{PM},i}^{(j)}(\lambda_i^{(\mathrm{q})})$ according to (4.3) using $Q_{\Lambda_i^{(\mathrm{unq})}\Lambda_i^{(\mathrm{q})}}$. The resulting lookup-tables were used in EPMU enhanced $\mathcal{L}_3$-SCL and $\mathcal{L}_3$-SCL-ML decoders. Code simulation results of EPMU enhanced $\mathcal{L}_3$-SCL and $\mathcal{L}_3$-SCL-ML decoders are compared to their non-enhanced counterparts and to the unquantized $\mathcal{L}_\infty$-SCL decoder in Figure 4.5.

The $\mathcal{L}_3$-SC performance is not improved using EPMU (cf. —+— vs. —+—). Similarly, the effect on PM-FER of $\mathcal{L}_3$-SCL is minor (cf. ⋯◦⋯ vs. ⋯◦⋯). For $\mathcal{L}_3$-SCL-ML however, which is what we set out to enhance the list construction for, EPMU leads to 0.25 dB gain in $\mathrm{E_b/N_0}$ at $\mathrm{P_{e,B}} = 10^{-3}$ for list size $L = 32$ (cf. -▲- vs. -▲⋯). For list size $L = 128$, the gain in $\mathrm{E_b/N_0}$ is 0.2 dB at $\mathrm{P_{e,B}} = 10^{-3}$ (cf. -□- vs. -□⋯), bringing $\mathcal{L}_3$-SCL-ML with EPMU enhancement up to a 0.2 dB gap close to the unquantized $\mathcal{L}_\infty$-SCL decoder at $\mathrm{P_{e,B}} = 10^{-3}$ (cf. -□⋯ vs. ⋯◦⋯). Note that the comparison is appropriate despite the different list sizes $L \in \{32, 128\}$ since $\mathcal{L}_\infty$-SCL saturates to its ML-LB at $L = 32$ already (cf. ⋯◦⋯ vs. —+—) and thus would not benefit from larger list size $L = 128$. As a rule of thumb, the EPMU enhanced $\mathcal{L}_3$-SCL-ML achieves the same performance as the non-enhanced $\mathcal{L}_3$-SCL-ML at a quarter of the list size (cf. -□- vs. -▲⋯).

Figure 4.6 plots the list-FERs for $\mathcal{L}_3$-SCL with and without EPMU enhancement for various list sizes. It confirms that EPMU improves the list-FER, gaining 0.25 dB in $\mathrm{E_b/N_0}$ at $\mathrm{P_{e,B}} = 10^{-3}$ (cf. ⋯◦⋯ vs. ⋯◦⋯, -▲- vs. -▲⋯, -□⋯ vs. -□-). For small list sizes the improvement is minor (cf. -✳⋯ vs. -✳⋯). Figures 4.5 and 4.6 also show that $\mathcal{L}_3$-SCL-ML is able to make good use of the enhanced list-FER from EPMU up to a list size in the range of $L = 64$ to $L = 128$, where it starts saturating to the unquantized $\mathcal{L}_\infty$-SCL decoder's performance.

## 4.2. Expected Path Metric Updates with Contraction Counting

In the previous section, we present a first technique (called expected path metric updates, EPMU) to obtain and exploit statistical reliability information about bit estimates to improve FERs of $\mathcal{L}_3$-SCL and $\mathcal{L}_3$-SCL-ML decoders. In this section, we refine this approach, by keeping contradiction counts in the decoder as a low-complexity measure of instantaneous reliability. We refer to the resulting enhancement as *expected path*

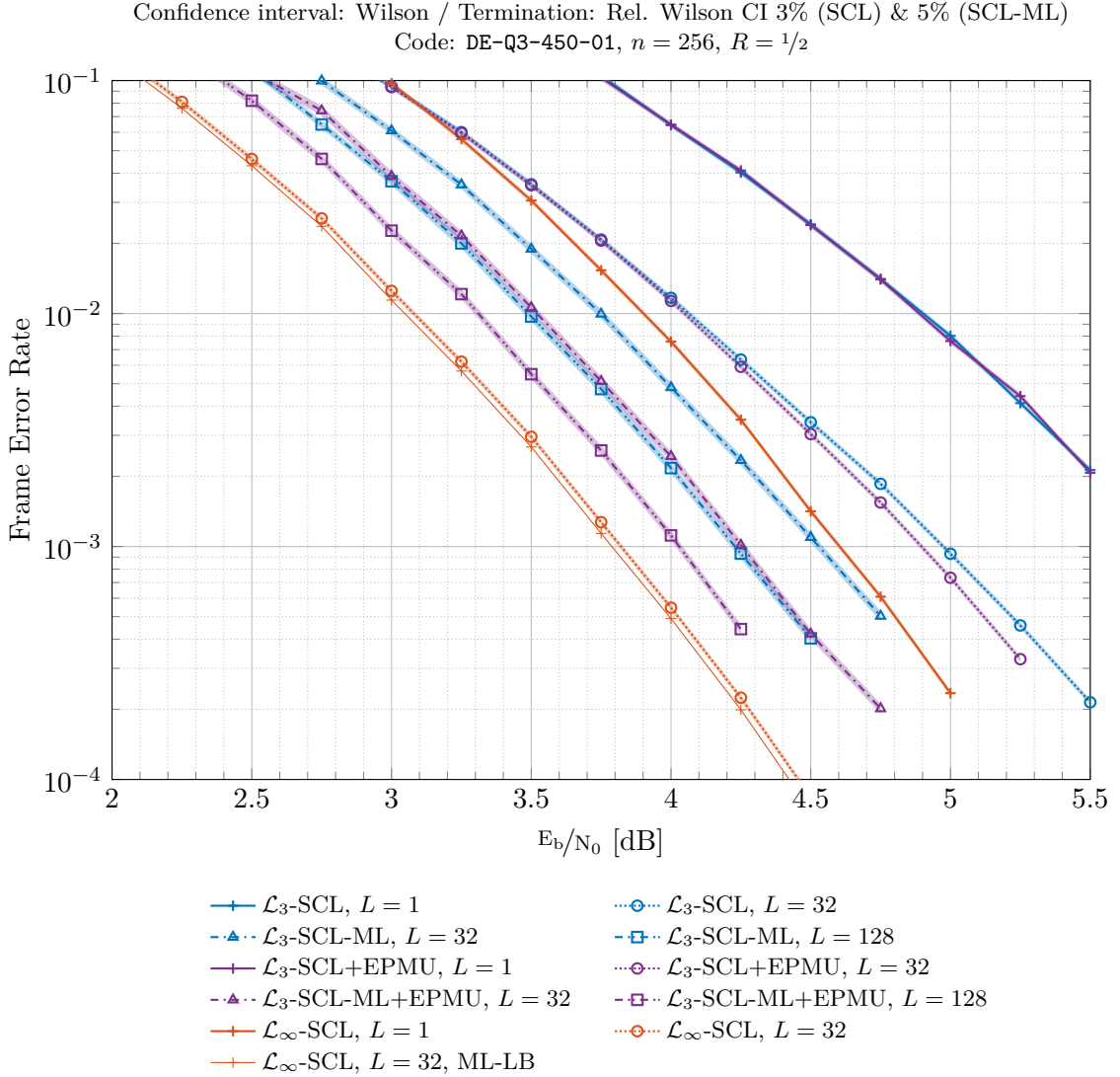

Figure 4.5.: PM-/LML-FER of $\mathcal{L}_\infty/\mathcal{L}_3$-SCL/-SCL-ML (varying list size $L \in \{1, 32, 128\}$) with optional EPMU enhancement, applied to $\mathrm{Q}(3, \delta^*_{\mathrm{cap}})$-BiAWGN($\sigma^2$)

*metric updates with contradiction counting* (EPMUCC). We proceed as follows: First, we introduce contradiction counts. Secondly, we highlight how contradiction counts are fundamentally different from LLRs from an implementation perspective, while the same tools (*e.g.*, density evolution) can be used to analyze them. Thirdly, we show how to exploit contradiction counts for further improved PM updates. Finally, we present simulation results and evaluate the utility of EPMUCC.

The objective of this section is to introduce a measure for instantaneous bit estimation

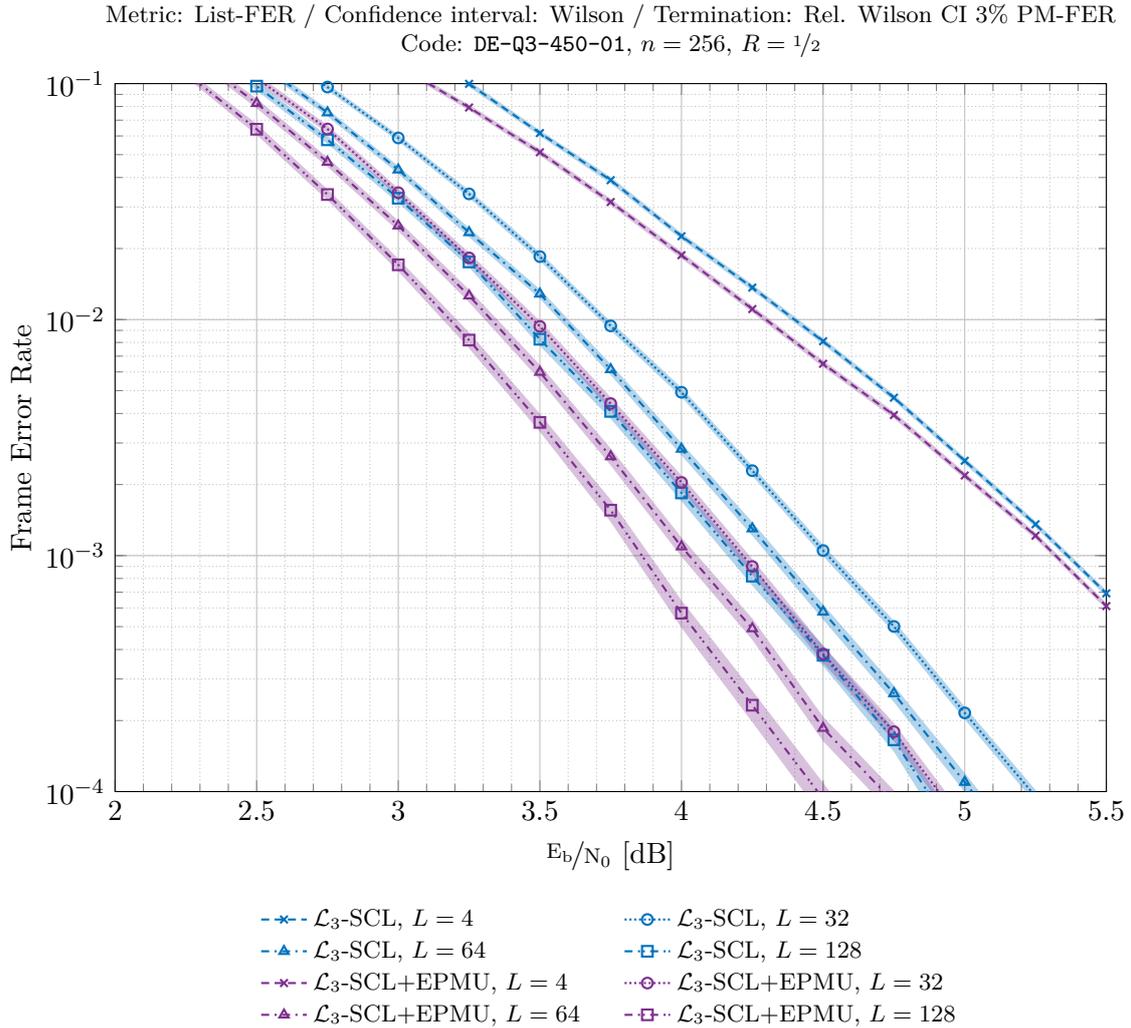

Metric: List-FER / Confidence interval: Wilson / Termination: Rel. Wilson CI 3% PM-FER
Code: `DE-Q3-450-01`, $n = 256$, $R = \frac{1}{2}$

Figure 4.6.: List-FER of $\mathcal{L}_3$-SCL (varying list size $L \in \{4, 32, 64, 128\}$) with optional EPMU enhancement, applied to Q($3, \delta^*_{\text{cap}}$)-BiAWGN($\sigma^2$)

reliability other than just increasing the level of detail of the LLR messages through finer quantization. *Contradiction counting* at variable nodes should intuitively provide such a measure: Whenever during decoding a variable node faces two contradicting inputs, *i.e.*, one of its inputs is `+1` while the other is `−1`, this suggests that the resulting bit estimation should have lower reliability. The total number of such contradiction events that occur while computing $\lambda_i^{(q)}$ via message passing over the respective decoding tree for the $i$-th bit, gives an indication for how reliable the result $\lambda_i^{(q)}$ probably is. For modelling and analyzing the resulting 3-level quantized SC decoder with contradiction counting, we treat the contradiction counters as part of the LLRs. Note that this is only for theoretical purposes. After formally introducing contradiction counts, we highlight their fundamental difference from LLRs which allows for low-complexity implementation.

A 3-level quantized SC decoder that counts contradiction events at variable nodes can be seen as a $\mathcal{L}_{(3,\text{CC})}$-SC decoder with set of LLRs $\mathcal{L}_{(3,\text{CC})} \triangleq \mathcal{L}_3 \times \mathbb{N}_0$. Check node and variable node operations in $\mathcal{L}_{(3,\text{CC})}$ are defined component-wise,

$$\lambda_1 \boxplus \lambda_2 = \left(\lambda_1^{(\text{q})}, c_1\right) \boxplus \left(\lambda_2^{(\text{q})}, c_2\right) \triangleq \left(\lambda_1^{(\text{q})} \boxplus \lambda_2^{(\text{q})}, c_1 + c_2\right), \tag{4.11}$$

$$\lambda_1 \bullet \lambda_2 = \left(\lambda_1^{(\text{q})}, c_1\right) \bullet \left(\lambda_2^{(\text{q})}, c_2\right) \triangleq \left(\lambda_1^{(\text{q})} \bullet \lambda_2^{(\text{q})}, c_1 + c_2 + 1_{(\lambda_1,\lambda_2)\in\{(+1,-1),(-1,+1)\}}\right), \tag{4.12}$$

for $\lambda_1, \lambda_2 \in \mathcal{L}_{(3,\text{CC})}$, where the first component performs plain $\mathcal{L}_3$-SC decoding operations and the second component counts contradictions at variable nodes. For the channel output LLRs $\lambda_{\text{ch}} \triangleq \left(\lambda_{\text{ch}}^{(\text{q})}, 0\right)$ fed into the $\mathcal{L}_{(3,\text{CC})}$-SC decoder, the second component is initialized to 0.

It might seem that contradiction counts are 'yet another piece of LLR information'. And in fact that is how we introduce them, how we analyze them, and how we implemented them in our prototype decoders to evaluate their FERs. But there is a fundamental difference which enables low-complexity implementation of contradiction counting: Throughout the algorithm it does not matter 'where' in the decoding tree the contradictions happen, only their total number is relevant. For an efficient implementation it suffices to have one contradiction counter per layer of the decoding tree. As SC decoding proceeds and LLR variables corresponding to the upper layers of the decoding tree are updated, so are the corresponding contradiction count variables. This process of incremental recalculations of LLRs and contradiction counters in the process of SC decoding is visualized in Figure 4.7. While in the order of $n$ units of memory are required to store LLRs, only in the order of $\log n$ units of memory are required to store contradiction counters. Similarly, computational and memory bandwidth requirements of contradiction counters are lower compared to LLRs. This fundamental difference between LLRs and contradiction counters justifies the special treatment of contradiction counters as something separate from LLRs.

To further enhance 3-level quantized decoders using contradiction counts, a similar construction as in Section 4.1 is undertaken: A super-decoder is formed out of a $\mathcal{L}_{(3,\text{CC})}$-SC and a $\mathcal{L}_{\widetilde{\infty}}$-SC decoder. This super-decoder is equivalent to a $\mathcal{L}_{(3,\text{CC},\widetilde{\infty})}$-SC decoder with LLR alphabet $\mathcal{L}_{(3,\text{CC},\widetilde{\infty})} \triangleq \mathcal{L}_{(3,\text{CC})} \times \mathcal{L}_{\widetilde{\infty}} = \mathcal{L}_3 \times \mathbb{N}_0 \times \mathcal{L}_{\infty}$. Check node and variable node operations are defined component-wise in direct analogy to (4.4), (4.5), (4.11), and (4.12); likewise for the channel output LLRs $\Lambda_{\text{ch}}$. Density evolution analysis of the $\mathcal{L}_{(3,\text{CC},\widetilde{\infty})}$-SC decoder and symmetrization of the resulting distributions (cf. (4.7), (4.8), (4.9), (4.10)) is used to obtain the joint distributions $Q_{\Lambda_i^{(\text{q})} C_i \Lambda_i^{(\text{unq})}}$. From the distributions, PM update functions $\hat{f}_{\mathrm{PM},i}^{(j)}(\lambda_i^{(q)}, c_i)$ can be derived in the spirit of (4.3) which use as input not only the quantized LLR $\lambda_i^{(q)}$ but also the contradiction count $c_i$. However, we investigate a slightly different approach here, which furthermore led to better empirical results: We use the marginal distribution $Q_{\Lambda_i^{(q)} \Lambda_i^{(unq)}}$ to design PM update functions $\hat{f}_{PM,i}^{(j)}(\lambda_i^{(q)})$ according to (4.3). The statistical knowledge about the contradiction counts is used in the following way: For density evolution it is assumed that the decoder has correctly decoded all $i-1$ bits preceding the current $i$-th bit. Hence, the distributions obtained by density evolution describe the LLRs and contradiction counts an SC decoder encounters while decoding the correct codeword (or respectively, the SCL decoder encounters while decoding the one path that will result in the correct codeword). Note that we have no information about the distributions of a decoder that has left the right path. From the marginal distributions $Q_{C_i}$ we learn that some numbers of contradiction events are extremely unlikely to happen in the correct path. We can therefore classify the paths into 'plausible' and 'implausible' paths based on their contradiction counts. If a path exhibits contradiction counts that are unlikely to occur in the correct path according to the density evolution analysis, then we assume that this path is probably incorrect and we temporarily add a big penalty to its PM for the next round of PM comparisons and truncation of least likely paths. In our simulations, we penalized all paths with a value in the order of five times the largest EPMU value, once they exhibit a contradiction count that has a probability of less than $10^{-6}$ to happen in the correct path. This PM penalty is extremely large and almost certainly leads to the path's removal from the list, unless many other paths incur the same penalty.

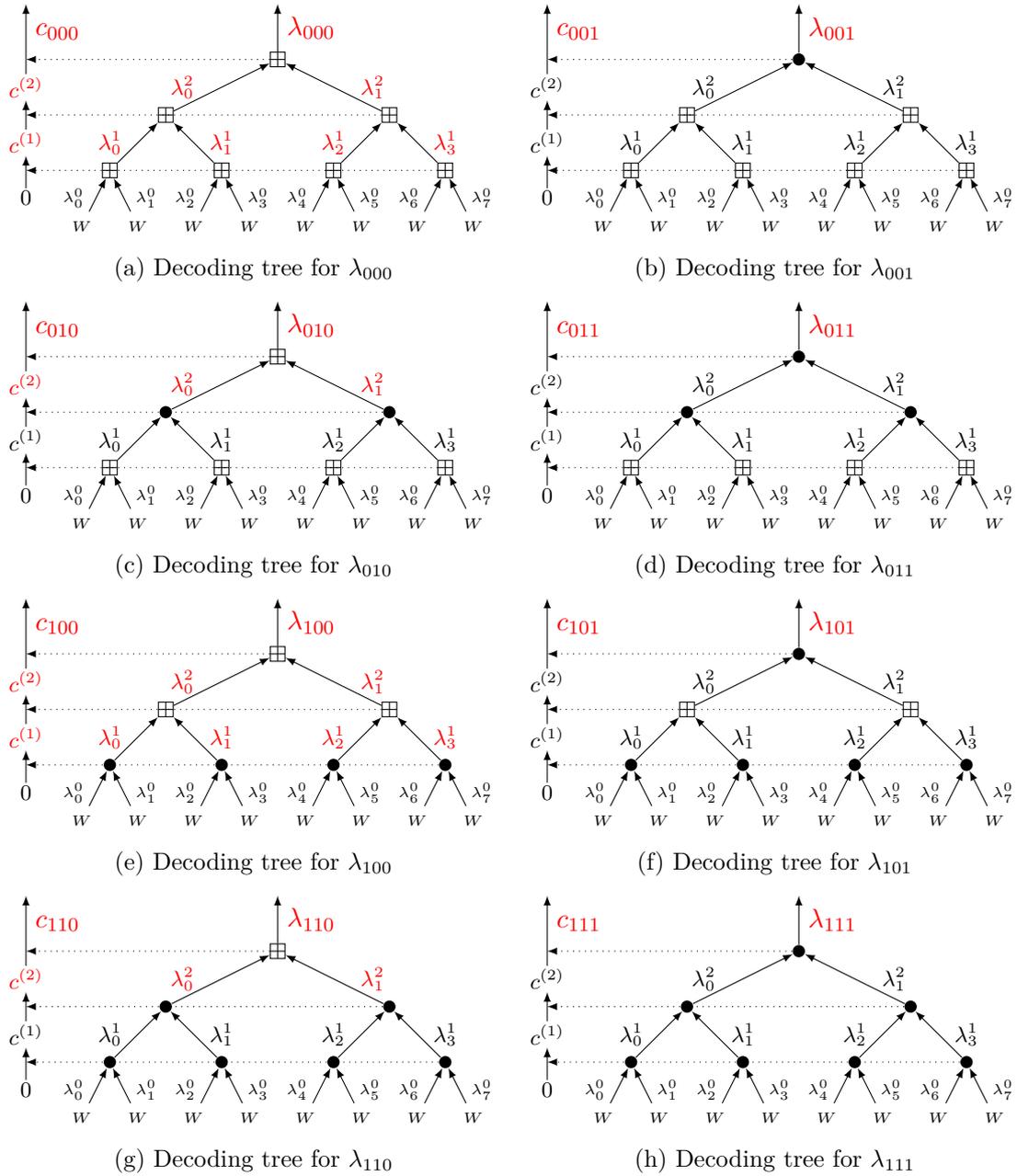

Figure 4.7.: Incremental updates/recalculations (red) of LLRs $\lambda_i^j$ and contradiction counters $c^{(j)}$ in the SC decoding process (dotted lines: contradiction events at the respective nodes are summed and added to the counter)

In Figure 4.8, code simulation results of EPMUCC enhanced $\mathcal{L}_3$-SCL and $\mathcal{L}_3$-SCL-ML decoders are compared to their non-enhanced and EPMU enhanced counterparts and to the unquantized $\mathcal{L}_\infty$-SCL decoder. The $\mathcal{L}_3$-SC performance is not improved using EPMUCC (cf. —+— vs. —+—). There is a slight improvement in the PM-FER of $\mathcal{L}_{(3,CC)}$-SCL vs. $\mathcal{L}_3$-SCL (cf. ··⊙·· vs. ··⊙··). Already for small list size $L = 16$, $\mathcal{L}_{(3,CC)}$-SCL-ML matches the FER of $\mathcal{L}_3$-SCL-ML with a four times larger list ($L = 64$) (cf. -·✳·- vs. -·✳·-). For large list size, $\mathcal{L}_{(3,CC)}$-SCL-ML with EPMUCC achieves the same FER at half the list size of EPMU enhanced $\mathcal{L}_3$-SCL-ML ($L = 64$ vs. $L = 128$, cf. -·▲·- vs. -·▲·-). For list size $L = 128$, $\mathcal{L}_{(3,CC)}$-SCL-ML with EPMUCC enhancement lacks $0.15\,\mathrm{dB}$ behind in $E_b/N_0$ compared to the unquantized $\mathcal{L}_\infty$-SCL decoder at $P_{e,B} = 10^{-3}$ (cf. ··□·· vs. ··⊙··). Note again that the comparison is appropriate despite the different list sizes $L \in \{32, 128\}$ since $\mathcal{L}_\infty$-SCL saturates to its ML-LB at $L = 32$ already (cf. ··⊙·· vs. —+—) and thus would not benefit from larger list size $L = 128$.

Yet, in particular compared to EPMU and its gains over non-enhanced decoders, the improvements from EPMUCC are sobering given the additional complexity over EPMU. Among the reasons could be that contradiction counts are not informative in early decoding stages where the decoding trees contain a large number of check nodes and only few variable nodes. By the time the decoder reaches later stages with many variable

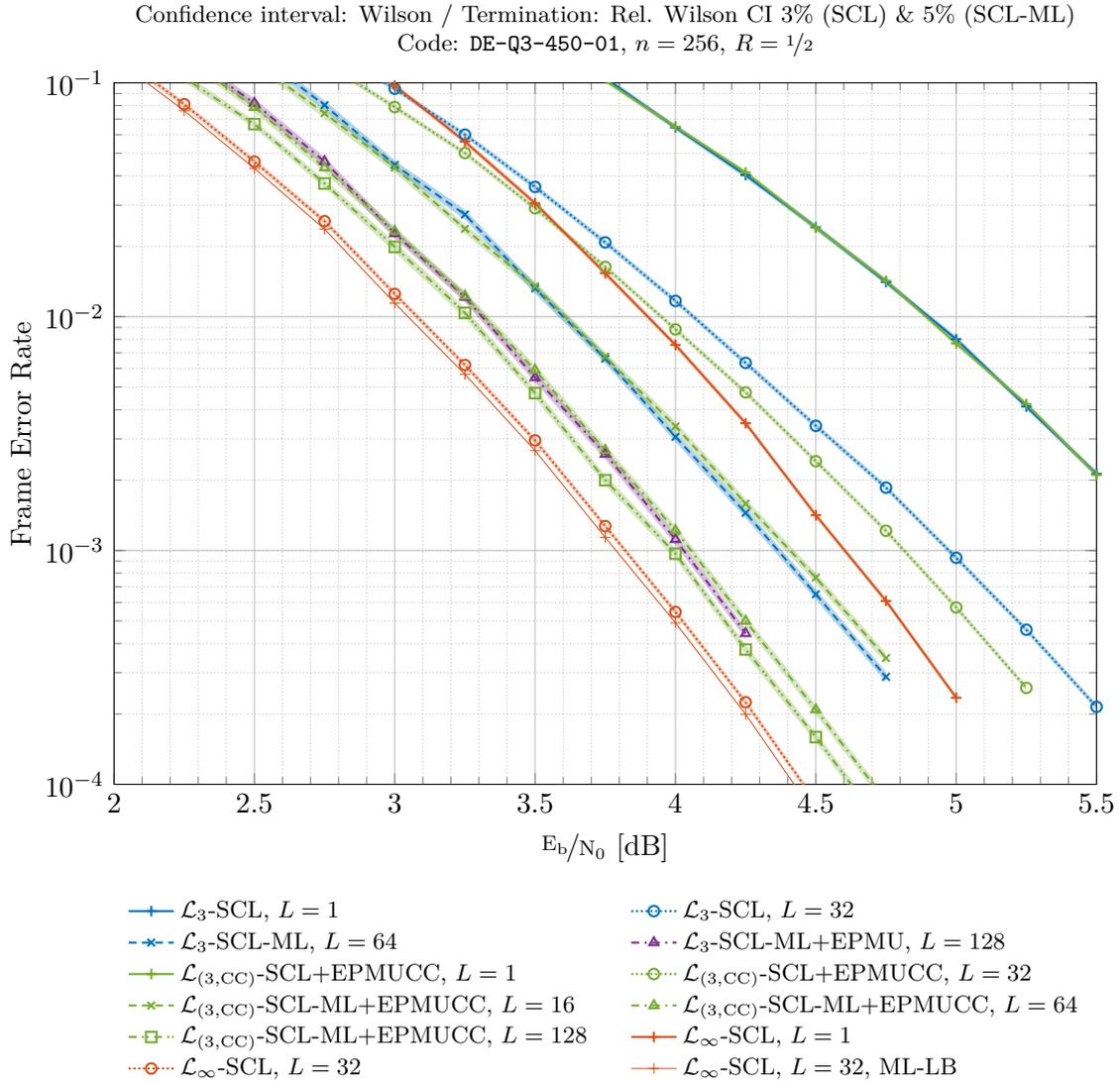

Figure 4.8.: PM-/LML-FER of $\mathcal{L}_\infty/\mathcal{L}_3/\mathcal{L}_{(3,\text{CC})}$-SCL-/-SCL-ML (varying $L$) with optional EPMU/EPMUCC enhancement, applied to Q(3, $\delta^*_{\text{cap}}$)-BiAWGN($\sigma^2$)

nodes, the correct path might have already been removed from the list.

## 4.3. Ideas for Further Enhancements

Many enhancements following the philosophy of the previous section can be envisioned, *e.g.*, erasure or double-erasure counting at check nodes (one or both of the inputs are `0`, respectively). While contradiction counts are informative in later decoding stages with many variable nodes, check node based counts are informative in earlier decoding stages. Furthermore, better compact representations and approximations (*e.g.*, Gaussian approximation) of joint distributions would simplify the density evolution analysis.

# 5. Robustness of Results for 3-Level Quantization

The aim of this chapter is to put our findings from previous chapters into perspective, (a) w.r.t. the original objective of improving energy efficiency of coarsely quantized coding schemes at low rates, and (b) w.r.t. 'competing approaches' (comprising use of finer quantization) a practitioner might consider over rather 'radical' 3-level quantization.

In Section 5.1 we evaluate the performance of our methods at low code rates. We show substantial gains over 'plain' 3-level quantized decoders in this regime.

Some perhaps consider 3-level quantization extreme and wonder about the improvements one can obtain from spending one additional bit on quantization. We investigate this concern in Section 5.2, where we study the performance of a 7-level quantized scheme. For PM-FER $P_{e,B} = 10^{-3}$, a 7-level quantized SCL decoder looses only $0.5\,\mathrm{dB}$ in $E_b/N_0$ w.r.t. an unquantized decoder, both over 7-level quantized channel output. The gap to unquantized channel output and decoder is only an additional $0.2\,\mathrm{dB}$. This leads to the conclusion that from a practical point of view coarse but not super coarse quantization might achieve the best cost-benefit tradeoff.

Lastly, in Section 5.3, we gauge the performance impact of one quantized layer vs. one unquantized layer in the SC decoding tree. When performing one unquantized decoding step on unquantized channel output and 3-level quantizing only thereafter, a $0.5\,\mathrm{dB}$ gain in $E_b/N_0$ at PM-FER $P_{e,B} = 10^{-3}$ is observed. We highlight that such a scheme can be viewed as 'virtual' multi-stage decoding where the unquantized part is 'demodulation' of virtual 'super-symbols' and all decoding happens 3-level quantized thereafter. This connects our topic with recent advancements in coded modulation.

## 5.1. Comparison at Low Code Rates

Our work on enhancing (3-level) quantized decoding of polar codes was in part motivated by the pessimistic outlook uncovered in Figure 1.2(b) in particular for the low code rate regime. Simulation results in prior chapters, especially Chapter 4, showed rate $R = 1/2$ codes only. Hence an important robustness test for our enhancement techniques is whether they enable substantial gains for low rate codes as well.

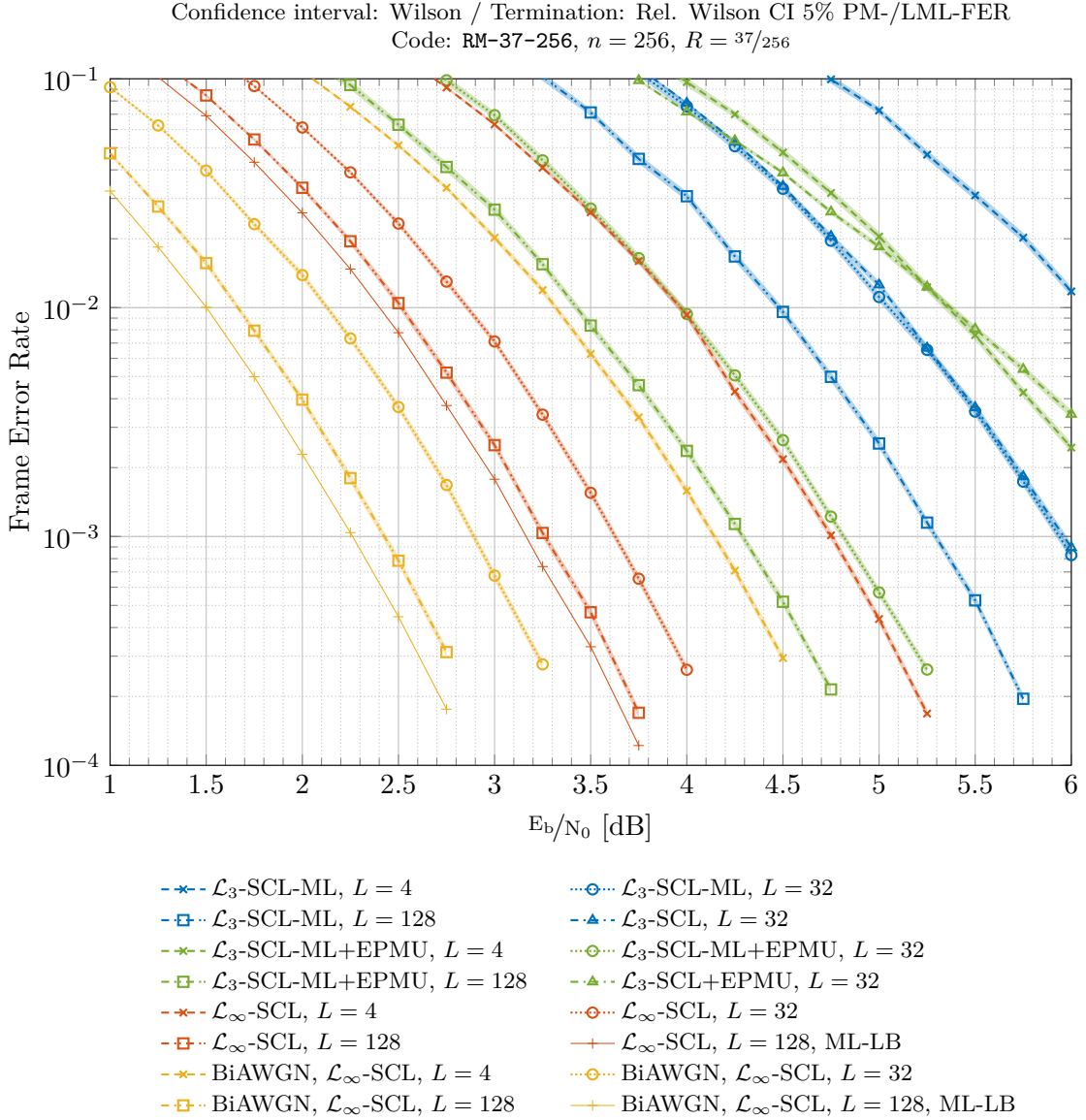

Figure 5.1.: PM-/LML-FER of $\mathcal{L}_\infty/\mathcal{L}_3$-SCL-/-SCL-ML (varying list size $L \in \{4, 32, 128\}$) with optional EPMU enhancement, applied to $Q(3, \delta^*_{cap})$-BiAWGN($\sigma^2$) (default) or BiAWGN($\sigma^2$) (where stated)

Figure 5.1 shows simulation results for the Reed-Muller code with $n = 256$, $k = 37$, $R = {}^{37}\!/_{256} \approx 0.145$. This is well in the low rate regime where the bounds shown in Figure 1.2 predict a considerable gap between capacity and 3-level quantized SC decoding. Considering that capacity is eventually achieved by an unquantized decoder (at least in the large block length regime) while 3-level quantized decoders are bound far away from capacity, these bounds also suggest a considerable gap between unquantized and quantized decoder FER.

We find this prediction confirmed in Figure 5.1. The gap in $E_b/N_0$ for FER $10^{-3}$ between $\mathcal{L}_3$-SCL-ML and $\mathcal{L}_\infty$-SCL with $L = 32$ is 2.3 dB (cf. ⊶ vs. ⊶). With $L = 128$ the gap is still 2.0 dB (cf. ⊟ vs. ⊟). Note that Reed-Muller codes exhibit poor SC but good ML decoding performance [MHU14]. A considerable list size even larger than $L = 128$ is necessary for $\mathcal{L}_\infty$-SCL to start saturating towards its ML-LB (cf. ⊟ vs. ——, ⊟ vs. ——). We remark also that there is no noticeable gain from the additional ML-among-list step of $\mathcal{L}_3$-SCL-ML vs. $\mathcal{L}_3$-SCL at $L = 32$ (cf. ⊶ vs. ▲). This is due to the fact that PM-FER and list-FER are close for $\mathcal{L}_3$-SCL at $L = 32$. At $L = 128$, there is a minor gap ($\approx 0.15$ dB) between PM-FER and list-FER of $\mathcal{L}_3$-SCL (not shown in Figure 5.1) and thus gains from ML-among-list stay limited.

Considering the empirical results in Chapter 4, it is little surprising that again EPMU enhancement does not positively affect the PM-FER of $\mathcal{L}_3$-SCL (cf. ▲ vs. ▲). However, EPMU enhanced $\mathcal{L}_3$-SCL-ML gains 1.1 dB in $E_b/N_0$ over 'plain' $\mathcal{L}_3$-SCL-ML at FER $10^{-3}$ for $L = 32$ (cf. ⊶ vs. ⊶) and 1.0 dB for $L = 128$ (cf. ⊟ vs. ⊟). This effectively halves the gap between quantized and unquantized decoder (cf. ⊶ vs. ⊶ vs. ⊶, ⊟ vs. ⊟ vs. ⊟).

The performance of EPMU enhanced $\mathcal{L}_3$-SCL-ML with $L = 32$ is close to that of $\mathcal{L}_\infty$-SCL with $L = 4$ (cf. ⊶ vs. ✳), suggesting a rule of thumb factor eight in list size blowup as cost for coarse quantization. A list size $L = 128$ is required for EPMU enhanced $\mathcal{L}_3$-SCL-ML to get close to the performance of $\mathcal{L}_\infty$-SCL with $L = 4$ over the unquantized BiAWGN output (cf. ⊟ vs. ✳).

We conclude that the enhancement techniques for quantized decoders devised in Chapters 3 and 4 are useful also in the low code rate regime.

## 5.2. Comparison to 7-Level Quantization

As motivated in Section 1.1, we focussed our work on 3-level quantization because we seek insights into fundamental properties and effects of quantized polar decoding which we assume to be more pronounced in this 'extreme' regime. From a practical point of view one might consider spending a little more resources on quantization to gain performance and thus achieve a better cost-benefit tradeoff. Two bits are required to store 3-level quantized LLR values. Since we only consider mid-tread uniform quantization (cf. Sections 2.8 and 6) the finest LLR quantization fitting into three bits is 7-level.

Hence in the following we examine the effect of 'spending one bit more on quantization' with the example of 7-level quantization.

The 7-level quantized decoder $\mathcal{L}_7$-SC/-SCL/-SCL-ML uses the LLR alphabet $\mathcal{L}_7 \triangleq \{0, \pm 1, \pm 2, \pm 3\}$ with check node and variable node operations defined as

$$\lambda_1 \boxplus \lambda_2 \triangleq \mathrm{sign}(\lambda_1)\,\mathrm{sign}(\lambda_2)\,\min\{|\lambda_1|, |\lambda_2|\}, \tag{5.1}$$

$$\lambda_1 \bullet \lambda_2 \triangleq \max\{\min\{\lambda_1 + \lambda_2, +3\}, -3\}, \tag{5.2}$$

where $\lambda_1, \lambda_2 \in \mathcal{L}_7$. Channel LLRs are computed for the BiAWGN and then quantized uniformly into seven LLR values in $\mathcal{L}_7$, cf. Section 2.8. We defer further technical details concerning quantization and how to obtain a good quantization threshold to Section 6, in particular Section 6.3. We denote the optimum quantization threshold as $\delta_{\mathrm{de}}^*$.

Numerical evaluation results of a 7-level quantized scheme are depicted in Figure 5.2. The loss from 7-level quantization of channel output LLRs is only 0.2 dB in $\mathrm{E_b/N_0}$ at FER $10^{-3}$ (cf.  vs. ). Both unquantized decoders saturate to their ML-LBs with list size $L = 32$ (cf.  vs. ,  vs. ). The $\mathcal{L}_7$-SCL with $L = 32$ has a gap of only 0.5 dB to the $\mathcal{L}_\infty$-SCL at PM-FER $10^{-3}$ (cf.  vs. ). Furthermore, note that the list-FER of $\mathcal{L}_7$-SCL with $L = 32$ is again 0.4 dB better than its PM-FER, and within 0.1 dB from the PM-FER of the unquantized decoder (cf.  vs.  vs. ). In light of the observations in previous chapters we assume that an additional ML-among-list step after $\mathcal{L}_7$-SCL would be able to claim some of the gain from PM-FER toward list-FER. In terms of SC performance, there is a gap of 0.15 dB due to channel output LLR quantization (cf.  vs. ) and an additional gap of 0.4 dB due to LLR quantization (cf.  vs. ) at FER $10^{-3}$.

A performance comparison of 7-level and 3-level quantization is shown in Figure 5.3. The coarser 3-level quantization of channel output alone causes a loss of more than 0.8 dB in $\mathrm{E_b/N_0}$ at $10^{-3}$ compared to 7-level quantization even when using the unquantized $\mathcal{L}_\infty$-SCL (cf.  vs.  vs. ). In addition to that, there is the loss due to LLR quantization which is 1.2 dB for 3-level quantization (cf.  vs. ) compared to 0.5 dB for 7-level quantization (cf.  vs. ). Furthermore looking at the list-FERs, an additional ML-among-list step after SCL decoding could reclaim up to 75% of the gap to the unquantized decoder for 7-level quantization (cf.  vs.  vs. ) but only up to 40% for 3-level quantization (cf.  vs.  vs. ).

We conclude that from a practical point of view the 'one additional bit spent on quantization' in 7-level vs. 3-level quantization leads to huge performance improvements and brings the quantized close to the unquantized decoder. This is in line with prior results which show that few quantization bits are sufficient to achieve close-to-unquantized per-

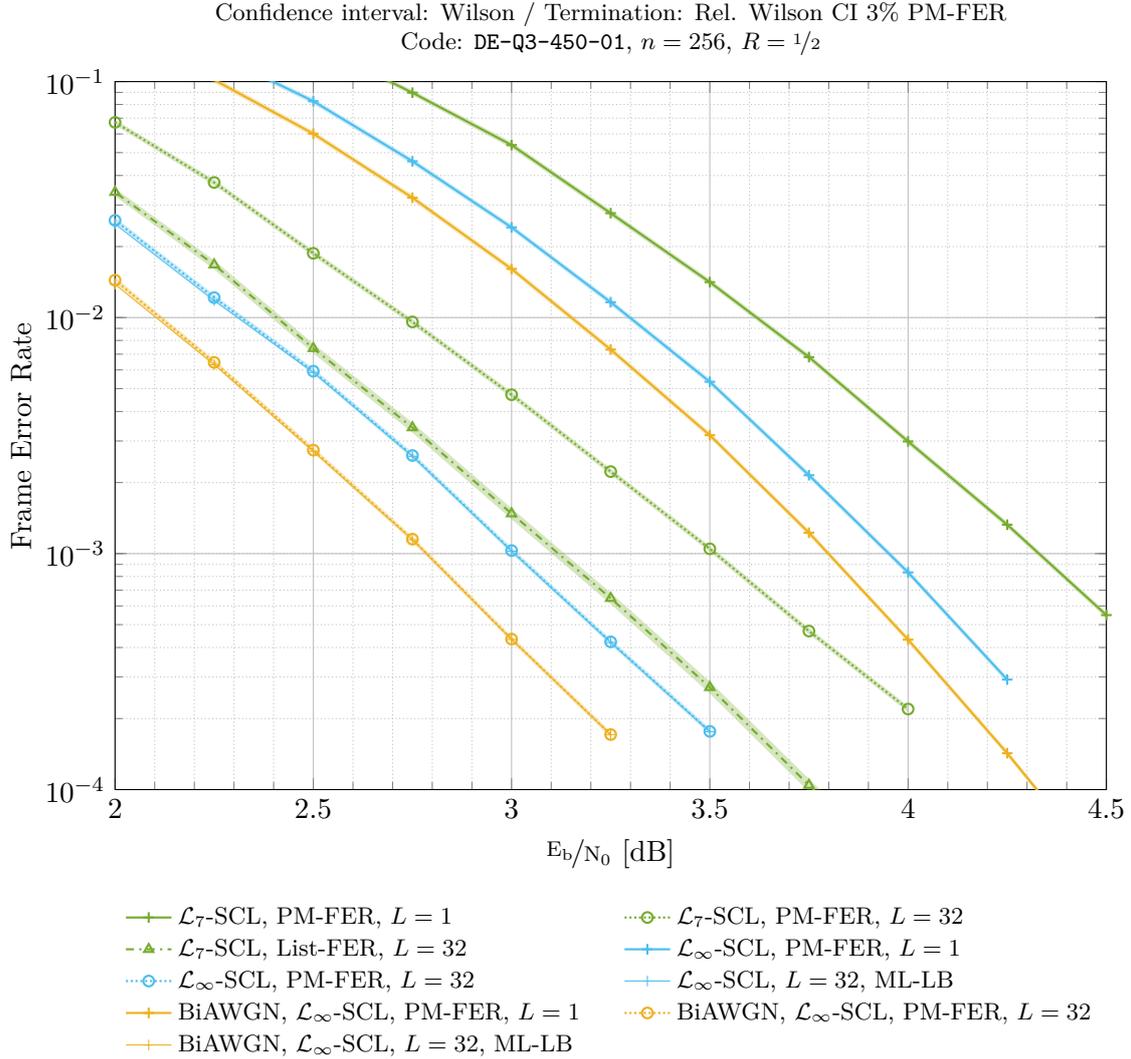

Confidence interval: Wilson / Termination: Rel. Wilson CI 3% PM-FER
Code: `DE-Q3-450-01`, $n = 256$, $R = \frac{1}{2}$

Legend:
- $\mathcal{L}_7$-SCL, PM-FER, $L = 1$
- $\mathcal{L}_7$-SCL, List-FER, $L = 32$
- $\mathcal{L}_\infty$-SCL, PM-FER, $L = 32$
- BiAWGN, $\mathcal{L}_\infty$-SCL, PM-FER, $L = 1$
- BiAWGN, $\mathcal{L}_\infty$-SCL, $L = 32$, ML-LB
- $\mathcal{L}_7$-SCL, PM-FER, $L = 32$
- $\mathcal{L}_\infty$-SCL, PM-FER, $L = 1$
- $\mathcal{L}_\infty$-SCL, $L = 32$, ML-LB
- BiAWGN, $\mathcal{L}_\infty$-SCL, PM-FER, $L = 32$

Figure 5.2.: PM-/List-FER of $\mathcal{L}_\infty/\mathcal{L}_7$-SCL (varying list size $L \in \{1, 32\}$), applied to Q(7, $\delta_{\mathrm{de}}^*$)-BiAWGN($\sigma^2$) (default) or BiAWGN($\sigma^2$) (where stated)

formance [BPB15; Mei+15; Ler+13]. The cost-benefit tradeoff probably favors 7-level over 3-level quantization, and 7-level quantized schemes provide a good starting point for practical low-complexity yet performant implementation.

## 5.3. Comparison to 3-Level Quantization in 'Virtual' Multi-Stage Decoding

Finally, we seek to gauge the effect of a single layer of quantized/unquantized operations in the decoding tree. At later stages of message passing over a decoding tree, LLR values

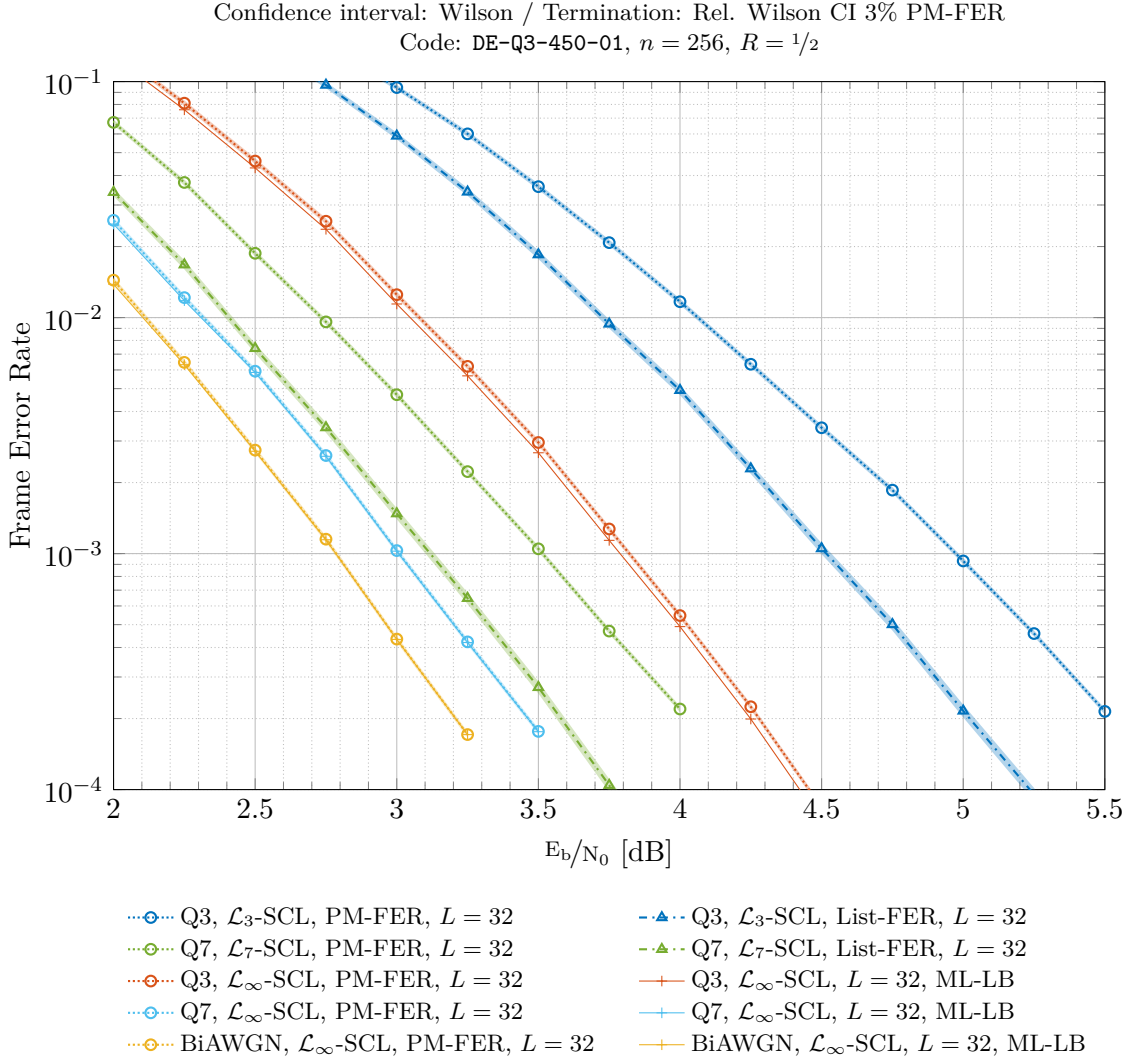

Confidence interval: Wilson / Termination: Rel. Wilson CI 3% PM-FER
Code: `DE-Q3-450-01`, $n = 256$, $R = \frac{1}{2}$

Figure 5.3.: PM-/List-FER of $\mathcal{L}_3/\mathcal{L}_7/\mathcal{L}_\infty$-SCL (list size $L = 32$), applied to Q($3, \delta^*_{\text{cap}}$)-BiAWGN($\sigma^2$) (legend Q3), Q($7, \delta^*_{\text{de}}$)-BiAWGN($\sigma^2$) (legend Q7), or BiAWGN($\sigma^2$) (legend BiAWGN)

are usually well polarized such that a small number of quantization levels is sufficient to capture them. Fine quantization at this point would not have a big positive impact. In low layers of the decoding tree, however, it is helpful to still be able to distinguish between nuances. Thus, finer quantization leads to bigger performance improvements here. In the following we take this insight to the 'extreme' and investigate a polar coding scheme where the first layer of decoding tree operations is unquantized and performed on unquantized LLRs. Only then, LLRs are 3-level quantized, and the remaining upper layers of the decoding tree are performed in $\mathcal{L}_3$.

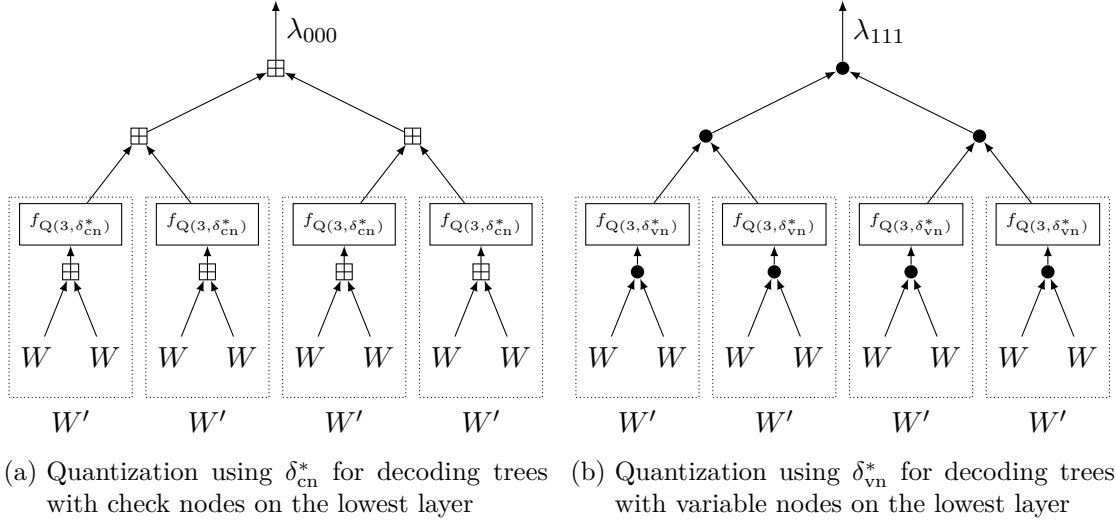

(a) Quantization using $\delta_{\mathrm{cn}}^*$ for decoding trees with check nodes on the lowest layer

(b) Quantization using $\delta_{\mathrm{vn}}^*$ for decoding trees with variable nodes on the lowest layer

Figure 5.4.: Example decoding trees for 'first layer unquantized then 3-level quantized' scheme; note that two different quantization thresholds are used for trees with check nodes (a) and variable nodes (b) at the lowest layer

Two examples of decoding trees resulting from the aforementioned construction are shown in Figure 5.4. Note that two different quantization thresholds $\delta_{\mathrm{cn}}^*$ and $\delta_{\mathrm{vn}}^*$ are used for quantization after the first layer in decoding trees where the lowest layer comprises check nodes (cf. Figure 5.4(a)) and variable nodes (cf. Figure 5.4(b)), respectively. Jointly optimal quantization thresholds for the decoder and information bits of the polar code have been determined using the technique devised in Section 6.3 and are here referred to as $\delta_{\mathrm{cn}}^*$ and $\delta_{\mathrm{vn}}^*$. The resulting code is `DE-MSD-450-01`, cf. Section 2.6.3.

Equivalently, this construction can be viewed as a 'virtual' multi-level coding/multi-stage decoding system (cf. Figure 5.5, [Hub94; FHW96]), where two two-valued $W$ channel accesses are combined into one virtual four-valued $W'$ super-channel access. The respective mapping/'modulation' is the polar transform. The super-channel output is component-wise quantized and processed in two 3-level quantized decoders implementing polar codes $\mathcal{C}_1$ and $\mathcal{C}_2$ with half the original block length and information bits

$$\mathcal{I}_1 = \mathcal{I} \cap \left[0 : \frac{n}{2} - 1\right], \quad k_1 = |\mathcal{I}_1|, \quad \mathcal{I}_2 = \mathcal{I} \cap \left[\frac{n}{2} : n - 1\right], \quad k_2 = |\mathcal{I}_2|, \quad k = k_1 + k_2. \quad (5.3)$$

This connects our work with coded modulation [Sei+13; Sei15; Böc16; Böc+17; PY18]. Because of the similarity to multi-stage decoding (MSD), we use the abbreviation 'MSD' to refer to this 'first layer unquantized then 3-level quantized' scheme.

To be able to analyze the MSD scheme with density evolution and to obtain a prototype SCL decoder, we define an equivalent $\mathcal{L}_{\mathrm{MSD}}$-SC decoder again: Let $\mathcal{L}_{\mathrm{MSD}} \triangleq (\mathcal{L}_\infty \cup \mathcal{L}_3) \times$

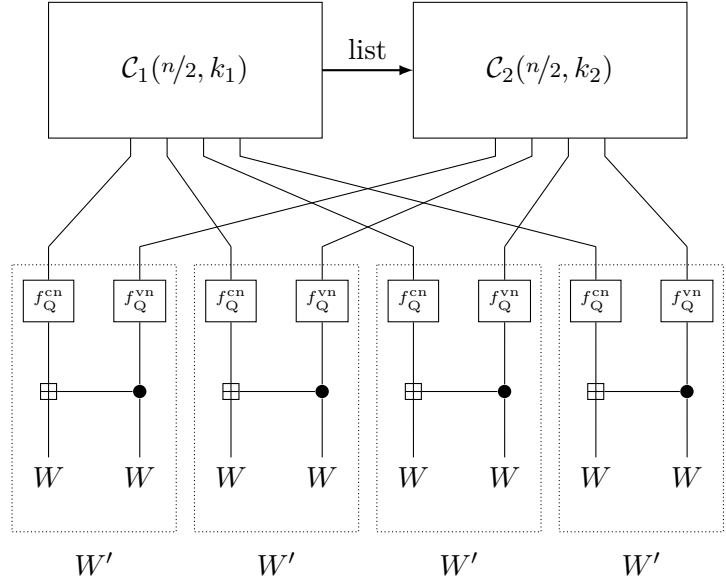

Figure 5.5.: 'Virtual' multi-level coding/multi-stage decoding system were two two-valued $W$ channel accesses are combined into one four-valued $W'$ super-channel access and two successive stages are used to decode $\mathcal{C}_1$ and $\mathcal{C}_2$

$\{\mathfrak{U}, \mathfrak{Q}\}$, where the first component contains the actual LLR value which can be from either $\mathcal{L}_\infty$ or $\mathcal{L}_3$, and the second component is a label $\{\mathfrak{U}, \mathfrak{Q}\}$ denoting whether the LLR value is still unquantized ($\mathfrak{U}$) or has already been quantized ($\mathfrak{Q}$). Consequently, check and variable node operations on the first (unquantized) decoding layer are defined as

$$\lambda_1 \boxplus \lambda_2 : \quad \left(\lambda_1^{(\mathrm{unq})}, \mathfrak{U}\right) \boxplus \left(\lambda_2^{(\mathrm{unq})}, \mathfrak{U}\right) \triangleq \left(f_{\mathrm{Q}(3,\delta_{\mathrm{cn}}^*)}\left(\lambda_1^{(\mathrm{unq})} \boxplus \lambda_2^{(\mathrm{unq})}\right), \mathfrak{Q}\right), \tag{5.4}$$

$$\lambda_1 \bullet \lambda_2 : \quad \left(\lambda_1^{(\mathrm{unq})}, \mathfrak{U}\right) \bullet \left(\lambda_2^{(\mathrm{unq})}, \mathfrak{U}\right) \triangleq \left(f_{\mathrm{Q}(3,\delta_{\mathrm{vn}}^*)}\left(\lambda_1^{(\mathrm{unq})} \bullet \lambda_2^{(\mathrm{unq})}\right), \mathfrak{Q}\right), \tag{5.5}$$

and on following (quantized) layers as

$$\lambda_1 \boxplus \lambda_2 : \quad \left(\lambda_1^{(\mathrm{q})}, \mathfrak{Q}\right) \boxplus \left(\lambda_2^{(\mathrm{q})}, \mathfrak{Q}\right) \triangleq \left(\lambda_1^{(\mathrm{q})} \boxplus \lambda_2^{(\mathrm{q})}, \mathfrak{Q}\right), \tag{5.6}$$

$$\lambda_1 \bullet \lambda_2 : \quad \left(\lambda_1^{(\mathrm{q})}, \mathfrak{Q}\right) \bullet \left(\lambda_2^{(\mathrm{q})}, \mathfrak{Q}\right) \triangleq \left(\lambda_1^{(\mathrm{q})} \bullet \lambda_2^{(\mathrm{q})}, \mathfrak{Q}\right), \tag{5.7}$$

where $\lambda_1, \lambda_2 \in \mathcal{L}_{\mathrm{MSD}}$. The combinations $(\mathfrak{U}, \mathfrak{Q})$ and $(\mathfrak{Q}, \mathfrak{U})$ cannot occur due to the structure of the decoding trees. The operations are defined such that unquantized LLR values fall back to the unquantized operations but get quantized after the first application, and quantized LLR values fall back to the quantized operations and stay quantized. The channel output LLR consists of the LLR of the unquantized BiAWGN together with the $\mathfrak{U}$ label,

$$\lambda_{\text{ch}} = f_{\text{LLR}}(y) \triangleq \left(\frac{2}{\sigma^2}y, \mathfrak{U}\right) \in (\mathcal{L}_\infty \times \{\mathfrak{U}\}). \tag{5.8}$$

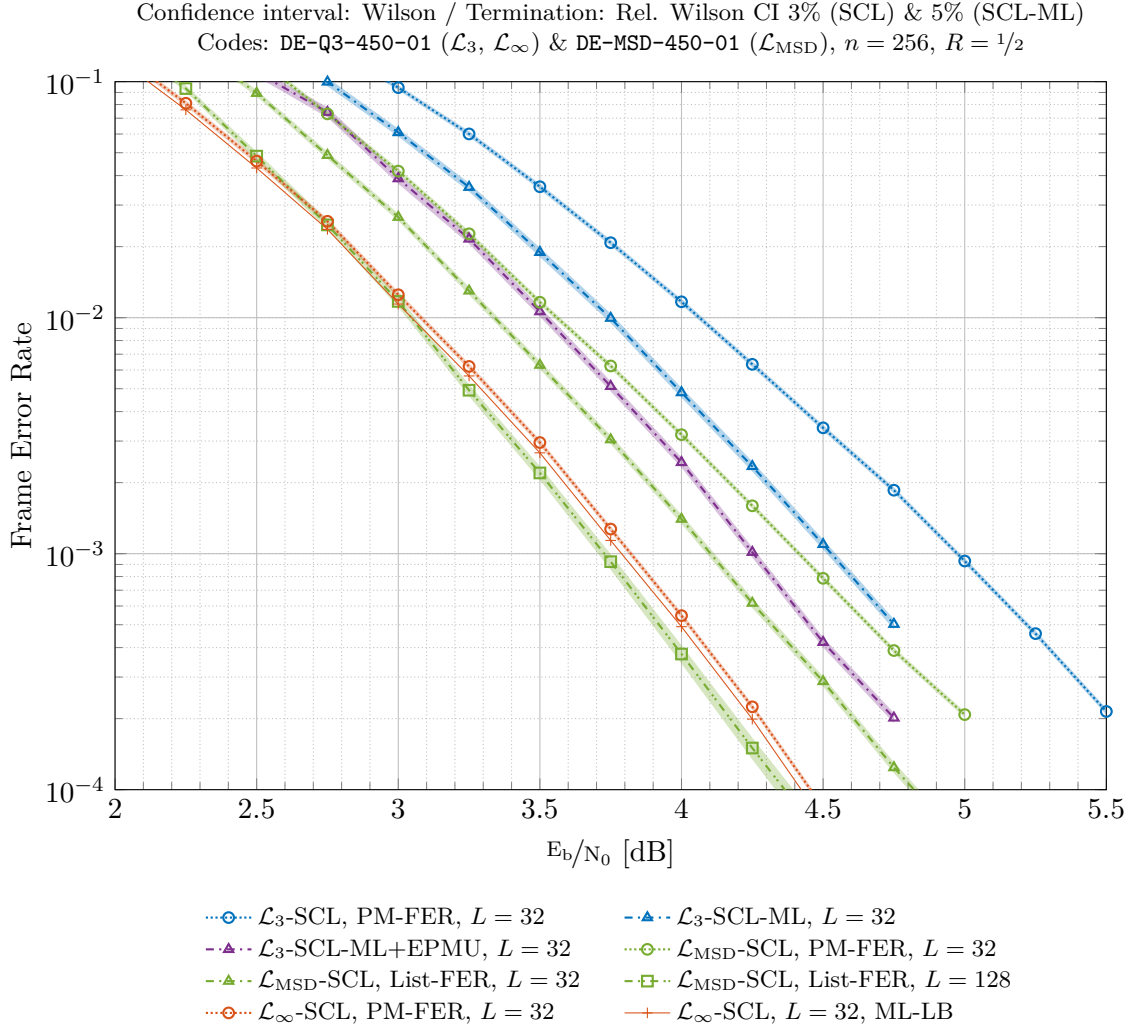

Figure 5.6.: PM-/LML-FER of $\mathcal{L}_\infty/\mathcal{L}_3/\mathcal{L}_{\text{MSD}}$-SCL-/-SCL-ML (varying list size $L$), applied to Q(3, $\delta^*_{\text{cap}}$)-BiAWGN($\sigma^2$) ($\mathcal{L}_3$, $\mathcal{L}_\infty$) or BiAWGN($\sigma^2$) ($\mathcal{L}_{\text{MSD}}$)

Simulation results for the 'first layer unquantized then 3-level quantized' scheme ($\mathcal{L}_{\text{MSD}}$-SCL) are shown in Figure 5.6. At FER $10^{-3}$ and list size $L = 32$ it gains 0.6 dB in $\text{E}_\text{b}/\text{N}_0$ over full 3-level quantization ($\mathcal{L}_3$-SCL) (cf. ···◦··· vs. ···◦···), leaving a gap of 0.6 dB in $\text{E}_\text{b}/\text{N}_0$ to the unquantized $\mathcal{L}_\infty$-SCL decoder (cf. ···◦··· vs. ···◦···). As such, $\mathcal{L}_{\text{MSD}}$-SCL performs midway between $\mathcal{L}_3$-SCL-ML with and without EPMU enhancement at FER $10^{-3}$ with $L = 32$ (cf. -·-▲-·- vs. ···◦··· vs. -·-▲-·-). In light of the observations in previous chapters and the gaps of 0.3 dB and 0.7 dB between PM-FER of $\mathcal{L}_{\mathrm{MSD}}$-SCL with $L = 32$ and its list-FER for $L = 32$ and $L = 128$, respectively, we suspect that further gains could be achieved using an ML-among-list step after SCL decoding (cf. ⋯▲⋯ vs. ⋯○⋯ and ⋯□⋯ vs. ⋯○⋯, respectively). This would bring $\mathcal{L}_{\mathrm{MSD}}$-SCL-ML close to the performance of the unquantized decoder. Note that $\mathcal{L}_{\infty}$-SCL saturates to its ML-LB for $L = 32$ and hence would not benefit from larger list sizes (cf. ⋯○⋯ vs. —+—).

While the 'first layer unquantized then 3-level quantized' scheme is again an extreme, it gives some indication of how much performance gain can be expected from choosing finer quantization for early decoding stages. For practical implementation one would perhaps use increasingly coarser quantization across the layers of the decoding tree.

Overall, we find our results robust to the scenarios examined in this chapter. In Section 5.1 we showed that EPMU can offer substantial gains in the low code rate regime which particularly suffers from quantization, according to previous studies. We showed in Section 5.2 using a 7-level quantized scheme that few quantization bits suffice for performance close to the unquantized decoder while still keeping low-complexity. The results of Section 5.3 suggest that selective use of finer quantization on lower layers of the decoding tree yields most of the performance benefits and that heterogeneous quantization across decoding stages should be considered to achieve the best cost-benefit tradeoff. In this chapter we occasionally bypassed the topic of quantization threshold selection which is treated extensively in the following chapter.

# 6. Quantization Threshold Selection

In previous sections we hint to the question of how to determine 'the most suitable' quantization. In this section we present and compare the results of optimizing three different criteria, namely channel capacity, a finite block length bound, and the union bound resulting from density evolution. We show that the suboptimality of quantized polar decoders is only accounted for in the last method. If the quantization parameters are determined according to this third method, the combination of channel with quantized LLRs and decoder with quantized messages exhibits lower error probabilities than when determined using the first or second method.

Recall briefly our notation introduced in Section 2.8: For the purposes of this thesis we restrict ourselves to mid-tread uniform quantization, *i.e.*, schemes with an odd number of levels that are spaced uniformly and symmetrically around and including 0. An $N$-level quantization scheme is associated with a set of labels $\mathcal{Q}_N \triangleq \left\{ \mathtt{0}, \pm \mathtt{1}, \dots, \pm \frac{\mathtt{N-1}}{\mathtt{2}} \right\} \subseteq \mathbb{Z}$ (denoted in monospaced font), $|\mathcal{Q}_N| = N$, each of which has a reconstruction value $x_q \triangleq 2\delta q$ according to the chosen quantization threshold $\delta$. Any $x \in \mathbb{R}$ is quantized to the nearest reconstruction value using the quantizer

$$f_{Q(N,\delta)} \colon \mathbb{R} \to \mathcal{Q}_N, x \mapsto \underset{q \in \mathcal{Q}_N}{\arg \min} |x - x_q|, \tag{6.1}$$

with ties broken towards 0. This results in decision boundaries at $\pm(2i-1)\delta$ for $i \in \left\{ 1, \dots, \frac{N-1}{2} \right\}$. For examples of 3-level and 7-level quantization, cf. Figure 2.15.

As the number $N$ of quantization levels is usually predetermined, the goal is to pick the quantization threshold $\delta$ that minimizes the block error probability after decoding, for a predefined polar code with information bits $\mathcal{I}$.

## 6.1. Maximization of Channel Capacity

An information theorist's first 'reflex' is perhaps to pick the quantization threshold that maximizes the capacity of the concatenation of the BiAWGN channel, the LLR computation, and the quantizer, cf. Figure 2.16. This intuition stems from the data processing inequality: information lost in an earlier step cannot be recovered in any later step. Following this intuition, the approach minimizes the bottleneck imposed by the resulting $\mathrm{Q}(N,\delta)$-BiAWGN$(\sigma^2)$ channel. However, it ignores the effects of the second step, the quantized decoder, which should not be neglected, as becomes clear in the course of the following analysis.

Under channel capacity maximization, the quantization threshold is chosen as

$$\delta_{\mathrm{cap}}^* = \arg\max_\delta \mathrm{I}\Big(\mathrm{Q}(N,\delta)\text{-BiAWGN}\big(\sigma^2\big)\Big). \tag{6.2}$$

Note that the $\mathrm{Q}(N,\delta)$-BiAWGN$(\sigma^2)$ channel is a discrete BMS channel whose channel law $P\big(\lambda_{\mathrm{ch}}^{(\mathrm{q})}\big|x\big)$ can be derived from $\sigma^2$, $N$ and $\delta$ using the cumulative density function of a Gaussian distribution, cf. Section 2.1. The mutual information is easily computed from the channel law. Note that w.l.o.g. $\delta \geq 0$. Furthermore, the optimum quantization threshold cannot be too large, otherwise almost all LLR values are quantized to `0` which is an unfavorable starting point for any subsequent decoder. Recall that

$$f_{\mathrm{LLR}}(y) = \frac{2}{\sigma^2} y \tag{6.3}$$

for a BiAWGN$(\sigma^2)$. Therefore,

$$p_{\Lambda_{\mathrm{ch}}^{(\mathrm{unq})}\big|X}(\,.\,|+1) \sim \mathcal{N}_{\mathbb{R}}\left(\frac{2}{\sigma^2}, \frac{4}{\sigma^2}\right). \tag{6.4}$$

As a result, it can be assumed that the optimum quantization threshold lies somewhere between 0 and a small integer multiple of $\frac{2}{\sigma^2}$. Hence, the RHS of (6.2) can be efficiently computed to arbitrary precision using grid search over $\delta$.

Is $\delta_{\mathrm{cap}}^*$ 'optimal'? The PM-FERs of applying $\mathcal{L}_7$-/$\mathcal{L}_\infty$-SCL (with varying list size $L \in \{1, 32\}$) to a $\mathrm{Q}(7,\delta)$-BiAWGN$(\sigma^2)$ (with $N = 7$ and varying quantization threshold $\delta \in \left\{\delta_{\mathrm{cap}}^*, \delta_{\mathrm{de}}^*\right\}$) are depicted in Figure 6.1. When using an unquantized decoder (cf. 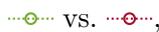 vs. 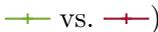, 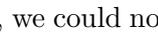 vs. 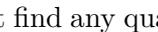), we could not find any quantization threshold that would have led to lower error probabilities. Using a 7-level quantized decoder however, quantization thresholds other than the channel capacity maximizing one can lead to lower error probabilities (cf. 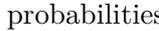 vs. 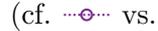, 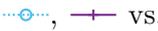 vs. 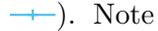). Note that we first made this observation by manually tuning $\delta$. Later, we devised a new method for picking $\delta$ which we present in Section 6.3 and whose results are reported in Figure 6.1.

For any $N$ and $\delta$, the $\mathrm{Q}(N,\delta)$-BiAWGN$(\sigma^2)$ belongs to the class of binary-input memoryless symmetric channels for which polar codes achieve capacity under (unquantized) SC decoding in the large block length regime [Arı09]. It is therefore not surprising that an unquantized decoder is able to keep up with the $\mathrm{Q}(N,\delta)$-BiAWGN$(\sigma^2)$ (at least it should for sufficiently large block lengths), such that the capacity of the concatenation of channel and decoder is dominated by the capacity of the channel only. Thus one should pick the channel capacity maximizing quantization threshold in order to maximize the capacity of the overall system. Quantized decoders on the other hand do not have this asymptotic 'optimality guarantee'. And in fact, from the previous observations we can conclude that their suboptimality (and thus impact on the performance of the overall system) needs to be taken into account when determining the quantization threshold.

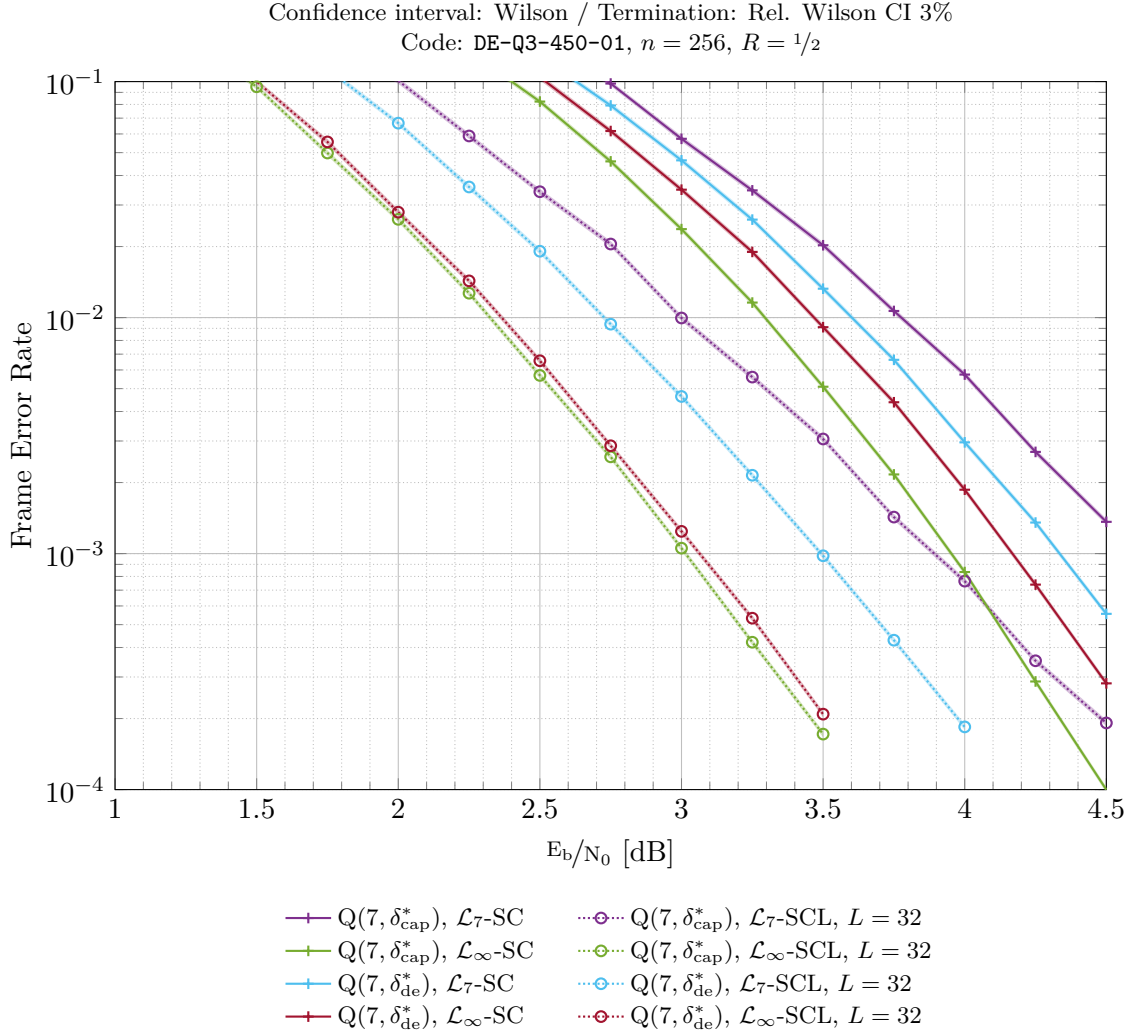

Figure 6.1.: FER/PM-FER of $\mathcal{L}_7$-/$\mathcal{L}_\infty$-SC/-SCL (varying list size $L \in \{1, 32\}$) applied to $Q(7, \delta)$-BiAWGN($\sigma^2$) (varying quantization threshold $\delta \in \left\{\delta_{\mathrm{cap}}^*, \delta_{\mathrm{de}}^*\right\}$)

Note that strictly speaking the aforementioned argument only holds in the large block length regime, not for fairly short block length $n = 256$ as in Figure 6.1. In the next section we provide evidence that the observed is not due to short block length effects, and in the section thereafter we show how to pick $\delta$ to improve the performance of quantized decoders despite their suboptimality.

## 6.2. Maximization of a Finite Block Length Bound

In the finite short and medium block length regime the so called *normal approximation* [Str62; PPV10; Pol10; TT13; Ers16] provides better guidance than the asymptotic Shannon capacity regarding which information rates can be sustained by a channel $P_{Y|X}$ with input distribution $P_X$ at a given block length $n$ and block error probability $\mathsf{P_{e,B}}$. Let

$$\mathsf{i}(x,y) \triangleq \log \frac{P_{X,Y}(x,y)}{P_X(x)P_Y(y)} = \log \frac{P_{Y|X}(y|x)}{P_Y(y)} \qquad \text{'information density'} \qquad (6.5)$$

denote the information conveyed through the channel in an instance where $x$ was transmitted and $y$ was received. The normal approximation takes into account not only the average information content of a channel output symbol

$$\mathsf{C}\big(P_{Y|X}\big) = \mathsf{E}[\mathsf{i}(X,Y)] = \mathsf{I}(X;Y), \qquad \text{'channel capacity'} \qquad (6.6)$$

but also the variability of the information content

$$\mathsf{V}\big(P_{Y|X}\big) = \mathsf{Var}[\mathsf{i}(X,Y)]. \qquad \text{'channel dispersion'} \qquad (6.7)$$

The latter captures the effect that not all channel outputs might be equally informative about the channel input. In the large block length regime, these fluctuations in information density level out, but for a short block it might as well happen that overproportionally many channel outputs for a particular block are little informative about the input, causing a block error and drop in information rate.

Using the normal approximation, the sustainable rate is characterized [PPV10; Ers16, eq. (1)] as

$$\mathsf{R}\big(P_{Y|X}, n, \mathsf{P_{e,B}}\big) \approx \mathsf{C}\big(P_{Y|X}\big) - \sqrt{\frac{\mathsf{V}\big(P_{Y|X}\big)}{n}} \, Q^{-1}(\mathsf{P_{e,B}}), \qquad (6.8)$$

where $Q^{-1}(.)$ is defined in (2.1). For finite $n$, the second term on the RHS of (6.8) captures the effect of channel variability; for $n \to \infty$, this term vanishes as expected.

For a discrete channel $P_{Y|X}$ with input distribution $P_X$ both $\mathsf{C}\big(P_{Y|X}\big)$ and $\mathsf{V}\big(P_{Y|X}\big)$ are easily calculated, allowing to approximate $\mathsf{R}\big(P_{Y|X}, n, \mathsf{P_{e,B}}\big)$ as in (6.8). This can be

done for any $Q(N,\delta)$-BiAWGN$(\sigma^2)$ as a function of $N$ and $\delta$. Maximizing this finite block length bound leads to the following design rule for the quantization threshold,

$$\delta_{\mathrm{fbl}}^* = \arg\max_{\delta} \mathsf{R}\Big(Q(N,\delta)\text{-BiAWGN}\big(\sigma^2\big), n, \mathsf{P}_{\mathrm{e,B}}\Big). \tag{6.9}$$

As argued before, the optimum $\delta$ lies somewhere between 0 and a small integer multiple of $\frac{2}{\sigma^2}$. Hence, the RHS of (6.9) can be evaluated using grid search over $\delta$.
However, for our parameters of interest ($n \in \{128, 256\}, N \in \{3, 7\}, \mathsf{P}_{\mathrm{e,B}} = 10^{-3}$),

$$\mathsf{R}\Big(Q(N,\delta_{\mathrm{cap}}^*)\text{-BiAWGN}\big(\sigma^2\big), n, \mathsf{P}_{\mathrm{e,B}}\Big) \approx \mathsf{R}\Big(Q(N,\delta_{\mathrm{fbl}}^*)\text{-BiAWGN}\big(\sigma^2\big), n, \mathsf{P}_{\mathrm{e,B}}\Big), \tag{6.10}$$

and thus $\delta_{\mathrm{cap}}^* \approx \delta_{\mathrm{fbl}}^*$. We conclude that finite block length effects likely not account for (all) the observed suboptimality of the capacity maximization rule. In the next section we show how the suboptimality of the quantized decoder can be taken into account in determining the quantization threshold $\delta$.

## 6.3. Minimization of the Union Bound from Density Evolution

Recall from Figure 2.16 that given the quantization parameters $N$ and $\delta$ and the channel parameter $\sigma^2$, we can derive the channel law $P\big(\lambda_{\mathrm{ch}}^{(\mathrm{q})}\big|x\big)$ of the $Q(N,\delta)$-BiAWGN$(\sigma^2)$. Under the all-zero codeword assumption we obtain the channel output distribution $P\big(\lambda_{\mathrm{ch}}^{(\mathrm{q})}\big) = P\big(\lambda_{\mathrm{ch}}^{(\mathrm{q})}\big|+1\big)$. Density evolution (cf. Section 2.5) using the appropriate check node and variable node operations employed by an $\mathcal{L}$-SC decoder (for some LLR alphabet $\mathcal{L}$, cf. Section 3.1) yields the distributions $P\big(\lambda_i^{(\mathrm{q})}\big)$, of LLRs computed by the decoder for decoding the $i$-th bit $u_i$. The 'probability of erroneously decoding the $i$-th bit', $\mathsf{P}_{\mathrm{e}}\Big(P_{\Lambda_i^{(\mathrm{q})}}\Big)$, is given as

$$\mathsf{P}_{\mathrm{e}}\Big(P_{\Lambda_i^{(\mathrm{q})}}\Big) \triangleq \mathsf{Pr}\Big[\big\{\Lambda_i^{(\mathrm{q})} < 0\big\}\Big] + \frac{1}{2}\mathsf{Pr}\Big[\big\{\Lambda_i^{(\mathrm{q})} = 0\big\}\Big], \tag{6.11}$$

since under all-zero codeword assumption the decoder errs if either the LLR suggests $\hat{u}_i = 1$ (negative LLR) or the LLR is undecided (LLR zero) and thus the decoder flips a coin. Recall from Section 2.6.1 the following union upper bound on the block error probability $\mathsf{P}_{\mathrm{e,B}}$ [MT09, eq. (3)],

$$\mathsf{P}_{\mathrm{e,B}}^{(\mathrm{ub})}\Big(Q(N,\delta)\text{-BiAWGN}\big(\sigma^2\big)\Big) \triangleq \sum_{i \in \mathcal{I}} \mathsf{P}_{\mathrm{e}}\Big(P_{\Lambda_i^{(\mathrm{q})}}\Big) \geq \mathsf{P}_{\mathrm{e,B}}, \tag{6.12}$$

where $P_{\Lambda_i^{(\mathrm{q})}}$ resulting from density evolution is implicitly a function of $N$, $\delta$ and $\sigma^2$.

Assuming that the union bound $\mathsf{P}_{e,B}^{(ub)}$ is at least a reasonably good proxy for the actual block error probability $\mathsf{P}_{e,B}$ w.r.t. its dependency on $\delta$, we choose the quantization threshold via density evolution as

$$\delta_{de}^* = \arg\min_{\delta} \mathsf{P}_{e,B}^{(ub)}\Big(Q(N,\delta)\text{-BiAWGN}\big(\sigma^2\big)\Big). \tag{6.13}$$

Again, the RHS of (6.13) is evaluated using grid search over $\delta$.

The quantization threshold obtained from minimizing the union bound computed via density evolution can lead to sizable improvements in the performance of both $\mathcal{L}$-SC and $\mathcal{L}$-SCL decoders, cf. Figure 6.1 for $\mathcal{L}_7$. For $L = 32$, we observed a 0.4 dB gain in $\mathsf{E}_b/\mathsf{N}_0$ at $\mathsf{P}_{e,B} = 10^{-3}$ between $\delta_{de}^*$ and $\delta_{cap}^*$ (cf. —⊙— and —⊙—).

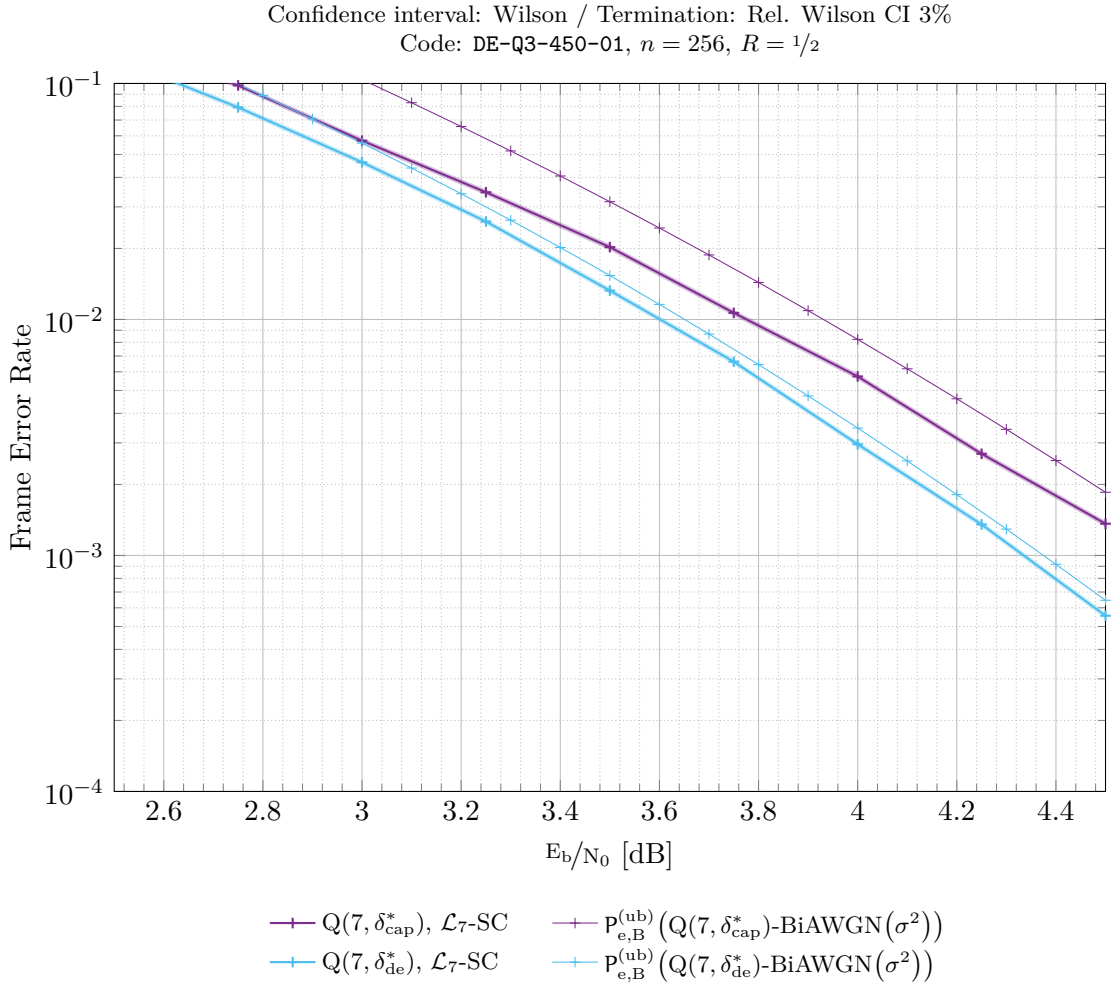

Figure 6.2.: FER and union upper bound on block error probability $\mathsf{P}_{e,B}^{(ub)}$ of $Q(7,\delta)$-BiAWGN$(\sigma^2)$ with varying $\delta$ decoded using $\mathcal{L}_7$-SC

Note from Figure 6.2 that the union upper bound $P_{e,B}^{(ub)}$ tracks the trend of the error probability well but does not lie on top of it (cf. ─+─ vs. ─+─, ─+─ vs. ─+─).

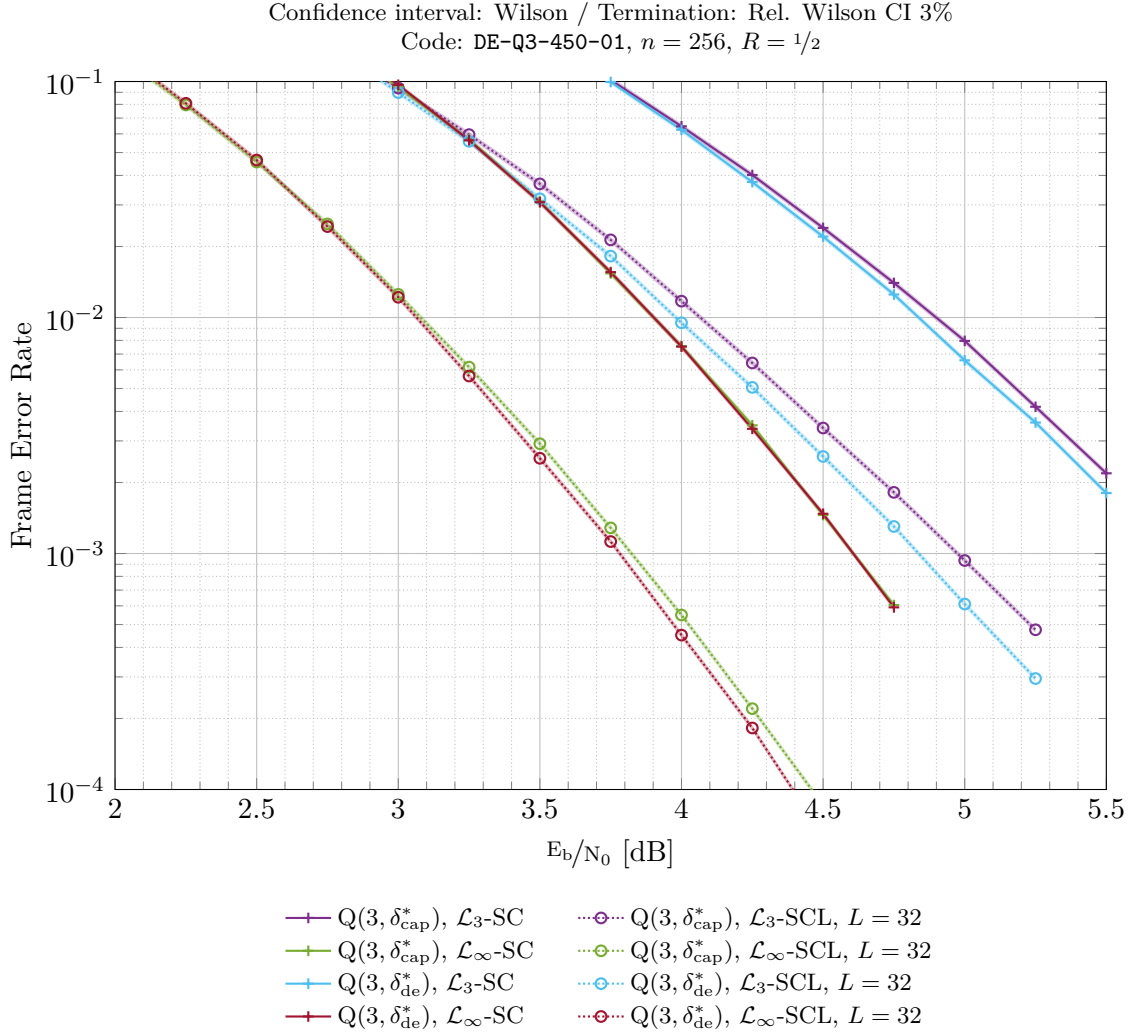

Figure 6.3.: FER/PM-FER of $\mathcal{L}_3$-/$\mathcal{L}_\infty$-SC/-SCL (varying list size $L \in \{1, 32\}$) applied to Q(7, $\delta$)-BiAWGN($\sigma^2$) (varying quantization threshold $\delta \in \left\{\delta_{\mathrm{cap}}^*, \delta_{\mathrm{de}}^*\right\}$)

Figure 6.3 shows the performance of $\mathcal{L}_3/\mathcal{L}_\infty$-SCL with $L \in \{1, 32\}$ and quantization thresholds $\delta \in \left\{\delta_{\mathrm{cap}}^*, \delta_{\mathrm{de}}^*\right\}$ determined using capacity maximization and union bound minimization, respectively. While $\delta_{\mathrm{de}}^*$ yields some minor improvements over $\delta_{\mathrm{cap}}^*$ for $\mathcal{L}_3$-SCL with $L = 32$ (cf. ─○─ vs. ─○─), overall the difference in $P_{e,B}$ for $\delta_{\mathrm{de}}^*$ and $\delta_{\mathrm{cap}}^*$ is rather small for 3-level quantized schemes. And certainly so if compared to the impact on 7-level quantized schemes (cf. Figure 6.1 vs. 6.3). This observation together with the fact that $\delta_{\mathrm{de}}^*$ needs to be optimized for every code separately using density evolution is the reason and justification why we use the capacity maximizing quantization threshold $\delta_{cap}^*$ throughout this work for 3-level quantized systems.

In this chapter we compared three different approaches to determining quantization thresholds, maximization of channel capacity (Section 6.1), of a finite block length bound (Section 6.2), and minimization of the union bound obtained from density evolution (Section 6.3). Thresholds determined using the last approach can lead to considerable gains because it takes into account the suboptimality of quantized decoders. Finite block length effects could not account for the observed suboptimality of the capacity maximization approach.

# 7. Conclusion

Our work leads to insights and recommendations for communication system design practitioners (Section 7.1), as well as interesting directions for future research (Section 7.2).

## 7.1. Recommendations for Practitioners

Based on our findings, we offer practitioners the following key take-away messages:

- The cost-benefit tradeoff, between complexity/cost/energy consumption and error correcting performance, of small-list fine-quantization vs. large-list coarse-quantization is worth careful examination.

- LLR quantization causes PM quantization in $\mathcal{L}$-SCL, which has numerous negative effects because decisions based on quantized PMs do not preserve the ML criterion. A subsequent ML-among-list step (as in $\mathcal{L}$-SCL-ML) can be used to reduce the gap between PM-FER and list-FER, but the list-FER remains poor. Without ML-among-list, larger list sizes can negatively impact the error correcting performance.

- Expected path metric updates can be used to improve the list-FER. 'Unorthodox' indicators of instantaneous reliability of bit estimates, such as counting contradiction events, can refine statistical reliability knowledge at a complexity strictly less than that of finer LLR quantization.

- Three-level quantization is perhaps 'extreme' for implementation. Our observations corroborate that as little as three quantization bits can suffice to closely match the performance of unquantized decoders. Furthermore, early decoding steps benefit more from finer quantization than later steps for which coarse quantization suffices. This could be utilized to maximize cost-benefit.

- In determining quantization thresholds, the suboptimality of quantized decoders needs to be taken into account. Density evolution provides the techniques to do so. Capacity maximization of the resulting quantized channel alone is insufficient.

- Using a contemporary scientific programming language (such as Julia [Bez+17]) with a sufficiently sophisticated concept of data types enables versatile implementation and comprehensive code reuse.

## 7.2. Directions for Future Research

The following research directions are not covered in this thesis and left for future work:

- 'Contradiction counting at variable nodes' is but one of many low-complexity measures for instantaneous reliability of bit estimates that can be used to enhance expected path metric updates. Others that come to mind immediately are, *e.g.*, 'double erasure counting at check nodes' or 'clipping events counting at variable nodes'. Insights are still missing as to which of these indicators are informative and hence improve performance, and why.

- Density evolution is the primary tool used for analyzing and designing expected path metric updates. Due to fine discretization of probability distributions, the computational complexity explodes as more low-complexity instantaneous reliability measures are added. More work on compact representations or approximations of the involved distributions is required to render the analysis tractable. The commonly used Gaussian approximation can be a starting point for this inquiry.

- Polar codes could be designed specifically for enhanced quantized decoders (which employ, *e.g.*, EPMU/EPMUCC). As these modifications concern list decoding in particular, density evolution cannot be the primary design tool. Instead, the genetic algorithm we sketched could be of use, in particular for small block lengths. Furthermore, lookup tables for the check node and variable node operations in quantized decoders could be subject to optimization as well.

- What can be learned from the codes designed using the genetic algorithm about how to design codes for parameters for which no Reed-Muller code exists?

- 'Traditionally', CRCs are used to guide an SCL decoder in the selection of the correct codeword from its list at the end of the decoding process. At this point, the correct codeword might already have been accidentally removed from the list. Several approaches exist that try to 'spread' the protecting effect across the decoding process [WQJ16; TM16; Yua+18]. This would lead to better decisions as to which paths to keep in the list and which to remove, similar to EPMU. We have not investigated this effect for quantized decoders, and how the two techniques compare and/or complement/substitute each other.

# A. List Decoding on the Binary Erasure Channel

Polar SC and SCL decoding on the binary erasure channel (BEC) has the property that all LLR values occurring in the decoder are $-\infty$, $+\infty$, or 0. Hence, the decoder is inherently 3-level quantized. This links the study of polar decoding on the BEC to our topic. Furthermore, if the list size $L$ of the SCL decoder is chosen sufficiently large, then SCL essentially implements ML decoding. Recall that ML decoding on the BEC can be performed in number of operations proportional to $n^3$ via Gaussian elimination.

The interplays of BEC, SCL with finite list size, SCL with unbounded list size (which is equivalent to ML), and ML are interesting and might enable further analysis and insights into the behavior of SCL for channels other than the BEC. In this chapter, we briefly document an experimental observation we made in the course of our work.

Throughout the chapter, we consider polar codes with rate $R = \frac{1}{2}$ for different block lengths $n = 2^m$, $m \in \{3, 5, 6, 8\}$. All are designed for the BEC with erasure probability $\varepsilon_{\mathrm{design}} = \frac{3}{8}$ using density evolution. All frozen bits are set to 0. Due to channel and decoder symmetry, we assume w.l.o.g. that the all-zero codeword is transmitted. In this case, all LLRs $\lambda_i$ are either 0 or $\infty$, where 0 means that the decoder is uncertain about the bit's value ('erased bit'), and $\infty$ means that the bit is certainly 0. Hence, under the all-zero codeword assumption, the LLRs $\lambda_i$ in the genie-aided SC decoder (which is assumed by density evolution, cf. Sections 2.5 and 2.6.1) are distributed as

$$\Lambda_i = \begin{cases} 0 & \text{w.p.} \quad \varepsilon_i, \\ \infty & \text{w.p.} \quad 1 - \varepsilon_i, \end{cases} \tag{A.1}$$

where $\varepsilon_i$ is obtained via density evolution. After resolving the all-zero codeword assumption, every $i$-th bit is associated with a $\mathrm{BEC}(\varepsilon_i)$. Hence, the LLR $\Lambda_i$ reflects the correct bit value with probability $1 - \varepsilon_i$ and is erased with probability $\varepsilon_i$. Recall the mutual information and capacity of the BEC under uniform input,

$$\mathrm{C}(\mathrm{BEC}(\varepsilon_i)) = \mathrm{I}(\mathrm{BEC}(\varepsilon_i)) = 1 - \varepsilon_i. \tag{A.2}$$

The mutual information $\mathrm{I}(\mathrm{BEC}(\varepsilon_i))$ of the synthetic channels resulting from polarization for transmission over $\mathrm{BEC}(\varepsilon_{\mathrm{channel}})$, with $m = 6$ and $\varepsilon_{\mathrm{channel}} = \varepsilon_{\mathrm{design}} = \mathrm{{}^3\!/\!_8}$, is plotted in Figure A.1 for information bits (□) and frozen bits (■).

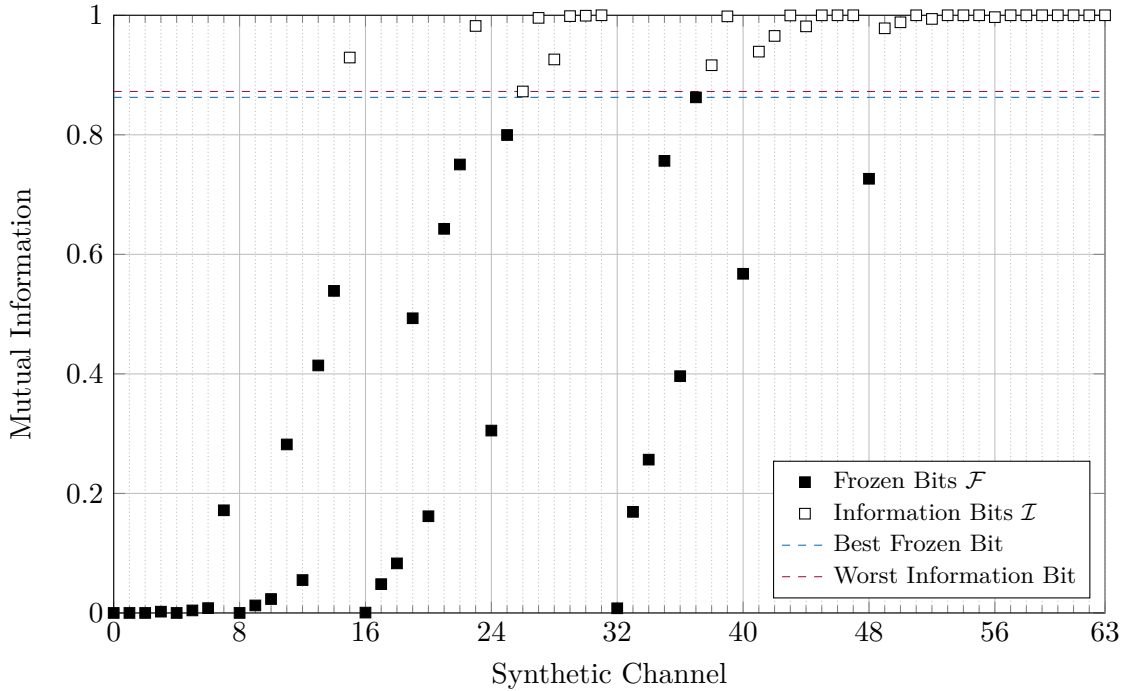

Figure A.1.: Mutual information $\mathrm{I}(\mathrm{BEC}(\varepsilon_i))$ of synthetic channels for transmission over $\mathrm{BEC}(\varepsilon_{\mathrm{channel}})$, with $m = 6$ and $\varepsilon_{\mathrm{channel}} = \varepsilon_{\mathrm{design}} = \mathrm{{}^3\!/\!_8}$

It is a well-known fact that for $n \to \infty$, a fraction $\mathrm{C}(\mathrm{BEC}(\varepsilon_{\mathrm{channel}}))$ of synthetic channels has mutual information $\mathrm{I}(\mathrm{BEC}(\varepsilon_i)) \approx 1$, and a fraction $1 - \mathrm{C}(\mathrm{BEC}(\varepsilon_{\mathrm{channel}}))$ has mutual information $\mathrm{I}(\mathrm{BEC}(\varepsilon_i)) \approx 0$. Based on the notion of mutual information of synthetic channels, visualized in Figure A.1, we define the *mutual information loss*,

$$\mathrm{I}_{\mathrm{loss}}(\mathcal{I}, \varepsilon_{\mathrm{channel}}) \triangleq \sum_{i \in \mathcal{I}} (1 - \mathrm{I}(\mathrm{BEC}(\varepsilon_i))) = \sum_{i \in \mathcal{I}} \varepsilon_i, \tag{A.3}$$

which intuitively measures the information (about the input to a $\mathrm{BEC}(\varepsilon_{\mathrm{channel}})$) 'lost' because of information bits that have not fully polarized due to finite $n$. Note that $\mathrm{I}(\mathrm{BEC}(\varepsilon_i))$ has an implicit dependency on $\varepsilon_{\mathrm{channel}}$. After sorting the synthetic channels, the mutual information loss corresponds to the area between the mutual information of the information bits and the horizontal line at the ideal mutual information 1. This is illustrated in Figure A.2, for $m = 6$ and $\varepsilon_{\mathrm{channel}} = 0.4$.

In the following, we run an SCL decoder with unbounded list size $L = \infty$. Hence, we

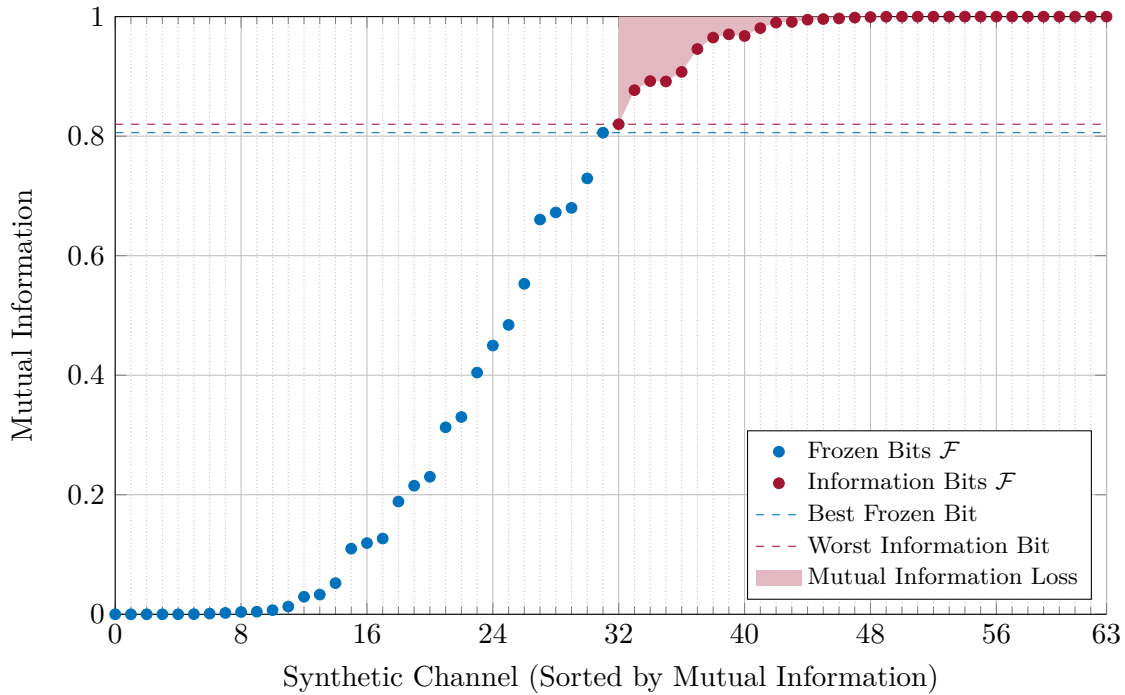

Figure A.2.: Mutual information loss of information bits for transmission over BEC($\varepsilon_{\text{channel}}$), with $m = 6$ and $\varepsilon_{\text{channel}} = 0.4$

'reuse' the symbol $L$ from now on: The random variable $L_i$ denotes the number of paths in the SCL decoder's list before decoding the $i$-th bit, $l_i$ is the corresponding realization, $L_n$ denotes the length[1] of the list after decoding the last $((n-1)$-th) bit. Again, we assume w.l.o.g. that the all-zero codeword was transmitted. The decoder starts with a list of length $l_0 = 1$ containing only the empty path. For each $i$-th bit, one of the following three events occurs in the SCL decoder:

$\mathcal{E}_{\text{E}}$: For all paths, $\lambda_i = 0$.

$\mathcal{E}_{\text{A}}$: For all paths, $\lambda_i = \infty$.

$\mathcal{E}_{\text{H}}$: For half of the paths, $\lambda_i = +\infty$, and for the other half of the paths, $\lambda_i = -\infty$.

Here, $\mathcal{E}_{\text{E}}$, $\mathcal{E}_{\text{A}}$, and $\mathcal{E}_{\text{H}}$ are reminiscent of *erased*, *all* and *half*, respectively. If $i \in \mathcal{I}$, the SCL decoder proceeds as follows:

$\mathcal{E}_{\text{E}}$: All paths are duplicated, one copy continues with $\hat{u}_i = 0$, and the other copy continues with $\hat{u}_i = 1$. We call this a *branching event*.

---
[1] *List length* refers to the actual number of paths in the list, while conventionally *list size* is the maximum number of paths and in the scenario at hand infinity.

$\mathcal{E}_A$: All paths continue with $\hat{u}_i = 0$.

$\mathcal{E}_H$: Half of the paths, those with $\lambda_i = +\infty$, continue with $\hat{u}_i = 0$, and the other half of the paths, those with $\lambda_i = -\infty$, continue with $\hat{u}_i = 1$.

Note that for information bits $i \in \mathcal{I}$, in the case of $\mathcal{E}_E$ the list length doubles ($l_{i+1} = 2l_i$), in the case of $\mathcal{E}_A$ and $\mathcal{E}_H$ the list length is unaltered ($l_{i+1} = l_i$). For $i \in \mathcal{F}$, the SCL decoder proceeds as follows:

$\mathcal{E}_E$: All paths continue with the frozen bit value, $\hat{u}_i = 0$.

$\mathcal{E}_A$: All paths continue with the correctly decoded frozen bit value $\hat{u}_i = 0$.

$\mathcal{E}_H$: Half of the paths, those with $\lambda_i = -\infty$, receive an infinitely large penalty on their path metric. This reflects the fact that these paths are invalid due to contradicting bit values. Hence, the paths with $\lambda_i = -\infty$ are removed from the list, while the paths with $\lambda_i = +\infty$ continue with $\hat{u}_i = 0$. We call this a *consolidation event*.

Note that for $i \in \mathcal{F}$, in the case of $\mathcal{E}_E$ and $\mathcal{E}_A$ the list length is unaltered ($l_{i+1} = l_i$), in the case of $\mathcal{E}_H$ the list the list length is halved ($l_{i+1} = \frac{1}{2}l_i$). An immediate consequence of the above is that the list length $l_i$ for all $i$ is a non-negative power of 2.

Let $B$ be a random variable denoting the number of branching events that occur during a run of the SCL decoder with unbounded list size. Hence, $\mathsf{E}[B]$ is the number of branching events averaged over many independent runs of the SCL decoder. Interestingly,

$$\mathrm{I}_{\mathrm{loss}}(\mathcal{I}, \varepsilon_{\mathrm{channel}}) \triangleq \sum_{i \in \mathcal{I}}(1 - \mathrm{I}(\mathrm{BEC}(\varepsilon_i))) = \mathsf{E}[B], \tag{A.4}$$

where $\mathsf{E}[B]$ and $\mathrm{I}(\mathrm{BEC}(\varepsilon_i))$ implicitly depend on $\varepsilon_{\mathrm{channel}}$, and $\mathsf{E}[B]$ also on $\mathcal{I}$. This is an empirical observation based on the simulation results depicted in Figure A.3.

Three aspects are worth pointing out: *First*, in hindsight, the observation is not very surprising for the BEC. ML decoding (and thus also SCL decoding with unbounded list size) on the BEC cannot produce a wrong codeword. Rather, a decoding error is declared if the codeword cannot be determined uniquely, *i.e.*, the final list has length $l_n > 1$. The true codeword is always in the list. Under the all-zero codeword assumption, this means the all-zero path is always in the list. Density evolution can be used to obtain the erasure probabilities $\varepsilon_i$ of $\Lambda_i$ for that path, and since either all paths decode to $\lambda_i = 0$ or no path does so, $\varepsilon_i = \mathsf{Pr}[\mathcal{E}_{E,i}]$. Let $B_i$ be a random variable indicating whether a branching event occurs in the $i$-th decoding step. Clearly,

$$B_i = \begin{cases} 1 & \text{w.p.} \quad \mathsf{Pr}[\mathcal{E}_{E,i}] = \varepsilon_i, \\ 0 & \text{w.p.} \quad 1 - \mathsf{Pr}[\mathcal{E}_{E,i}] = 1 - \varepsilon_i, \end{cases} \qquad \text{and} \qquad B = \sum_{i \in \mathcal{I}} B_i. \tag{A.5}$$

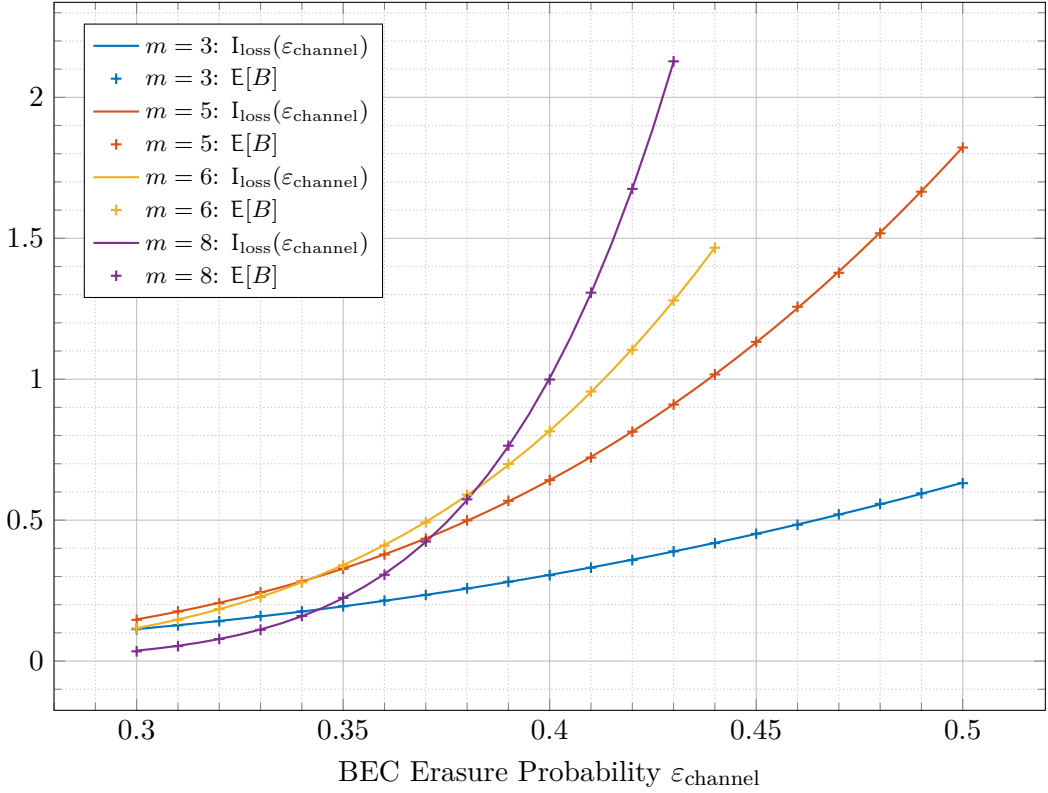

Figure A.3.: Mutual information loss $I_{\text{loss}}(\varepsilon_{\text{channel}})$ equals average number of branching events $\mathtt{E}[B]$ for SCL decoding with unbounded list size on the $\text{BEC}(\varepsilon_{\text{channel}})$

Hence,

$$\mathtt{E}[B] = \sum_{i \in \mathcal{I}} \mathtt{E}[B_i] = \sum_{i \in \mathcal{I}} \varepsilon_i = I_{\text{loss}}(\mathcal{I}, \varepsilon_{\text{channel}}). \tag{A.6}$$

*Second*, however, the observation, and in particular the formulation as an information loss, prompts new research questions. It is intuitive that on average $I_{\text{loss}}(\mathcal{I}, \varepsilon_{\text{channel}})$ hypotheses need to be tested to resolve an ambiguity of $I_{\text{loss}}(\mathcal{I}, \varepsilon_{\text{channel}})$ bits about the input. In that sense, a branching event is nothing but posing a hypothesis test. Is there any equivalent of this intuition for channels other than the BEC?

*Finally*, while branching events are characterized rather easily, this seems not to be the case for consolidation events. A reason is that the probability for a consolidation depends on the list length in the respective situation. To see this, consider that consolidations can only occur if $l_i > 1$. Hence, density evolution for the all-zero path is likely insufficient to analyze this phenomenon. A characterization of consolidations would yield a full characterization of SCL decoding (with unbounded list size) on the BEC. Subsequently, new insights on SCL decoding with finite list size could be derived as well.

XIII